\documentclass[preprint, 3p, authoryear]{article} 
\usepackage[a4paper,left=2.5cm, right=2.5cm, top=2.5cm, bottom=2.5cm]{geometry}
\usepackage[hyphens]{url}
\usepackage{makecell}
\usepackage{amssymb}
\usepackage{csquotes}
\usepackage{amsmath}
\usepackage{graphicx}
\usepackage{tabularx}
\usepackage{enumitem}
\usepackage{pdfpages}
\usepackage{authblk}

\usepackage[
    backend=biber,
    uniquename=false,
    style=authoryear,
    sorting=nyt,
    maxcitenames=2
]{biblatex}
\usepackage{xcolor}
\usepackage{fancyhdr}
\definecolor{WNEcolor}{HTML}{C00000}

\newlist{questions}{itemize}{1}
\setlist[questions]{label=\textbf{Q:}}

\title{Informer In Algorithmic Investment Strategies on High Frequency Bitcoin Data}
\date{\vspace{-5ex}}

\author[1]{Filip Stefaniuk}
\author[2]{Robert Ślepaczuk}
\affil[1]{\small University of Warsaw, Faculty of Economic Sciences, Ul. Długa 44/50, 00-241 Warsaw, Poland, ORCID: https://orcid.org/0009-0004-4968-8704, email: filip.stefaniuk@gmail.com}
\affil[2]{\small University of Warsaw, Faculty of Economic Sciences, Department of Quantitative Finance and Machine Learning, Quantitative Finance Research Group, Ul. Długa 44/50, 00-241 Warsaw, Poland, ORCID: https://orcid.org/0000-0001-5527-2014, Corresponding author: rslepaczuk@wne.uw.edu.pl }

\begin{document}

\maketitle
\begin{abstract}
    The article investigates the usage of Informer architecture for building automated trading strategies for high frequency Bitcoin data. Three strategies using Informer model with different loss functions: Root Mean Squared Error (RMSE), Generalized Mean Absolute Directional Loss (GMADL) and Quantile loss, are proposed and evaluated against the Buy and Hold benchmark and two benchmark strategies based on technical indicators. The evaluation is conducted using data of various frequencies: 5 minute, 15 minute, and 30 minute intervals, over the 6 different periods. Although the Informer-based model with Quantile loss did not outperform the benchmark, two other models achieved better results. The performance of the model using RMSE loss worsens when used with higher frequency data while the model that uses novel GMADL loss function is benefiting from higher frequency data and when trained on 5 minute interval it beat all the other strategies on most of the testing periods. The primary contribution of this study is the application and assessment of the RMSE, GMADL and Quantile loss functions with the Informer model to forecast future returns, subsequently using these forecasts to develop automated trading strategies. The research provides evidence that employing an Informer model trained with the GMADL loss function can result in superior trading outcomes compared to the buy-and-hold approach.
\end{abstract}

\section{Introduction}
Developing automated trading strategies is a challenging task that has captivated both institutional investors and individual researchers for a long time. With the rise of computing power and development of machine learning (ML), more and more automated algorithms are being developed and deployed into the market. It is estimated that currently 70\% to 80\% of all market transactions are carried out by automated trading software \parencite{Yadav2015} and this number is expected to increase in the coming years.

Although numerous systems built in the past have handled daily market data, this is insufficient in the age of High-Frequency Trading and advanced computing power. This is especially true in the context of trading assets such as Bitcoin. Bitcoin (BTC) is a decentralized digital currency that allows peer-to-peer transactions over the Bitcoin network, which is a secure, public ledger using blockchain technology. It was first introduced in an anonymous 2009 white paper (\cite{btc2009}), by a person or group of people using the pseudonym Satoshi Nakamoto. Since then, Bitcoin has grown in popularity and many cryptocurrency exchanges that facilitate bitcoin sales and purchases  have emerged. This asset is known for its high volatility, for example, in 2017 the Bitcoin price went up nearly 20 times, reaching \$19,497 and then dropping by 84\%, in 2021 Bitcoin reached a new all-time high of \$63,314 but then the price fell to \$34,770,  almost 55\%\footnote{https://blog.obiex.finance/6-biggest-bitcoin-crashes-that-have-happened-in-crypto-history/}. Since then, Bitcoin has become somewhat more stable, especially after approval of Bitcoin Spot ETFs, although price movements are still greatly exaggerated compared to more classical investment instruments. 

The study explores the idea of building an automated trading strategy for Bitcoin. Five strategies are proposed and evaluated on the historical Bitcoin data of high frequencies: 5 minutes, 15 minutes, and 30 minutes; from a period of 21.08.2019 to 24.07.2024. The first two strategies are treated as benchmarks and are based on classical technical indicators, namely the Moving Average Convergence Divergence (MACD) and Relative Strength Index (RSI). The other three employ the Informer \parencite{zhou2021informerefficienttransformerlong}, a state-of-the-art attention-based neural network model designed to efficiently handle long time series, to predict the returns and subsequently choose positions according to the model's forecasts.

The work aims to answer the following research questions:
\begin{questions}
    \item \textit{Is it possible to create an algorithmic strategy for trading Bitcoin, that is more efficient than the Buy\&Hold approach?}
    \item \textit{Does signal from Informer model allow to create strategies that are more efficient on trading Bitcoin than strategies based on technical indicators?}
    \item \textit{How does selection of the machine learning model loss function influence the strategy performance?}
    \item \textit{Does usage of higher frequency data allow to create more efficient strategies?} 
\end{questions}
Answering those questions required designing and implementing a method to compare various trading strategies. Such comparison is a main contribution of this study. To the best of current knowledge, no other research has yet been performed where an Informer model is trained with the Quantile or GMADL \parencite{michankow2024102375} loss function, followed by the utilization of its forecasts in buy/sell signals generation to develop automated trading strategies. The study involves an exhaustive analysis of the approach including the comparison to the benchmark strategies: buy-and-hold and two technical indicator-based strategies, usage of data with different time intervals and multiple time periods. Finally, a sensitivity analysis is conducted to show how changing the parameters affect the perfomance of the tested strategies.  An additional contribution of the research is an open sourced implementation of framework for efficient comparison of trading strategies, that is available on Gitlab and enables reproducing the results\footnote{https://gitlab.com/FilipStefaniuk/wne-msc-thesis}.

The structure of the thesis is as follows: Chapter \ref{chapter:review} surveys the associated literature and gives a brief overview of prior similar studies. Chapter \ref{chapter:data} explains the acquisition, preprocessing, and covers the analysis of the datasets used in the study. In Chapter \ref{chapter:methodology}, the notion of trading strategy is formally defined, the metrics for comparison are introduced, and the details of each examined strategy are outlined. Chapter \ref{chapter:experiments} displays the experimental results, detailing the selection of strategy hyperparameters and presenting the outcomes of the strategy evaluation along with sensitivity analysis. Lastly, Chapter \ref{chapter:conclusion} draws the conclusions of the study.

\section{Literature Review}\label{chapter:review}
The study of financial market efficiency has a rich history. For years, researchers have been trying to develop algorithms capable of predicting the price movements of financial assets as well as creating automated trading strategies that rely on such signals. The concept of developing an automated system capable of regularly outperforming the market is so attractive that it leads to the publication of thousands of papers on that topic annually.
\subsection{Efficient Market Hypothesis}
These endeavors stand in contrast to the Efficient Market Hypothesis (EMH), also known as the efficient market theory, which argues that the stock prices already reflect all accessible information. EMH suggests that it is impossible to consistently achieve higher returns than average market returns, thus the best investing strategy is a passive portfolio. The hypothesis was formulated in three forms: \textit{weak}, \textit{semi-strong}, and \textit{strong}. The \textit{weak} form challenges the foundation of technical analysis by suggesting that stock prices move in a random manner. The \textit{semi-strong} form argues that stock prices immediately incorporate all public information, arguing for the ineffectiveness of the fundamental analysis. The \textit{strong} form extends this argument, implying that the price of the stock includes both public and private information. The EMH, even in its weakest form, implies that the unpredictability of short-term price fluctuations makes efforts to create automated trading systems ineffective \parencite{Malkiel1973}. 

The EMH was acknowledged in the early review of theoretical and empirical market models by \textcite{fama1970}. The author put forth certain evidence indicating the presence of a statistically significant positive correlation in daily returns, suggesting that such patterns might be exploited to develop profitable trading strategies. However, they ultimately concluded that this evidence might not be robust enough to challenge the validity of the EMH. This research was continued in \textcite{RePEc:bla:jfinan:v:46:y:1991:i:5:p:1575-617}, which was published two decades later. In this publication, the author presented additional evidence for predicting daily and weekly returns based on historical returns, and the findings align with those of the earlier study. 

The argument in favor of EMH was presented in \textcite{malkiel2005}. The publication demonstrated that over a long period, professional investment managers do not outperform their index benchmarks. The study showed that in a 20-year period 90\% of managed index funds, was outperformed by the S\&P 500 index, asserting that the only valid long-term strategy is the passive portfolio. 

Conversely, the most persistent criticisms of the EMH relate to the preferences and actions of market participants. Psychologists and experimental economists recorded numerous specific behavioral biases that are prevalent in human decision making when faced with uncertainty. They argue that irrational behaviors such as overconfidence \parencite{fischoff78,RePEc:oup:qjecon:v:116:y:2001:i:1:p:261-292., odean2001}, overreaction \parencite{28e39847-87a7-3356-8a2d-ed0ea430ab56}, loss aversion \parencite{RePEc:ecm:emetrp:v:47:y:1979:i:2:p:263-91,RePEc:ucb:calbrf:rpf-269}, herding \parencite{RePEc:bla:jfinan:v:56:y:2001:i:1:p:387-396}, and psychological accounting \parencite{doi:10.1126/science.7455683} are predictable and can be utilized to develop profitable trading strategies that exceed passive benchmarks. 

Despite extensive research, the validity of the EMH continues to be debated without a definitive conclusion. More modern approaches explore the idea of time-varying weak-form market efficiency (\cite{RePEc:bla:jecsur:v:25:y:2011:i:1:p:69-108}) and the Adaptive Markets Hypothesis (\cite{RePEc:eee:empfin:v:18:y:2011:i:5:p:868-879}) arguing that the predictability of markets can change over time and depend on various conditions such as the composition of market participants.

\subsection{Early Statistical Approach}
Nevertheless, a number of statistical systems and techniques that attempt to analyze and forecast asset prices have been developed. The Box-Jenkins method applies autoregressive moving average (ARMA) or autoregressive integrated moving average (ARIMA) models to find the best fit of a time series model to past values of a time series \parencite{boxjen76}. The model combines autoregression, differencing, and moving averages to predict stock prices, and despite its simplicity, it has been widely applied and successful in financial forecasting. Another method called the exponential smoothing model (ESM) employs the exponential window function to smooth the data in which exponential smoothing parameters are estimated \parencite{billah2006}. Furthermore, those approaches were often combined with several other classical Machine Learning non-linear models. 

In the study by \textcite{ZHANG2003159}, authors combined the ARIMA model with Artificial Neural Networks (ANNs) to predict the prices of pairs of currencies. Similar approach was further explored in 
\textcite{KHASHEI20112664}, in which authors use ARIMA-ANN hybrid model to further improve predictions on the aforementioned currencies datasets. In \textcite{PAI2005497}, the authors conducted a comparison between the ARIMA model and Support Vector Machines (SVMs) across diverse stock prices, demonstrating that both methodologies successfully forecasted the prices. 
In \textcite{WANG2012758} the authors developed a hybrid approach that uses ESM, ARIMA, and ANN to predict daily prices of the Dow Jones Industrial Average and Shenzhen index. \textcite{ayo2014} utilized the ARIMA model in data sets from the New York and Nigeria Stock Exchanges, illustrating its effectiveness in forecasting short-term prices, and compared its results with ANN. \textcite{azari2018} illustrated the effectiveness of the ARIMA model in forecasting Bitcoin's price by examining its predictions on a time series over a three-year period. However, they admitted that the approach was unable to predict abrupt price fluctuations, such as volatility observed in late 2017.
\textcite{nguyen2022} utilized a combination of ARIMA and GARCH family models to predict S\&P500 log returns, employing these predictions to develop a trading strategy. Their findings indicated that strategies incorporating hybrid models outperformed those relying only on the ARIMA model.

\subsection{Early Machine Learning Systems}
Advancements in computing power lead to wider adoption of ML models and development of more complex neural network architectures. Models such as Convolutional Neural Networks (CNNs) \parencite{lecun_deep_2015} or Recurrent Neural Networks (RNNs) (and their variants Long-Short Term memory Networks (LSTMs) \parencite{10.1162/neco.1997.9.8.1735} and Gated Recurrent Units(GRUs) \parencite{DBLP:journals/corr/ChoMGBSB14}) were developed. Initially, they were used mainly in the computer vision and natural language processing domains, but they got quickly adapted and successfully implemented for financial forecasting. 

\textcite{Chen_2015} modeled and predicted China stock market daily returns with LSTM, showing it outperforms previous classic statistical approaches.
\textcite{honchar2016} compared how different RNN architectures are able to forecast price movements. They used Google stock price index data and applied multi-layer RNN, LSTM and GRU, to predict prices at different time horizons. The study showed that LSTM outperformed other architectures for long sequences. In the other publication \textcite{Persio_2017}, the same authors compared Multi Layer Perceptron (MLP), CNN with LSTM architectures for predicting daily S\&P500 returns and minute-by-minute FOREX EUR/USD returns. They reported that CNN can model financial time series better than other architectures.
\textcite{Fischer_2017} utilized an LSTM model for selecting S\&P 500 stocks in portfolio construction. The data came from 1992 until 2015, and the goal was formulated as a classification problem. The researches reported that the LSTM outperformed classical ML methods like random forest and logistic regression classifier. Moreover, they attempted to disentangle the black-box of LSTMs predictions by analyzing the selected stocks. They discovered that the selected stocks consist mainly of stocks with below-mean momentum, strong short-term reversal characteristics, and high volatility. \textcite{RePEc:war:wpaper:2021-23}
applied several classical Machine Learning algorithms to technical analysis indicators for the WIG20, DAX, S\&P 500, and a few selected CEE indices. The findings of the study demonstrate that quantitative methods outperform passive strategies when considering risk-adjusted returns, with the Bayesian Generalized Linear Model and Naive Bayes emerging as the leading models for the examined indices.

\textcite{Selvin_2017} predicted prices for ~2000 NSE listed companies with  LSTM, RNN, and CNN models. They identified CNN as the best model that turned out to be best in predicting the changes in trends. \textcite{Nelson_2017} used LSTM in combination with technical analysis indicators. They reported an average accuracy of 55.9\% when predicting if the price of a particular stock is going to rise or not in the near future. \textcite{hossain2018} proposed model integrating LSTM and GRU architectures to predict S\&P 500 prices.  In this method, the input data is initially fed into the LSTM network to produce a preliminary prediction, after which the output from the LSTM layer is forwarded to the GRU layer for generating the final prediction. The research demonstrates that the suggested hybrid model surpasses the performance of standalone LSTM and GRUs. 
\textcite{Siami‐Namini_2018} examined how deep learning-based time series forecasting algorithms such as LSTM compare to traditional-based algorithms such as the ARIMA model. The average reduction in error rates obtained by LSTM was found to be between 84 and 87 percent compared to ARIMA, indicating the superiority of LSTM over ARIMA.
Similarly \textcite{RePEc:war:wpaper:2020-27} evaluated the effectiveness of investment strategies based on traditional models and an LSTM on the S\&P 500 index time series, spanning 20 years of data from 2000 to 2020. The combination of signals from several methods doubled the returns on the same level of risk of the passive strategy benchmark. The study, however, concluded that the LSTM model exhibited a considerably lower robustness to parameter variations compared to traditional methods.

\subsection{Modern Machine Learning Systems}
Although RNNs became widely adopted and proved to be effective for sequential data, the architecture struggled with processing longer sequences due to exploding and vanishing gradient phenomena \parencite{9142152}. In order to overcome this limitation, the Attention mechanism \parencite{bahdanau2014neural, luong-etal-2015-effective} was developed. 

While initially it was used in combination with the RNNs, \textcite{DBLP:journals/corr/VaswaniSPUJGKP17} introduced a neural network architecture based solely on the Attention mechanism called Transformer. The newly introduced architecture exhibited a significant enhancement in the performance, leading to its widespread adoption across multiple domains. Nowadays, it lies at the heart of state-of-the-art large-scale, multimodal models such as GPT-4 \parencite{openai2024gpt4technicalreport} or Gemini \parencite{geminiteam2024geminifamilyhighlycapable}. 

The time series forecasting domain also adopted the new architecture. The \textcite{benidis_deep_2023} collects general purpose state-of-the-art time-series forecasting architectures developed in recent years. Among the first of the modern deep forecasting models is DeepAR \parencite{salinas2019deeparprobabilisticforecastingautoregressive} which still is primarily based on RNNs. The model outputs parameters of a previously chosen family of distributions. Samples from this distribution can be fed back into the model during prediction. However, most of the other modern models fully adapt the new Transformer architecture. 
Temporal Fusion Transformer (TFT) developed by \textcite{lim2020temporalfusiontransformersinterpretable}, incorporates Transformer architecture with novel components for embedding static covariates, performing “variable selection”, and gating components that skip over irrelevant parts of the context. The TFT is trained to predict forecast quantiles, and promotes forecast interpretability by modifying self-attention and learning input variable importance.
\textcite{DBLP:journals/corr/abs-2009-14799} proposed MQ-Transformer, a Transformer architecture that employs attention mechanisms in the encoder and decoder separately, and consider learning positional embeddings from event indicators. The authors discussed improvements not only on forecast accuracy but also on excess forecast volatility where their model improves over the state-of-the-art. 
\textcite{zhou2021informerefficienttransformerlong} proposed the Informer, a computationally efficient Transformer architecture, that specifically targets applications with long forecast horizons. These robust novel architectures can be specifically utilized for forecasting financial time series, particularly when dealing with higher-frequency data. 

\textcite{Barez_2023} explored the application of deep learning Transformers architectures for high-frequency Bitcoin-USDT log-return forecasting and compared them to the LSTM model. The results indicate that the model based on Transformer achieves a higher cumulative PnL than the LSTM when trading with multiple signals during backtesting. \textcite{Hu_2021}
applied TFT for stock price predictions comparing it with LSTM and again showing the superiority of Transformer model. \textcite{Zhao_2022} investigated the prediction capability of the Transformer model on Bitcoin and Ethereum price data, furthermore augmenting the model with sentiment data collected from Twitter. Similarly \textcite{Hájek_2024} integrated news sentiment and investor attention metrics with a Temporal Fusion Transformer framework. The effectiveness of the model was demonstrated by analyzing stock price data for 41 of the largest market capitalization companies over the period from 2010 to 2021. 

\textcite{Penmetsa_2023} analyzed the effectiveness of deep learning to predict  Bitcoin, Ethereum and Litecoin prices, by integrating momentum and volatility technical indicators into network's input. They conducted a study using LSTM and Transformer architectures, concluding that Transformers tend to outperform LSTM models in price prediction and trends in cryptocurrency data. \textcite{Wang_2023} introduced hybrid approach combining Transformer architecture with bi-LSTM. The authors argued that Transformer is better at obtaining long distance information, while LSTM can better capture short distance signals. They presented the effectiveness of their method using 5 index stocks and 14 Shanghai and Shenzhen stocks. 

The Informer architecture was used by \textcite{10280785} to predict intra-day Chinese stock price. They also did a comparison with LSTM model showing that Informer outperformed the RNN network. Another such comparison was done by \textcite{Duan_2024} who systematically compared the effectiveness of Informer and LSTM model for price predictions concluding that while in general Informer outperforms LSTM, the fusion of the two combines the advantages of both models and  enhances the prediction accuracy. Lastly \textcite{lu2023stockmarketindexprediction} designed three experiments to compare Informer with the commonly used networks of LSTM, Transformer and BERT \parencite{DBLP:journals/corr/abs-1810-04805} on 1-minute and 5-minute frequencies for four different stocks and market indices. The prediction results are measured by three evaluation criteria: MAE, RMSE and MAPE. Informer obtained the best performance among all the networks on every dataset.

\vspace{0.5cm}
The chapter provided an overview of the historical development of research aimed at predicting financial asset movements. It emphasized several significant publications within this continually expanding area of study. Although financial forecasting has consistently held popularity, recent years have witnessed an intensified focus on predictions conducted at higher frequencies, aided by the utilization of increasingly sophisticated systems akin to the one presented in this study.

\section{Data}\label{chapter:data}
The data used in the research consider the BTC/USDT cryptocurrency pair from a period of 21.08.2019 to 24.07.2024 (5 years). Bitcoin (BTC) is a decentralized digital currency that allows peer-to-peer transactions. Tether (USDT) is a type of stablecoin, with a value that closely mirrors that of the USD. It actively works to keep its valuation stable through market mechanisms: each Tether issued is backed by one US dollar worth of assets. The pair closely mirrors the behaviour of BTC/USD however it has the higher trading volume, as it is used by investors who want to hedge against the inherent volatility of their cryptocurrency investments while still keeping value within the crypto market.

\begin{figure}[h]
    \centering
    \caption{Price of the BTC/USDT}
    \label{fig:data-btcusdt}
    \includegraphics[width=\linewidth]{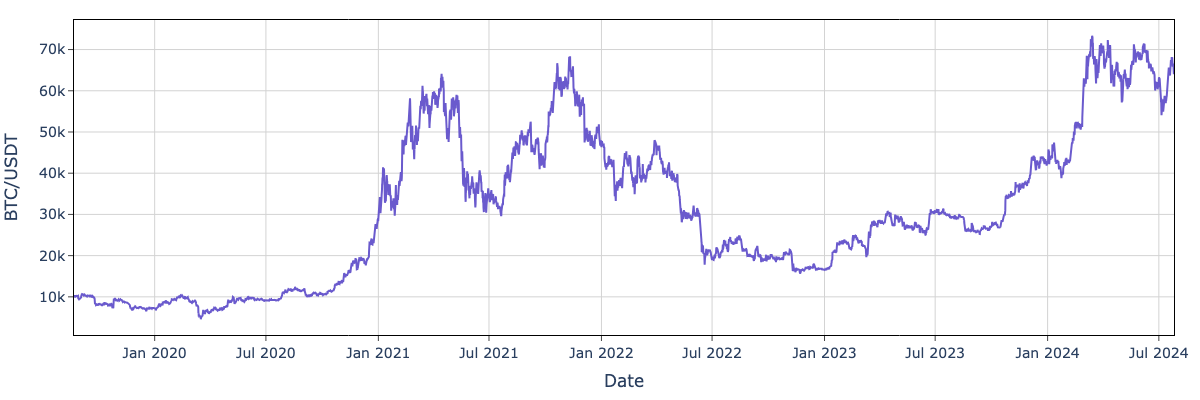}
    \raggedright \tiny Note: Price of BTC/USDT cryptocurrency pair in a period from 21.08.2019 to 24.07.2024. The data was obtained from https://www.binance.com/en-NG/landing/data.
\end{figure}

\noindent
The selection of this data for the research was driven by the following motivations:
\begin{itemize}
    \item Availability of high quality historical data. Many cryptocurrency exchanges provide free access to historical data, with fine-grained intervals, even up to 1 second. The data is easily accessible through either a dedicated API or can be downloaded as csv files. This is a significant difference compared to availability of historical stock prices, which are not easily accessible even for hourly interval data, and usually the access requires a fee. Additionally, this concrete pair BTC/USDT was one of the earliest pairs available on the exchanges, so it has the longest historical data available. 

    \item Cryptocurrency exchanges operate continuously, 24 hours a day, 7 days a week. Hence, there is no need for special handling of holidays and overnight price fluctuations during stock market closures. The historical price data forms a consistent time series with equally spaced intervals.
    
    \item Bitcoin stands as the most established cryptocurrency, with the BTC/USDT pair being the most widely traded and one of the highest in trading volume \footnote{https://coinranking.com/exchange/-zdvbieRdZ+binance/markets}, which makes it more liquid and stable compared to the other cryptocurrency pairs.
\end{itemize}

The data was obtained through the Binance API\footnote{https://www.binance.com/en-NG/landing/data}. Binance is one of the largest global cryptocurrency exchanges that provides a platform to trade various cryptocurrencies. It was established in 2017 and offers a comprehensive range of services, including trading, listing, fundraising, and delisting or withdrawal of cryptocurrencies. The API allows downloading the so-called k-line data, the candlestick data of various intervals, consisting of: \texttt{open time}, \texttt{close time}, \texttt{open price}, \texttt{high price}, \texttt{low price}, \texttt{close price} and \texttt{volume}. The research uses k-line intervals of \texttt{5min}, \texttt{15min} and \texttt{30min}.

After downloading, the data was checked for missing data points. Even though exchange operates continuously, there were maintenance windows and the exchange sometimes was down. 8 gaps in the data were detected, with missing data of over 101 hours (4.2 days) in total. This constitutes to 0.002 of all data points in the period, thus it shouldn't have significant impact into the experiment. For the simplicity sake and to make the time series consistent, the gaps were filled by copying the price and volume values from the last available data point.
Then, returns from each interval are computed as:

\begin{equation}\label{eq:ret}
\texttt{returns} = \frac{\texttt{close price} - \texttt{open price}}{\texttt{open price}}
\end{equation}

Table \ref{tab:data-stats} presents the descriptive statistics of the BTC/USDT returns for all three intervals. We can observe, that the standard deviation of the returns grows with the interval length, which is expected. The minimum and maximum returns for each interval are quite alike, suggesting that significant price fluctuations occur abruptly and swiftly. 
Interestingly, the maximum return was observed for the \texttt{15min} interval, indicating that 
\begin{table}[h!]
	\begin{center}
        \caption{Descriptive statistics of BTC/USDT}
            \vspace{0.3cm}
		\begin{tabular}{lrrr}
			\textbf{Statistic} & \makecell{\textbf{BTC/USDT} \\ \textbf{5min}} & \makecell{\textbf{BTC/USDT} \\ \textbf{15min}} & \makecell{\textbf{BTC/USDT} \\ \textbf{30min}} \\
			\hline
			count & 518400 & 172800 & 86400 \\
			mean & 0.0000060 & 0.0000176 & 0.0000346 \\
			std & 0.0021843 & 0.0036712 & 0.0050768 \\
			\hline
			min & -0.1022537 & -0.1191688 & -0.1662924 \\
			25\% percentile & -0.0007716 & -0.0013018 & -0.0017677 \\
			50\% percentile & 0 & 0.0000094 & 0.0000280 \\
			75\% percentile & 0.0007855 & 0.0013443 & 0.0018575 \\
			max & 0.1842885 & 0.2262878 & 0.1460125 \\
			\hline
			kurtosis & 203.12 & 140.22 & 58.61 \\
			skewness & 0.57 & 0.97 & -0.29 \\
			\hline
			KS test stat. & 0.49 & 0.49 & 0.49 \\
			KS test p-value & 0.00e+00 & 0.00e+00 & 0.00e+00 \\
			\hline
                \multicolumn{4}{p{.7\textwidth}}{\tiny Note:  Descriptive statistics of returns with intervals of \texttt{5min}, \texttt{15min} and \texttt{30min}. The statistics are not annualized. Null hypothesis of Kolmogorov-Smirnov (KS) test is that the distribution is normal.}
		\end{tabular}
	\end{center}
        \label{tab:data-stats}
\end{table}
the price must first have quickly increased and then dropped (in the span of 30 min). The distributions are leptokurtic, with kurtosis significantly higher than the normal distributions, that increases with smaller intervals. In all cases skewness is between -1 and 1 which indicates a relatively symmetrical distribution. Futhermore, in all three cases Kolmogorov-Smirnov test for normality has a test statistic equal to 0.49 and a p-value of 0.0. Therefore, a null hypothesis that the data is normally distributed is rejected.
The given analysis underscores the significance of investigating and creating trading systems that function in shorter time intervals. This is because rapid and abrupt price movements, which are critical for effective trading strategies, would go unnoticed by algorithms that operate on longer intervals.

\subsection{Additional Data}\label{section:additional_data}
The research utilizes data collected over several years, encompassing periods with varying market conditions and sentiments. To account for this, the data was enhanced with additional information that attempts to capture this information. 

Additional data that was added composes of:

    \paragraph{The Cboe Volatility Index (VIX Index)}\hspace{-3.5mm}\footnote{https://www.cboe.com/tradable\_products/vix/vix\_historical\_data/}, a benchmark index that measures the market’s expectation of future volatility. The VIX Index is based on options of the S\&P 500 Index and is considered the leading indicator of the broad U.S. stock market. The VIX Index is recognized as the world’s best gauge of U.S. equity market. The daily close value of the VIX index is used.
    \begin{figure}[h!]
    \centering
    \caption{Cboe Volatility Index (VIX)}
    \label{fig:data-vix}
    \includegraphics[width=\linewidth]{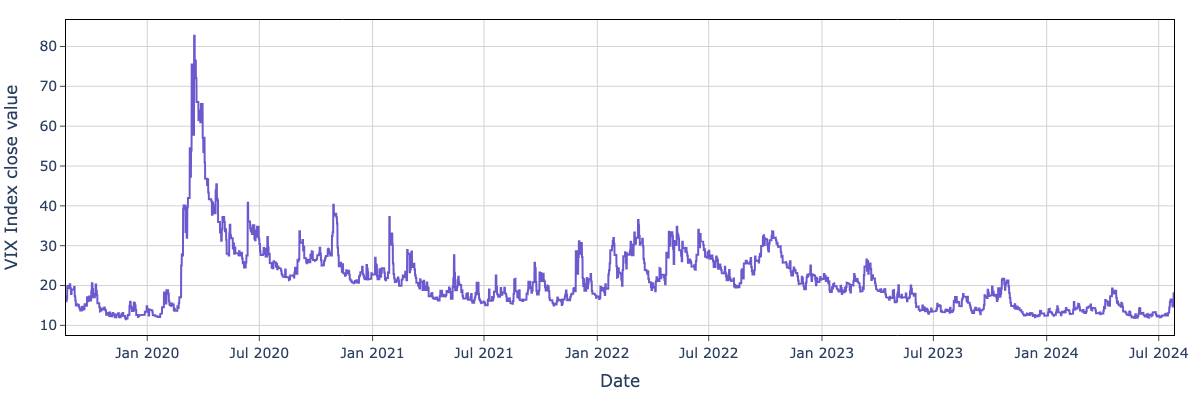}
    \raggedright \tiny Note: Cboe Volatility Index (VIX) value in a period from 21.08.2019 to 24.07.2024. Data was obtained from https://www.cboe.com/tradable\_products/vix/vix\_historical\_data/
    \end{figure}
    
    \paragraph{The Federal Funds effective rates}\hspace{-3.5mm}\footnote{https://fred.stlouisfed.org/series/FEDFUNDS}, that is the interest rate at which depository institutions trade federal funds (balances held at Federal Reserve Banks) with each other overnight. The federal funds rate indirectly influences longer-term interest rates such as mortgages, loans, and savings. The data was available at the monthly frequency.
    \newpage
     \begin{figure}[h!]
    \centering
    \caption{Note: The Federal Funds effective rates}
    \label{fig:data-fedfunds}
    \includegraphics[width=\linewidth]{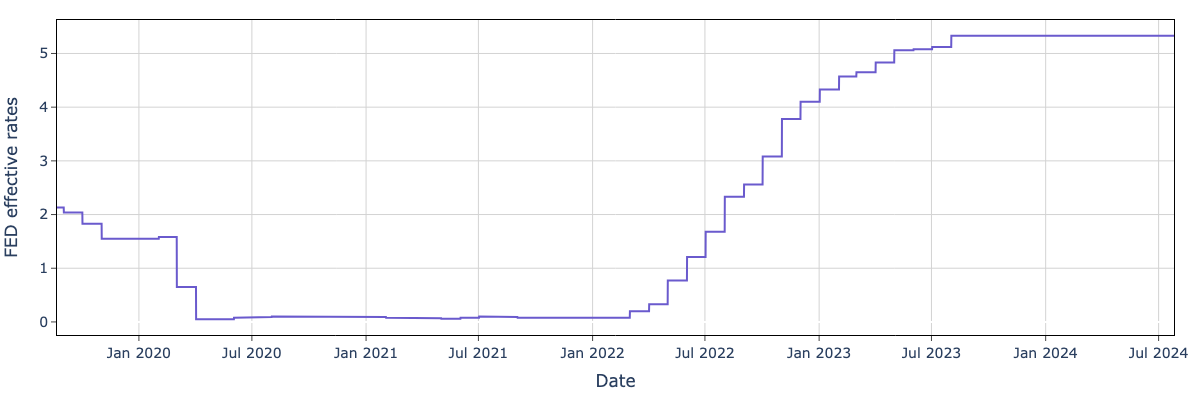}
    \raggedright \tiny Note: The federal funds effective rates in a period from 21.08.2019 to 24.07.2024. Data was obtained from https://fred.stlouisfed.org/series/FEDFUNDS
    \end{figure}
    
    \paragraph{Crypto Fear/Greed index}\hspace{-3.5mm}\footnote{https://alternative.me/crypto/fear-and-greed-index/}, which is founded on elements such as volatility, market momentum, reddit sentiment analysis, Google trends data and public surveys. The index is available at the daily frequency.
     \begin{figure}[h!]
    \caption{The Crypto Fear/Greed Index}
    \label{fig:data-feargreed}
    \centering
    \includegraphics[width=\linewidth]{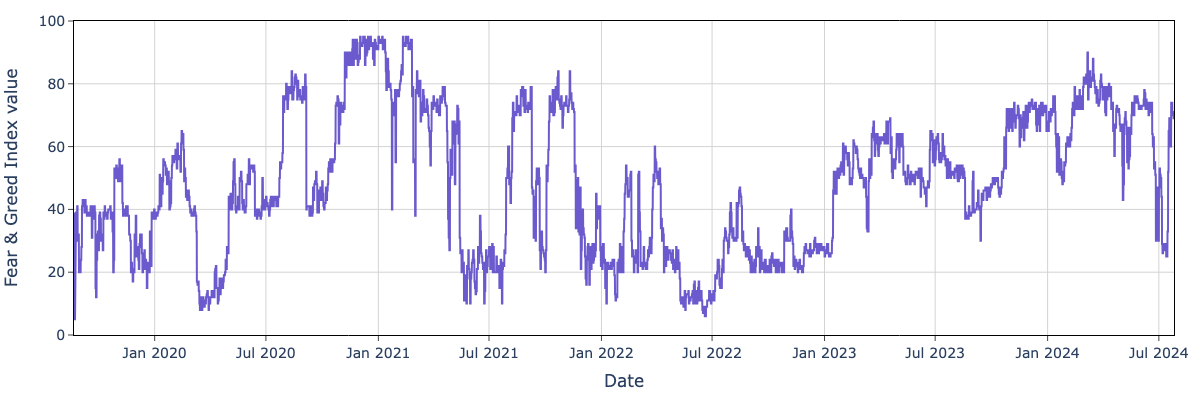}
    \raggedright \tiny Note: The Crypto Fear/Greed Index value in a period from 21.08.2019 to 24.07.2024. Data was obtained from https://alternative.me/crypto/fear-and-greed-index/
    \end{figure}

The aforementioned data was collected at a much lower frequency (daily/monthly) compared to the BTC/USDT data. Consequently, for every observation, a last known value had to be assigned, e.g. a \texttt{5min} observation with \texttt{open time} of 2022-08-17 04:05:00 was enriched with value of VIX Index and Crypto Fear/Greed Index from 2022-08-16 and effective rates value from 2022-07. Note that the values from the previous day/month are used as it is unknown at what hour/day the value was published. Using the information from the same day/month carries a potential risk of using the information that wasn't available at the time. 
\newpage
\subsection{Data Windows}
The research follows the approach of \cite{s22030917}, where strategies are independently evaluated on a rolling window that passes through the testing period. A rolling window consists of an \textit{in sample} part of the 24 months (2 years) and \textit{out of sample} part of 6 months. Each next window is moved forward by the length of the \textit{out of sample} part. In the end, the results of each subperiod are combined to provide testing results for the whole testing period. This method allows to better evaluate the robustness of the trading strategies, as it evaluates them independently on various periods, when the market conditions are different.

\begin{figure}[h]
    \centering
    \caption{Rolling data windows}
    \includegraphics[width=\linewidth]{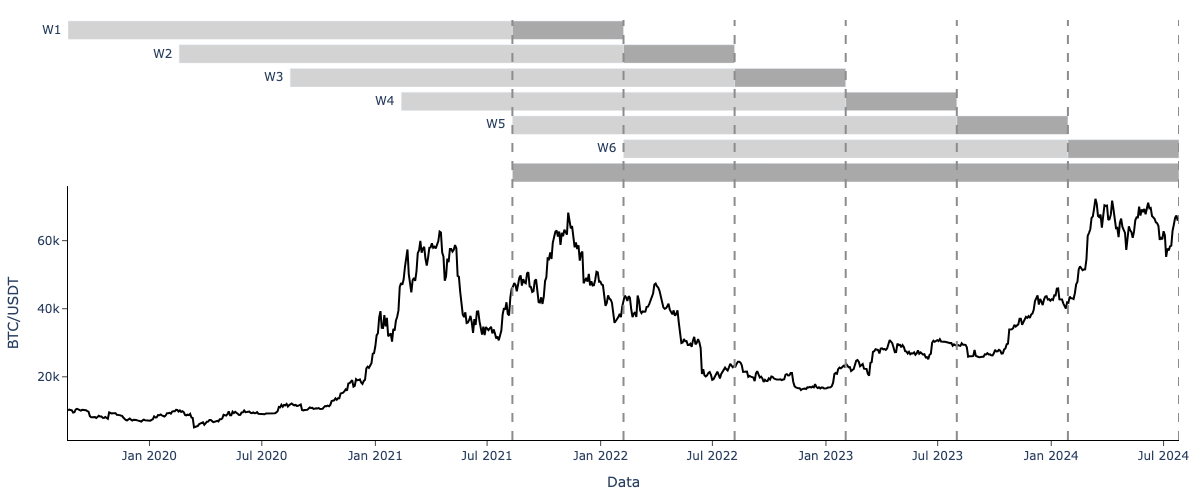}
    \label{fig:data-rolling}
    \raggedright \tiny Note: The figure presents how the dataset was split into rolling data windows consisting of \textit{in sample} and \textit{out of sample} parts.
\end{figure}

In total six windows with identical number of data points were created. The \textit{out of sample} is used for testing, further referenced as the \textit{test} part. The \textit{in sample} part is split into \textit{train} and \textit{validation} parts, with the \textit{validation} part being 20\% of the \textit{in sample} data.

\begin{table}[h]
        \caption{Number of data points}
	\begin{center}
		\begin{tabular}{lrrr}
			\textbf{Part} & \makecell{\textbf{BTC/USDT} \\ \textbf{5min}} & \makecell{\textbf{BTC/USDT} \\ \textbf{15min}} & \makecell{\textbf{BTC/USDT} \\ \textbf{30min}} \\
            \hline
            train & 165888 & 55296 & 27648 \\
            validation & 41472 & 13824 & 6912 \\
            test & 51840 & 17280 & 8640 \\
            \hline
         \multicolumn{4}{p{.7\textwidth}}{\tiny Note:  Number of data points in each of the dataset parts for the \texttt{5min}, \texttt{15min} and \texttt{30min} rolling windows.}
		\end{tabular}
	\end{center}
        \label{fig:data-windows-stats}
\end{table}

Tables \ref{tab:stats_5min}, \ref{tab:stats_5min} and \ref{tab:stats_30min} further detail the descriptive statistics for each \textit{out-of-sample} windows for \texttt{5min}, \texttt{15min} and \texttt{30min} interval data. The statistics are consistent with the ones computed for the whole period (Table \ref{tab:data-stats}).
Standard deviation is higher for the first two windows in all frequencies, suggesting that the market was more volatile during those periods.
Kurtosis is significantly higher than the normal distributions with the values largest for 1st and 5th window of \texttt{5min} data and 5th window for \texttt{15min} data, indicating many data points with returns close to zero in those windows. The Kolmogorov-Smirnov tests for normality indicate that none of the distributions is normal.

\begin{figure}[h!]
    \centering
    \caption{Distributions of returns for out-of-sample data}
    \label{fig:out_of_sample_dist}
    \includegraphics[width=0.85\linewidth]{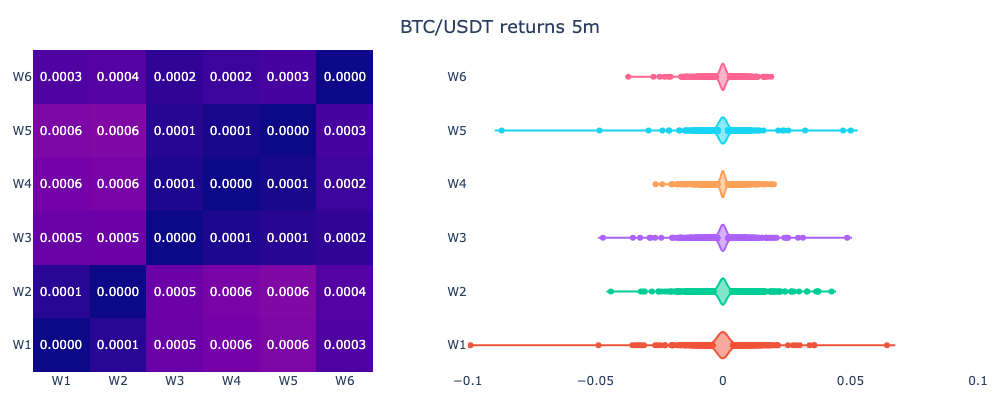}
    \includegraphics[width=0.85\linewidth]{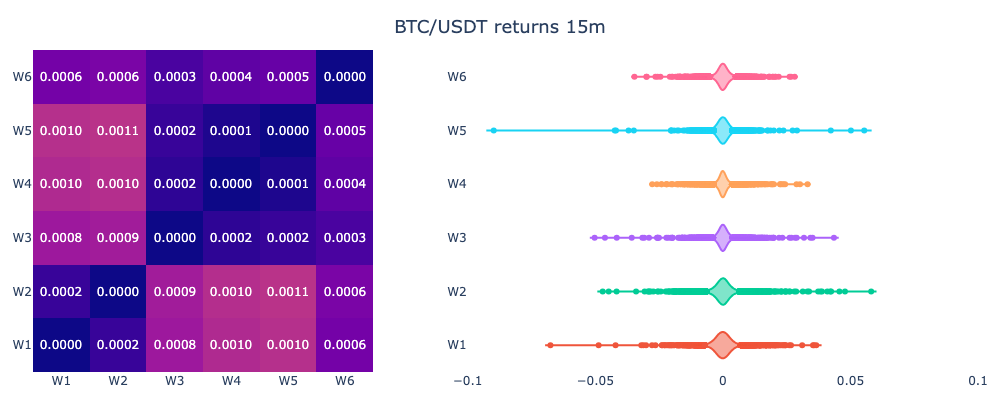}
    \includegraphics[width=0.85\linewidth]{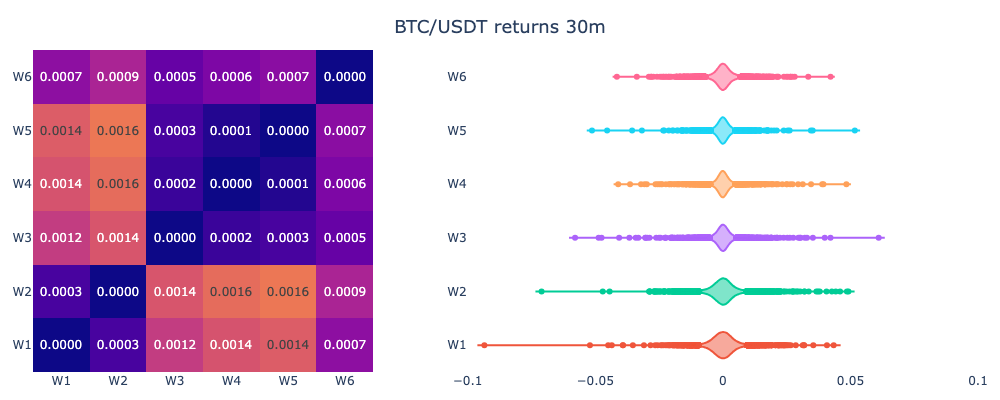}

    \raggedright \tiny Note: Visualizations of the returns distributions for the out-of-sample data for different windows from testing period, for 5min, 15min and 30min intervals. On the left, heatmaps visualize distances between each pair of distributions according to Wasserstein distance. On the right distributions are visualized as violin plots.
\end{figure}

\begin{table}
        \caption{Descriptive statistics for 5 min out-of-sample data}
        \vspace{-0.5cm}
        \label{tab:stats_5min}
	\begin{center}
            \small
		\begin{tabular}{lcccccc}
			\textbf{Statistic} & \textbf{W1} & \textbf{W2} & \textbf{W3} & \textbf{W4} & \textbf{W5} & \textbf{W6} \\
			\hline
			count & 51840 & 51840 & 51840 & 51840 & 51840 & 51840 \\
			mean & 0.0000010 & -0.0000089 & 0.0000016 & 0.0000050 & 0.0000080 & 0.0000099 \\
			std & 0.0021663 & 0.0023011 & 0.0015967 & 0.0013956 & 0.0013791 & 0.0016259 \\
			min & -0.0989054 & -0.0438821 & -0.0469833 & -0.0263642 & -0.0866874 & -0.0370693 \\
			25\% percentile & -0.0010410 & -0.0009298 & -0.0005471 & -0.0005338 & -0.0004952 & -0.0006930 \\
			50\% percentile & -0.0000139 & -0.0000029 & 0.0000070 & 0 & 0 & 0.0000064 \\
			75\% percentile & 0.0010127 & 0.0009137 & 0.0005638 & 0.0005350 & 0.0005185 & 0.0007262 \\
			max & 0.0643117 & 0.0426144 & 0.0487321 & 0.0200162 & 0.0501479 & 0.0189663 \\
			kurtosis & 124.60 & 27.33 & 72.17 & 27.74 & 419.46 & 18.82 \\
			skewness & -1.41 & 0.61 & -0.27 & -0.27 & -3.84 & -0.59 \\
			KS test stat. & 0.49 & 0.49 & 0.50 & 0.50 & 0.50 & 0.50 \\
			KS test p-value & 0.00e+00 & 0.00e+00 & 0.00e+00 & 0.00e+00 & 0.00e+00 & 0.00e+00 \\
            \hline
            \multicolumn{7}{p{\textwidth}}{\tiny Note:  Descriptive statistics of returns in data windows of \texttt{5min} interval data. The statistics are not annualized. Null hypothesis of Kolmogorov-Smirnov (KS) test is that the distribution is normal.}
		\end{tabular}
	\end{center}
\end{table}
\begin{table}
        \vspace{-1cm}
        \caption{Descriptive statistics for 15 min out-of-sample data}
        \label{tab:stats_15min}
        \vspace{-0.5cm}
	\begin{center}
            \small
		\begin{tabular}{lcccccc}
			\textbf{Statistic} & \textbf{W1} & \textbf{W2} & \textbf{W3} & \textbf{W4} & \textbf{W5} & \textbf{W6} \\
			\hline
			count & 17280 & 17280 & 17280 & 17280 & 17280 & 17280 \\
			mean & 0.0000024 & -0.0000276 & 0.0000048 & 0.0000149 & 0.0000240 & 0.0000296 \\
			std & 0.0035625 & 0.0038529 & 0.0027021 & 0.0023567 & 0.0023828 & 0.0027726 \\
			min & -0.0675796 & -0.0470951 & -0.0502659 & -0.0276761 & -0.0897967 & -0.0346516 \\
			25\% percentile & -0.0017774 & -0.0015974 & -0.0009356 & -0.0008720 & -0.0008726 & -0.0012156 \\
			50\% percentile & -0.0000367 & 0.0000399 & 0.0000138 & -0.0000080 & 0.0000222 & 0.0000141 \\
			75\% percentile & 0.0017230 & 0.0015821 & 0.0009610 & 0.0009004 & 0.0009122 & 0.0012940 \\
			max & 0.0366072 & 0.0581246 & 0.0435405 & 0.0331951 & 0.0554152 & 0.0281460 \\
			kurtosis & 19.08 & 19.40 & 38.81 & 21.94 & 176.31 & 11.54 \\
			skewness & -0.30 & 0.40 & -0.64 & 0.22 & -2.28 & -0.31 \\
			KS test stat. & 0.49 & 0.49 & 0.49 & 0.49 & 0.49 & 0.49 \\
			KS test p-value & 0.00e+00 & 0.00e+00 & 0.00e+00 & 0.00e+00 & 0.00e+00 & 0.00e+00 \\
               \hline
            \multicolumn{7}{p{\textwidth}}{\tiny Note:  Descriptive statistics of returns in data windows of \texttt{15min} interval data. The statistics are not annualized. Null hypothesis of Kolmogorov-Smirnov (KS) test is that the distribution is normal.}
		\end{tabular}
	\end{center}
\end{table}
\begin{table}
        \vspace{-1cm}
        \caption{Descriptive statistics for 30 min out-of-sample data}
        \vspace{-0.5cm}
        \label{tab:stats_30min}
	\begin{center}
            \small
		\begin{tabular}{lcccccc}
			\textbf{Statistic} & \textbf{W1} & \textbf{W2} & \textbf{W3} & \textbf{W4} & \textbf{W5} & \textbf{W6} \\
			\hline
			count & 8640 & 8640 & 8640 & 8640 & 8640 & 8640 \\
			mean & 0.0000043 & -0.0000540 & 0.0000101 & 0.0000296 & 0.0000473 & 0.0000595 \\
			std & 0.0049962 & 0.0054682 & 0.0038721 & 0.0034004 & 0.0031349 & 0.0039924 \\
			min & -0.0934631 & -0.0710468 & -0.0579109 & -0.0410247 & -0.0512786 & -0.0415287 \\
			25\% percentile & -0.0023584 & -0.0023627 & -0.0012400 & -0.0012067 & -0.0011561 & -0.0016455 \\
			50\% percentile & 0.0000360 & 0.0000073 & 0.0000117 & -0.0000114 & 0.0000311 & 0.0000521 \\
			75\% percentile & 0.0023458 & 0.0022461 & 0.0013298 & 0.0012539 & 0.0012576 & 0.0017785 \\
			max & 0.0434281 & 0.0492234 & 0.0610820 & 0.0484100 & 0.0517262 & 0.0422052 \\
			kurtosis & 22.73 & 12.11 & 34.48 & 25.83 & 34.78 & 10.52 \\
			skewness & -0.82 & 0.08 & -0.33 & 0.41 & -0.18 & -0.10 \\
			KS test stat. & 0.49 & 0.49 & 0.49 & 0.49 & 0.49 & 0.49 \\
			KS test p-value & 0.00e+00 & 0.00e+00 & 0.00e+00 & 0.00e+00 & 0.00e+00 & 0.00e+00 \\
               \hline
            \multicolumn{7}{p{\textwidth}}{\tiny Note:  Descriptive statistics of returns in data windows of \texttt{30min} interval data. The statistics are not annualized. Null hypothesis of Kolmogorov-Smirnov (KS) test is that the distribution is normal.}
		\end{tabular}
	\end{center}
\end{table}
The differences between \textit{out-of-sample} data windows are best visible in Figure \ref{fig:out_of_sample_dist}. Heatmaps on the left side of Figure \ref{fig:out_of_sample_dist} show the distances between the returns distributions between each \textit{out-of-sample} window, according to the Wasserstein metric (\cite{ctx69563056230003681}). For all the frequencies, the most significant shift in the distributions can be observed between first two and the next three windows, and then with the last, sixth, window. The violin plots of the distributions on the right-hand side of the figure provide additional information about the shapes of the distributions.

\section{Methodology}\label{chapter:methodology}
In this chapter, various trading strategies and the methodology for their evaluation are described. Fair comparison of strategies requires a formal definition of each strategy, clearly established comparison criteria, and metrics that compare different aspects of the strategies. Those are detailed in the initial section of this chapter \eqref{section:framework}. The latter section \eqref{section:strategies} describes the strategies that are compared in the study. For each strategy, a mechanism how the positions are selected is described as well as what are the hyperparameters that influence the strategy behavior. 

\subsection{Evaluation Framework}\label{section:framework}
Strategies are evaluated over certain periods of time. Consider a time series consisting of $T$ data points. The data point at each time step $t \in \{1..T\}$ represents a single time interval. During each time interval, one of the three possible positions can be taken:

\begin{itemize}
    \item \texttt{long position} - The asset is bought. The value of the portfolio increases if the price of an asset increases.
    \item \texttt{short position} - The asset is short sold and will need to be bought back when the position is closed. The value of the portfolio increases if the price of an asset decreases.
    \item \texttt{no position} - No asset in the portfolio. The value of the portfolio does not change, regardless of the price of the asset.

\end{itemize}
Additionally, the following assumptions are made:

\begin{itemize}
    \item When the strategy position changes, assets are always bought/sold at the close price of the previous interval. That is, if at the time $t-1$ the position was \texttt{ no position} and at the time $t$ the position is \texttt{ long position}, the assets are bought at the \texttt{close price} of the interval at time $t-1$. 
    \item Each such position change incurs an exchange fee $e$. In the experiments the value of $e$ is assumed to be $0.1\%$.\footnote{The value is equal to standard spot fee rate on Binance https://www.binance.com/en/fee/trading}.
    \item Any fraction of the asset can be traded\footnote{In practice the smallest fraction of the asset that can be bought on Binance is $0.00001$}.
    \item No splitting of portfolio allocation, the asset is always bought/sold with the full  portfolio value.
    \item At the end of the evaluation period all positions must be closed, that is, at time $T$ the position must be \texttt{no position}.
\end{itemize}

Formally, a trading strategy is a function $s_{\theta}: \mathbb{R}^{\lambda \times d} \rightarrow \{-1, 0, 1\}$, where $\theta$ are strategy hyperparameters, $\lambda$ is the size of the lookback window and $d$ is the dimensionality of the vector $x_t \in \mathbb{R}^d $ that represents new, known and relevant information at each time step $t$, which can be used by the strategy. The co-domain values $\{-1, 0, 1\}$ respectively, represent: \texttt{short position}, \texttt{no position}, and \texttt{long position}. Then, the recommended position $p_t$ according to the strategy $s$ at time $t$ is calculated in the following way:
\begin{equation}\label{eq:strategy}
s_{\theta}([x_{t-\lambda}; ...; x_{t-1}]) = p_t
\end{equation}
where $[\cdot ; \cdot]$ is a concatenation operation. Note that the strategy has access only to information prior to time $t$. The positions are computed for each time index $t$ over a given period of time. The portfolio value $E_t$ at any time $t$ can be calculated using the following formula:
\begin{equation}
E_0 = 1
\end{equation}
\begin{equation}
E_t = E_{t-1}(1+r_t \cdot p_t)(1 - (|p_t - p_{t-1}|e))
\end{equation}
where $r_t$ is a return in an interval $t$ computed as in \ref{eq:ret}, $p_t$ is a position recommended by the strategy and $e$ is the exchange fee rate. The portfolio value at the end of the evaluation period $T$ is equal to $E_T$.

Furthermore, following the methodology of \cite{s22030917}, the following metrics are computed:
\paragraph{Annualized Return Compounded (ARC)}
Metric that shows the annualized rate of return for the evaluation period, computed as \begin{equation}\label{eq:arc}
ARC = (\frac{E_T}{E_0})^{\frac{Y}{T}}
\end{equation}
where $E_0$ is the portfolio value at the beginning of the evaluation period, $E_T$ is the portfolio value at the end of the evaluation period, $T$ is number of the intervals in the evaluation period and $Y$ is the number of the intervals in a year\footnote{It is assumed that there are 365 days in a year, so for the datasets used in the research, this value is equal to $Y_{5m}=105120$, $Y_{15m}= 35040$ and $Y_{30m} = 17520$.}.
\paragraph{Annualized Standard Deviation (ASD)}
A metric that represents the yearly standard deviation of returns, assessing the strategy's volatility. The metric is determined as \begin{equation}
ASD = \sqrt{\frac{Y}{T}\sum_{t=0}^{T}(r_t - \bar{r})^2}
\end{equation}
where $r_t$ is the return at time $t$, $\bar{r}$ is the mean return in the evaluation period, $T$ is the number of intervals in the evaluation period and $Y$ is the number of intervals in a year the same as in \ref{eq:arc}.
\paragraph{Information Ratio (IR*)}
The Sharpie ratio (\cite{Sharpe+1998+169+178}) that is calculated under the assumption of a risk-free rate of zero. The metric becomes the annualized return divided by its annualized standard deviation: \begin{equation}\label{eq:ir}
    IR^{*}=\frac{ARC}{ASD}
\end{equation}
\paragraph{Maximum Drawdown (MD)}
The portfolio's maximum drawdown represents the greatest decline measurable between the portfolio's highest value and its consecutive lowest point. This metric acts as an indicator of possible percentage loss within a specified period
\begin{equation}\label{eq:md}
    MD = \max_{(t, \tau) \in \{[0..T]^2 : t < \tau\}}(\frac{E_t - E_{\tau}}{E_t})
\end{equation}
where $T$ represents the total number of intervals during the evaluation timeframe, with $E_t$ and $E_{\tau}$ denoting the portfolio values at times $t$ and $\tau$, respectively.
\paragraph{Modified Information Ratio (IR**)}\label{section:ir} is the Information Ratio \eqref{eq:ir} adjusted by the Maximum Drawdawn \eqref{eq:md}, a metric initially introduced by \cite{RePEc:war:wpaper:2018-09}:
\begin{equation}
    IR^{**} = IR^{*} \times \frac{|ARC|}{MD}
\end{equation}
\paragraph{Number of trades (N)}
The number of strategy position changes during the evaluation period. Calculated as
\begin{equation}
    N = \sum_{t=1}^{T}(|p_t - p_{t-1}|)
\end{equation}
where $p_t$ is the position at the time $t$. Note that while it is possible to change the position from \texttt{long position} to \texttt{short position} in a single time frame, it is counted as two trades: closing the long position and opening the short position.
\paragraph{Percentage of Long Position (LONG) / Short Position (SHORT)}
Metric that measures the percentage of how many time intervals during the entire evaluation period the strategy took the long / short position.
\begin{equation}
    LONG = \frac{1}{T}\sum_{t=0}^{T}(\frac{(p_t+1)p_t^2}{2}) \times 100\%
\end{equation}
\begin{equation}
    SHORT = \frac{1}{T}\sum_{t=0}^{T}(\frac{(1 - p_t)p_t^2}{2}) \times 100\%
\end{equation}

\subsection{Strategies}\label{section:strategies}
This study involves the evaluation and comparison of five strategies. Each strategy considered in the study consists of the complete mechanism that converts the signal, whether originating directly from the data, a technical indicator or a machine learning model, into the actual position. This is something that is often neglected, especially when evaluating the effectiveness of machine learning models. Although the precision of the model's predictions is crucial, a fair assessment of the strategy is feasible only after these predictions are converted into portfolio positions.

\subsubsection{Buy and Hold strategy}
The first strategy, the only viable one according to the Efficient Markets Hypothesis, is the "Buy and Hold" strategy. It is regarded as a benchmark that other strategies aim to surpass. Framed as a formal function as described in \eqref{eq:strategy}, strategy has no hyperparameters and at every time interval takes a \texttt{long position}:
\begin{equation}
    s^{\textit{BuyAndHold}}(\cdot) = 1 
\end{equation}

\subsubsection{Moving Average Convergence Divergence strategy}\label{section:macd_strat}
The strategy is based on the technical indicator of the Moving Average Convergence Divergence (MACD). The indicator was proposed by Gerald Appel almost 50 years ago and is still widely used by technical analysts today (\cite{10.5555/1408581})\footnote{Although the indicator was initially shared in the 1970s within the Systems \& Forecasts newsletter, the book detailing this indicator wasn't released until 2005.}. The standard MACD oscillator is a collection of three time series calculated from daily historical price data, most often the closing price. Time series are calculated using the Exponential Moving Average (EMA)\footnote{https://en.wikipedia.org/wiki/Exponential\_smoothing} with different window sizes. These are subsequently converted into positions as follows: when the MACD line moves above the signal line, it signifies an uptrend and acts as a buy signal. Conversely, when the MACD line falls below the signal line, it signals a downtrend, and one should exit the position:
\begin{equation}
    MACD_t(\cdot) = EMA_t^{\textit{fast}}(\cdot) - EMA_t^{\textit{slow}}(\cdot)
\end{equation}
\begin{equation}
    SIGNAL_t(\cdot) = EMA_t^{\textit{signal}}(MACD_t(\cdot))
\end{equation}
where the \textit{fast}, \textit{slow} and \textit{signal} are window sizes for Exponential Moving Average. Typically, these values are 12, 26, and 9 respectively, but in the context of this study, they are considered as strategy hyperparameters:
\begin{equation}
    s_{(\textit{fast}, \textit{slow}, \textit{signal}, \textit{short})}^{MACD}(\cdot) =
\begin{cases}
1 \quad if \; MACD_{t-1} \ge SIGNAL_{t-1} \\
0 \quad if \; MACD_{t-1} < SIGNAL_{t-1} \; \text{and}\; \textit{short} = 0 \\
-1 \quad if \; MACD_{t-1} < SIGNAL_{t-1} \; \text{and}\; \textit{short} = 1 \\
\end{cases}
\end{equation}
where $\textit{short} \in \{0, 1\}$ is an additional strategy hyperparameter that controls whether a strategy uses \texttt{exit position} or \texttt{short position} on sell signal.

\subsubsection{Relative Strength Index strategy}\label{section:rsi_strat}
The second strategy based on the technical indicator employs the Relative Strength Index (RSI) indicator. The RSI was developed by J.Welles Wilder Jr. and serves as a momentum oscillator designed to assess whether changes in asset prices suggest an overbought or oversold condition (\cite{wilder1978new}). The RSI is calculated as:
\begin{equation}
    RSI_t(\cdot) = \left( \frac{100}{1 + RS_t(\cdot)}\right) 
\end{equation}
\begin{equation}
    RS_t = \frac{SMMA^{\textit{window}}(U(\cdot))}{SMMA^{\textit{window}}(D(\cdot))}
\end{equation}
\begin{equation}
    U_t = \max(0, \texttt{close price}_t - \texttt{close price}_{t-1})
\end{equation}
\begin{equation}
    D_t = \max(0, \texttt{close price}_{t-1} - \texttt{close price}_{t})
\end{equation}
where $RS$ is a relative strength factor computed using Smoothed Moving Average (SMMA)\footnote{https://en.wikipedia.org/wiki/Moving\_average\#Modified\_moving\_average} calculated over the "Up period" ($U$) and "Down period" ($D$) time series, using $\textit{window}$ time intervals prior to $t$-th interval. Although usually the window is assumed to be 14, the study treats \textit{window} as a hyperparameter of the strategy.

To translate oscillator values into the positions, usually two threshold overbought/oversold for buy/sell signal are used. If the RSI crosses the over-sold threshold, it indicates a bullish sign, and when it crosses below the over-bought threshold, it is a bearish sign. Instead, in the study, four thresholds are used to indicate when: enter long position (\texttt{enter long}), exit long position (\texttt{exit long}), enter short position (\texttt{enter short}) and exit short position (\texttt{exit short}).
\begin{equation}
    s^{RSI}_{\theta_{RSI}}(\cdot) =
\begin{cases}
1 \quad if \; RSI_{t-1}(\cdot) > \textit{enter long} \\
0 \quad if \; RSI_{t-1}(\cdot) < \textit{exit long} \; \text{and} \; p_{t-1} = 1 \\
-1 \quad if \; RSI_{t-1}(\cdot) < \textit{enter short}\\
0 \quad if \; RSI_{t-1}(\cdot) > \textit{exit short}  \; \text{and} \; p_{t-1} = -1\\
p_{t-1} \quad else
\end{cases}
\end{equation}
where $\theta_{RSI} = (\textit{window}, \textit{enter long}, \textit{exit long}, \textit{enter short}, \textit{exit short})$ are strategy hyperparamethers and $p_{t-1}$ is a position taken on the previous time interval.

\subsubsection{Informer based strategies}
The last two strategies use the predictions of the state-of-the-art machine learning model for sequence modeling: Informer (\cite{zhou2021informerefficienttransformerlong}). The informer uses the modified Transformer architecture (\cite{DBLP:journals/corr/VaswaniSPUJGKP17}) to better handle long sequences. Several studies have shown that it surpasses other deep learning architectures in predicting financial time series (\cite{10280785}, \cite{lu2023stockmarketindexprediction}).
\paragraph{Informer Architecture}\label{section:informer}
 follows the encoder-decoder structure (\cite{chollet2017}, \cite{bahdanau2014neural}) where the encoder maps an input sequence of vectors $\mathbf{X} = (\mathbf{x}_1, ..., \mathbf{x}_n)$ to a sequence of continuous representations $\mathbf{Z} = (\mathbf{z}_1, ..., \mathbf{z}_n)$. Given $\mathbf{Z}$, the decoder then generates an output sequence $\mathbf{Y} = (\mathbf{y}_1, ..., \mathbf{y}_m)$ one element at a time. Both the encoder and decoder are built with a number of identical layers. Let $\textbf{X}_0 = \textbf{X}$ be an input to the first encoder layer, each $i$-th encoder layer transforms the sequence of $\textbf{X}_{i-1}$ input sequence to $\textbf{X}_{i}$ sequence, where $i \in \{1, .., L^{enc}\}$ and $L^{enc}$ is the number of encoder layers. The encoder output is equal to $\textbf{X}_{L^{enc}} = \textbf{Z}$. Similarly, each decoder layer transforms the sequence of previous predicted targets $\textbf{Y}_0$ and the encoder output $\textbf{Z}$ into the sequence ${\textbf{Y}_j}$, which is then consumed by the next decoder layer. The final output of the decoder layer is the output of the decoder $\textbf{Y} = \textbf{Y}_{L^{dec}}$, where $j \in \{1..L^{dec}\}$ and $L^{dec}$ is the number of decoder layers. \textit{Number of the encoder layers} and \textit{ the number of decoder layers} are model hyperparameters. The main building block of both those layers is \textit{multi-head probe-sparse self-attention}.

\paragraph{Multi-head probe-sparse self-attention:} The canonical self attention of Transformer is defined based on the tuple inputs, i.e, query (Q), key (K) and value (V), which performs the scaled dot-product as:
\begin{equation}
    \mathcal{A}(\mathbf{Q},\mathbf{K},\mathbf{V}) = \textit{softmax}\left(\frac{\mathbf{Q}\mathbf{K}^T}{\sqrt{d}}\right)\mathbf{V}
\end{equation}
where $\mathbf{Q} \in \mathbb{R}^{L_Q \times d}$, $\mathbf{K} \in \mathbb{R}^{L_K \times d}$, $\mathbf{V} \in \mathbb{R}^{V_Q \times d}$ and $L_Q$, $L_K$, $L_V$ are lengths of the query, key and value sequences respectively and the $d$ is the dimensionality of the model. Softmax\footnote{https://en.wikipedia.org/wiki/Softmax\_function} is a function that converts a vector of real numbers, so it can be interpreted as a probability distribution. In this formulation, computing attention requires the quadratic time dot-product computation. The authors of Informer proposed a measurement that transforms $\mathbf{Q}$ into a sparse $\mathbf{\bar{Q}}$ matrix, that contains only the top $u = c \cdot ln(L_Q)$ queries under the sparsity measurement, where $c$ is a constant sampling factor. The change makes the computation much more efficient and the architecture more applicable to the longer sequences.
\begin{equation}
    \mathcal{A}'(\mathbf{Q},\mathbf{K},\mathbf{V}) = \textit{softmax}\left(\frac{\mathbf{\bar{Q}}\mathbf{K}^T}{\sqrt{d}}\right)\mathbf{V}
\end{equation}

In each layer, attention is performed $h$ times with different linear projections of inputs. This mechanism is called multi-head attention and $h$ is the model hyperparameter that controls the number of so called attention heads.
\begin{equation}
    \mathcal{H}_i(\mathbf{Q},\mathbf{K},\mathbf{V}) = \mathcal{A}'(\mathbf{Q}\mathbf{W}_i^Q, \mathbf{K}\mathbf{W}_i^K,\mathbf{V}\mathbf{W}_i^V)
\end{equation}
\begin{equation}
    \mathcal{MH}(\mathbf{Q},\mathbf{K},\mathbf{V}) = [\mathcal{H}_1(\cdot); ...; \mathcal{H}_h(\cdot)]
\end{equation}
where the projections $\mathbf{W}_i^Q \in \mathbb{R}^{d \times L_Q}$, $\mathbf{W}_i^K \in \mathbb{R}^{d \times L_K}$, $\mathbf{W}_i^V \in \mathbb{R}^{d \times V_Q}$ are learned parameters.

The self-attention means that the layer "attends" to itself, that is, all queries, keys, and values consider the same input sequence i.e. $\mathbf{Q} = \mathbf{K} = \mathbf{V} = \mathbf{X}$, where $\mathbf{X}$ is an input to the \textit{multi-head probe-sparse self-attention layer} ($\mathcal{MHSA}$).
\begin{equation}
    \mathcal{MHSA}(\mathbf{X}) = \mathcal{MH}(\mathbf{X},\mathbf{X},\mathbf{X})
\end{equation}

\paragraph{Encoder:} The encoder layers of Informer consist of multi-head probe-sparse self-attention layer with distilling operation that trims the input’s time dimension. Given $\textbf{X}_{i-1}$ input to the $i$-th Informer layer, the output of the $i$-th layer is calculated as:
\begin{equation}
    \textbf{X}_{i} = MaxPool(ELU(\textit{Conv1d}(\mathcal{MHSA}(\textbf{X}_{i-1}))))
\end{equation}
where $\textit{Conv1d}$ performs a one dimensional convolutional filter over the time dimension (\cite{kiranyaz20191dconvolutionalneuralnetworks}), with $\textit{ELU}$ activation function (\cite{clevert2016fastaccuratedeepnetwork}) and max-pooling layer (\cite{6144164}) that down-samples the layer output.

\paragraph{Decoder:}
Informer uses standard Transformer decoder, which consists of two $\mathcal{MHSA}$ layers in which the first takes an input of past target outputs and the second output of the encoder. Those are followed with a position wise fully connected layer ($\mathcal{FFN}$). It consists of two linear transformations with a ReLU (\cite{agarap2019deeplearningusingrectified}) activation in between, and is applied to vectors at each position separately:
\begin{equation}
    \mathcal{FFN}(\mathbf{x}) = max(0, \mathbf{x}\mathbf{W}_1 + \mathbf{b}_1)\mathbf{W}_2 + \mathbf{b}_2
\end{equation}
where $\mathbf{W}_1 \in \mathbb{R}^{d\times f},\mathbf{W}_2 \in \mathbb{R}^{f \times d},\mathbf{b}_1 \in \mathbb{R}^f,\mathbf{b}_2 \in \mathbb{R}^d$ are trained parameters and $d$ is the model dimensionality and $f$ is the size of the fully connected layer, both of these values are model hyperparameters.
Around of each of those sub-layers there is a residual connection (\cite{5264952}), dropout regularization (\textit{Dropout}) and a layer normalization (\textit{LayerNorm}) (\cite{ba2016layernormalization}).
\begin{equation}
    \mathbf{Y}'_j = LayerNorm(Dropout(\mathcal{MHSA}(\mathbf{Y}_{j-1})) + \mathbf{Y}_{j-1})
\end{equation}
\begin{equation}
    \mathbf{Y}''_j = LayerNorm(Dropout(\mathcal{MHSA}(\mathbf{Z}) + \mathbf{Z})
\end{equation}
\begin{equation}
    \mathbf{Y}'''_j = \mathbf{Y}'_j + \mathbf{Y}''_j = [\mathbf{y}'''_{1j}; ...; \mathbf{y}'''_{mj}]
\end{equation}
\begin{equation}
    \mathbf{Y}_j = LayerNorm(Dropout([\mathcal{FFN}(\mathbf{y}'''_{ij}); ...; \mathcal{FFN}(\mathbf{y}'''_{}mj)]) + \mathbf{Y}''')
\end{equation}

\paragraph{Model input:}
The model input $\textbf{X}_0$ consists of sequence of $n$ $d$-dimensional input vectors. These vectors are derived by adding together three input components:
\begin{itemize}
    \item \textbf{Real input variables:} All real-value input variables are concatenated and linearly transformed into $d$-dimensional vector.
    \item \textbf{Categorical variables:} All categorical variables are mapped into the embeddings - real-value vectors that are model parameters. Then they are concatenated and linearly transformed into $d$-dimensional vector.
    \item \textbf{Positional embeddings:} Since Informer model contains no recurrence and convolution, the ordering of input vectors needs to be modeled explicitly. To achieve this, a method known as positional encodings is employed. For each position a vector of sine and cosine functions of different frequencies is computed, that is, each dimension of the positional encoding corresponds to a sinusoid.
\end{itemize}

\paragraph{Loss function:}\label{par:loss} As the first loss function used to train Informer model the study uses Root Mean Squared Error (RMSE) loss function which is commonly used for financial data forecasting\footnote{Another common loss function is Mean Squared Error (MSE), but the difference between the two is just that one is scaled version of the other.}. RMSE loss function is defined as:
\begin{equation}
    RMSE = \sqrt{\frac{1}{N}\sum_{i=1}^N(y_{i}-\hat{y_i})^2}
\end{equation}
where $y$ is observed value, $\hat{y}$ is model prediction, $N$ is number of observations.

However, optimizing model with such these loss function does not align with the objective of finding a strategy aimed at maximizing returns. Consider an asset, which price does not move for 99 out of 100 intervals and then suddenly drops. A model that predicts the same price every time would have  a 99\% accuracy, however a strategy based on such model predictions would have very poor results. This example illustrates that while the model using RMSE loss functions may be trained to have high accuracy of predictions, those may not necessarily be \textit{meaningful} predictions in terms of building the trading strategy. Moreover, with point-wise predictions, such as the ones coming from RMSE loss, there is no information of the certainty of the prediction.

To address the aforementioned issues the study introduces two other loss functions that may be better suited for training models that are later used in trading strategy. The first loss is a Quantile Loss, that is, instead of predicting the point, the model predicts a distribution, more precisely quantiles of the returns distribution. With such signal, a strategy can be constructed that considers confidence in the prediction of the model. The Quantile Loss is defined as:
\begin{equation}
    QuantileLoss = \frac{1}{N}\sum_{i=1}^{N}\sum_q max(q\cdot (y - \hat{y}), (1-q)\cdot(\hat{y} - y))
\end{equation}
where $y$ is observed value, $\hat{y}$ is model prediction, $N$ is number of observations. $q \in [0, 1]$ is a quantile, in the study the following quantiles were used
$q \in \{0.01$, $0.02$, $0.03$, $0.05$, $0.1$, $0.25$, $0.5$, $0.75$, $0.9$, $0.95$, $0.97$, $0.98$, $0.99$$\}$\footnote{A decision to train the model with many different quantiles was motivated by the fact, that then a strategy that bases on this model predictions can be fine-tuned with different quantile values without need for retraining the model.}.

The other loss function considered is a Generalized Mean Absolute Directional Loss (GMADL) (\cite{michankow2024102375}). It puts more emphasis on the direction of the returns, i.e. whether they were positive or negative rather than precision and rewards the model for correctly predicting larger return values. The loss function is defined as
\begin{equation}
    GMADL = \frac{1}{N}\sum_{i=1}^{N}(-1)\cdot(\frac{1}{1 + e^{-a\cdot y \cdot \hat{y}}} - \frac{1}{2}) \cdot (|y|)^b
\end{equation}
where $y$ is observed value, $\hat{y}$ is model prediction, $N$ is number of observations. $a$ and $b$ are loss function parameters that control the stepness of the function slope. In the study they are considered to be equal to $a=100$ and $b=2$. 
\begin{figure}[h]
    \caption{RMSE vs GMADL loss functions.}
    \vspace{0.5cm}
    \centering
    \includegraphics[width=\linewidth]{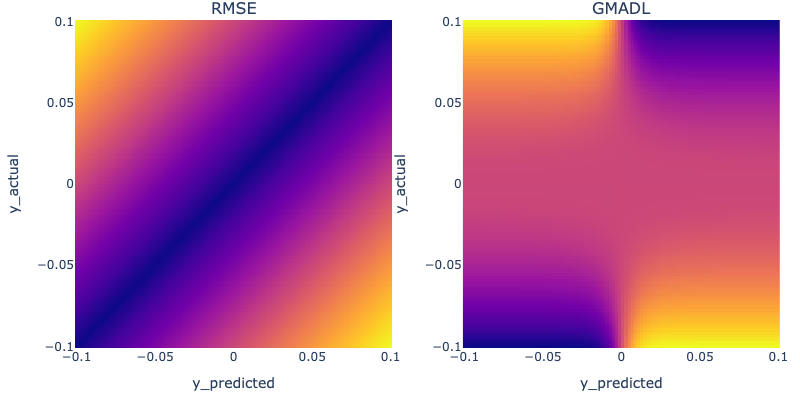}
    \label{fig:gmadl}
    \raggedright\tiny Note: Comparison of the two loss functions: Mean Absolute Error (MAE) on the left and Generalized Mean Absolute Directional Loss (GMADL) on the right. 
\end{figure}

\paragraph{Strategies}
Leveraging the predictions generated by the Informer model, one can formulate a trading strategy. Given that the models were trained using three distinct approaches, the RMSE loss, Quantile Loss and GMADL, each method provides the opportunity to develop a unique strategy that optimally utilizes the model's predictions.

\paragraph{RMSE Informer strategy:}\label{paragraph:gmadl_strat} Strategy that uses predictions of the Informer trained with RMSE is defined as the following: four threshold values are defined when to \textit{enter long}, \textit{exit long}, \textit{enter short} and \textit{exit short}. Those thresholds are compared directly with the return predicted by the model.
\begin{equation}
    s_{\theta_{RMSE}}^{RMSE}(\cdot) =
\begin{cases}
1 \quad if \; \hat{y}_{t} > \textit{enter long} \\
0 \quad if \; \hat{y}_{t} < \textit{exit long} \; \text{and} \; p_{t-1} = 1\\
-1 \quad if \; \hat{y}_{t} < \textit{enter short} \\
0 \quad if \; \hat{y}_{t} > \textit{exit short} \; \text{and} \; p_{t-1} = -1\\
p_{t-1} \quad else
\end{cases}
\end{equation}
where $\hat{y}_t$ is the model prediction for time step $t$ and $p_{t-1}$ is position on the previous timestep. The strategy hyperparameters are $\theta_{RMSE} =(\Theta_{RMSE}$, $\textit{enter long}$, $\textit{exit long}$, $\textit{enter short}$, $\textit{exit short})$ and $\Theta_{RMSE}$ are parameters of the Informer model trained with RMSE loss.

\paragraph{Quantile Informer strategy:}\label{paragraph:quantile_strat}
A strategy based on the predictions of the model trained with Quantile Loss is constructed in the following manner: probabilities are defined to \textit{enter long}, \textit{exit long}, \textit{enter short} and \textit{exit short} positions, as well as the \textit{threshold} value. If the return predicted by the model is above the \textit{threshold} with certain probability, i.e. $1 - \textit{enter long}$\footnote{Model prediction $\hat{y}_{t,q}$ for a quantile $q$ means that the value is below $\hat{y}_{t,q}$ with $q$ probability, and above $\hat{y}_{t,q}$ with $1-q$ probability.} quantile is above the threshold, a strategy enters \texttt{long position}. Conversly, if the \textit{enter short} quantile is below $-\textit{threshold}$ a \texttt{short position} is opened. The interpretation is that the model prediction has confidence of \textit{enter long}/\textit{enter short} that the returns are above/below the \textit{threshold}.
\begin{equation}
    s_{\theta_{Quantile}}^{Quantile}(\cdot) = 
\begin{cases}
1 \quad if \; \hat{y}_{t,1 - \textit{enter long}} > \textit{threshold} \\
0 \quad if \; \hat{y}_{t,\textit{exit long}} < -\textit{threshold}\; \text{and}\; p_{t-1} = 1 \\
-1 \quad if \; \hat{y}_{t,\textit{enter short}} < -\textit{threshold} \\
0 \quad if \; \hat{y}_{t,1-\textit{exit short}} > \textit{threshold}\; \text{and}\; p_{t-1} = -1 \\
p_{t-1} \quad else
\end{cases}
\end{equation}
where $\hat{y}_{t,q}$ is a model prediction at time $t$ for the $q$ quantile and $p_{t-1}$ is a position on the previous timestep. The strategy hyperparameters are $\theta_{Quantile} = (\Theta_{Quantile}, $ \textit{enter long}, \textit{exit long}, \textit{enter short}, \textit{exit short}, \textit{threshold}$)$.
 and $\Theta_{\textit{Quantile}}$ are parameters of the Informer model trained with Quantile loss.
 
\paragraph{GMADL Informer strategy:}\label{paragraph:gmadl_strat} Strategy that uses predictions of the Informer trained with GMADL is defined analogously to the one using RMSE: four threshold values are defined when to \textit{enter long}, \textit{exit long}, \textit{enter short} and \textit{exit short} and the thresholds are compared directly with the return predicted by the model. While in this case one does not know explicitly the confidence of those predictions, because of how GMADL is constructed, for the larger returns they should be more reliable than the predictions of the model trianed with RMSE loss.
\begin{equation}
    s_{\theta_{GMADL}}^{GMADL}(\cdot) =
\begin{cases}
1 \quad if \; \hat{y}_{t} > \textit{enter long} \\
0 \quad if \; \hat{y}_{t} < \textit{exit long} \; \text{and} \; p_{t-1} = 1\\
-1 \quad if \; \hat{y}_{t} < \textit{enter short} \\
0 \quad if \; \hat{y}_{t} > \textit{exit short} \; \text{and} \; p_{t-1} = -1\\
p_{t-1} \quad else
\end{cases}
\end{equation}
where $\hat{y}_t$ is the model prediction for time step $t$ and $p_{t-1}$ is position on the previous timestep. The strategy hyperparameters are $\theta_{GMADL} =(\Theta_{GMADL}$, $\textit{enter long}$, $\textit{exit long}$, $\textit{enter short}$, $\textit{exit short})$ and $\Theta_{GMADL}$ are parameters of the Informer model trained with GMADL loss.

\section{Experiments}\label{chapter:experiments}
The evaluation framework and strategies described in Chapter \ref{chapter:methodology} were implemented in Python\footnote{https://www.python.org/} and are publicly available in the Gitlab repository\footnote{https://gitlab.com/FilipStefaniuk/wne-msc-thesis}. Technical indicators were computed using TA-lib library\footnote{https://ta-lib.org/}, Informer model was trained using PytorchForecasting\footnote{https://pytorch-forecasting.readthedocs.io/en/stable/} and the Informer implementation was taken from the following Github repository\footnote{https://github.com/martinwhl/Informer-PyTorch-Lightning}. Section \ref{section:hparams} describes the hyperparameter selection process for each strategy and presents the values that were selected. The evaluation of the test data is then presented and analyzed in Section \ref{section:results}. The last Section \ref{section:sensitivity} presents the sensitivity analysis of how the selection of strategy hyperparameters influences the strategy performance.

\subsection{Training and hyperparameter selection}\label{section:hparams}
Each strategy was independently instanciated for each data window. Strategies that were based on machine learning model were trained using the \textit{training} part of the data window and the hyperparameters were selected using \textit{validation} part. Strategies that do not require training i.e. MACD based strategy and RSI based strategy had their hyperparameters selected using only the \textit{validation} part of the dataset. An alternative would be to use all in-sample (\textit{train} and \textit{validation} parts) to select hyperparameters for non-ml strategies, however a decision to use only \textit{validation} part was motivated by the fact that to make comparison fair, the same data should be used for similar strategy hyperparameter selection (thresholds) for ml and non-ml based strategies. The best hyperparameter combinations were selected by comparing the IR** \eqref{section:ir} metric. The reminder of this section details the hyperparameter search space for each strategy and lists which hyperparameter values were selected for the strategies for each data window.

\subsubsection{MACD Strategy hyperparameters}
As described in Section \ref{section:macd_strat}, the MACD strategy has four hyperparameters, three of which (\textit{fast}, \textit{slow}, \textit{signal}) control the size of the window for moving averages and the \textit{short} whether a sell signal should be interpreted to close \texttt{long position} or to additionally open \texttt{short position}. Table \ref{tab:macd_hspace} presents the values tested for the hyperparameters. A \textit{short} hyperparameter takes binary values. Window sizes take values of the 16 values of the Fibonacci sequence, the sequence has been chosen as it grows exponentially, thus allows to search efficiently the space of possible windows - using the equally distributed values would be less efficient as large window sizes are less sensitive to small window size changes. Note that even though combinations were searched exhaustively, an additional constraint that prevents the \textit{fast} window to be longer than \textit{slow} window has been applied.
\begin{table}[h!]
    \vspace{-0.5cm}
    \centering
    \caption{Hyperparameter search space for MACD strategy.}
    \begin{tabular}{l c}
         \textbf{Parameter} & \textbf{Values} \\
         \hline
         \textit{fast} &  [2, 3, 5, 8, 13, 21, 34, 55, 89, 144, 233, 377, 610, 987, 1597, 2584] \\
         \textit{slow} & [2, 3, 5, 8, 13, 21, 34, 55, 89, 144, 233, 377, 610, 987, 1597, 2584]\\
         \textit{signal} & [2, 3, 5, 8, 13, 21, 34, 55, 89, 144, 233, 377, 610, 987, 1597, 2584]\\
         \textit{short} & [0, 1] \\
         \hline
         \multicolumn{2}{p{0.9\textwidth}}{\tiny Note: Considered values of hyperparameters for MACD strategy, all 8192 combinations were searched exhaustively.}
    \end{tabular}
    \label{tab:macd_hspace}
\end{table}

Hyperparameters selected for each window for every data interval are presented in Table \ref{tab:macd_hparams}. A bias towards larger window sizes can be observed, suggesting that MACD indicator might not be appropriate for high frequencies such as 5, 15 or 30 minutes. Additionally for all three frequencies, the strategies for the last two windows do not use short positions. Setting the short position on the down trend should always be beneficial, so this can suggest that the data in the last two windows may be much less predictable.
\begin{table}[h!]
        \vspace{-0.5cm}
        \caption{Selected hyperparameters for MACD strategy.}
	\begin{center}
		\begin{tabular}{lcccc}
			\textbf{Window} & \textit{\textbf{fast}} & \textit{\textbf{slow}} & \textbf{\textit{signal}} & \textbf{\textit{short}} \\
			\hline
			W1-5min & 8 & 2584 & 987 & 1 \\
			W2-5min & 377 & 610 & 610 & 1 \\
			W3-5min & 987 & 2584 & 987 & 1 \\
			W4-5min & 610 & 2584 & 987 & 1 \\
			W5-5min & 987 & 1597 & 987 & 0 \\
			W6-5min & 1597 & 2584 & 377 & 0 \\
			\hline
			W1-15min & 8 & 377 & 1597 & 1 \\
			W2-15min & 144 & 987 & 2584 & 1 \\
			W3-15min & 377 & 2584 & 2584 & 1 \\
			W4-15min & 377 & 2584 & 610 & 1 \\
			W5-15min & 233 & 610 & 233 & 0 \\
			W6-15min & 144 & 377 & 2584 & 0 \\
			\hline
			W1-30min & 1597 & 2584 & 2584 & 1 \\
			W2-30min & 233 & 2584 & 2584 & 1 \\
			W3-30min & 13 & 2584 & 2584 & 1 \\
			W4-30min & 233 & 987 & 233 & 1 \\
			W5-30min & 144 & 233 & 233 & 0 \\
			W6-30min & 1597 & 2584 & 1597 & 0 \\
            \hline
            \multicolumn{5}{p{0.45\textwidth}}{\tiny Note: Table presents hyperparameter values selected for the MACD strategy for each window of 5min, 15min and 30min interval data.}
		\end{tabular}
	\end{center}
        \label{tab:macd_hparams}
\end{table}

\subsubsection{RSI strategy hyperparameters}
As detailed in Section \ref{section:rsi_strat} the RSI strategy has five hyperparameters. The \textit{window} parameter sets the window size for computing the moving averages. The other four set thresholds when \texttt{long position} and \texttt{short position} should be opened or closed. For the window size, similarly as with MACD strategy values of the Fibonacci sequence were used. Combinations of window sizes and various thresholds listed in Table \ref{tab:rsi_hspace} were searched exhaustively, "-" symbolize that the position will never be opened/closed regardless of the RSI oscillator value. Note that \textit{exit long} and \textit{exit short} hyperparameter values make sense to be different from "-" only if corresponding \textit{enter long} and \textit{enter short} hyperparameters are different than "-". However, the reverse situation can occur as the long/short position is automatically closed as the next short/long position is entered, that is e.g. a combination where \textit{enter long} is not equal to "-" and \textit{exit long} is equal to "-" can occur is \textit{enter short} is present as the \texttt{long position} will be closed if \textit{enter short} threshold is reached.
\begin{table}[h]
    \centering
    \vspace{-0.5cm}
    \caption{Hyperparameter space for RSI strategy.}
    \begin{tabular}{l c}
         \textbf{Parameter} & \textbf{Values} \\
         \hline
         \textit{window} &  [2, 3, 5, 8, 13, 21, 34, 55, 89, 144, 233, 377, 610, 987, 1597, 2584] \\
        \textit{enter long} & [-, 70, 75, 80, 85, 90, 95] \\
        \textit{exit long} & [-, 5, 10, 15, 20, 25, 30]\\
        \textit{enter short} & [-, 5, 10, 15, 20, 25, 30]\\
        \textit{exit short} & [-, 70, 75, 80, 85, 90, 95]\\
        \hline
         \multicolumn{2}{p{0.9\textwidth}}{\tiny Note: Considered values of hyperparameters for RSI strategy, all 38416 combinations were searched exhaustively.}
    \end{tabular}
    \label{tab:rsi_hspace}
\end{table}

The table lists selected hyperparameters. Contrary to the MACD strategy, the RSI strategy seems to benefit from higher frequencies, and the selected values for the \textit{window} hyperparameter were rather small. Only in few cases \textit{exit long} and \textit{exit short} hyperparameters were set, meaning that in most cases when the long/short position was closed, the short/long position was immediately opened.
\begin{table}[h!]
        \vspace{-0.5cm}
        \caption{Selected hyperparameters for the RSI strategy.}
	\begin{center}
		\begin{tabular}{lccccc}
			\textbf{Window} & \textbf{\textit{window}} & \textbf{\textit{enter long}} & \textbf{\textit{exit long}} & \textbf{\textit{enter short}} & \textbf{\textit{exit short}} \\
			\hline
			W1-5min & 21 & 80 & - & 25 & - \\
			W2-5min & 34 & 75 & - & 25 & - \\
			W3-5min & 5 & 95 & - & 15 & - \\
			W4-5min & 8 & 90 & 5 & - & - \\
			W5-5min & 13 & 90 & 20 & - & - \\
			W6-5min & 21 & 85 & 15 & - & - \\
			\hline
			W1-15min & 5 & 80 & - & 15 & - \\
			W2-15min & 34 & 75 & - & 30 & - \\
			W3-15min & 5 & 95 & - & 10 & - \\
			W4-15min & 34 & 75 & - & 30 & - \\
			W5-15min & 13 & 70 & - & 5 & - \\
			W6-15min & 21 & 85 & 15 & - & - \\
			\hline
			W1-30min & 5 & 95 & - & 5 & - \\
			W2-30min & 21 & 70 & - & 30 & - \\
			W3-30min & 5 & 95 & - & 15 & 85 \\
			W4-30min & 8 & 95 & 5 & - & - \\
			W5-30min & 13 & 70 & - & 5 & - \\
			W6-30min & 34 & 80 & 25 & - & - \\
               \hline
            \multicolumn{6}{p{0.9\textwidth}}{\tiny Note: Table presents hyperparameter values selected for the RSI strategy for each window of 5min, 15min and 30min interval data.}
		\end{tabular}
	\end{center}
 \label{tab:rsi_hparams}
\end{table}

\subsubsection{Informer Strategies training and hyperparameters}
The process of selecting the hyperparameters for Informer-based strategies was divided into two steps. First, a model hyperparameters were selected by training the Informer with different combinations of model hyperparameters, then the strategy hyperparameters were selected for the best model. In other words, first, the model that can produce the best predictions was selected and then, based on the predictions of the models, the best strategy was constructed. The reason for this split was motivated mostly by the heaviness of the training process and the fact that if both model and strategy parameters are changed at the same time, it is difficult to assess which has the biggest impact on the final strategy result. This divide carries the risk that the best model, in terms of the loss function, does not necessarily mean the best strategy, in terms of IR**. However, this is addressed, to some extent, by selecting a loss function that aligns more closely with the objective of maximizing returns \eqref{par:loss}.

\paragraph{Training and model hyperparameters}
A separate instance of Informer was trained for each data window and for each RMSE, Quantile and GMADL loss functions. The target variable in each case was \texttt{returns}. Each training run consisted of 40 epochs with an early stopping\footnote{The number of observations in each dataset (5min, 15min 30min) varied and was rather large for a 5 min dataset, so instead of running validation at the end of each epoch, it was run after every 300 batches. The \textit{patience} parameter that controls early stopping was set to 15, meaning that if in the next 15 consecutive validations the validation loss does not improve, the training is stopped.}. The input to the model consisted of
a sequence whose length was controlled by the \textit{past window} hyperparameter. The sequence consisted of two types of variables: \texttt{real} and \texttt{categorical}. The \texttt{real} variables were normalized before they were passed to the model. The \texttt{categorical} variables were mapped into real-value embeddings.
\begin{table}[h]
    \centering
    \caption{Input variables to Informer model.}
    \small
    \begin{tabular}{cc}
         \textbf{Variable type} & \textbf{Variable} \\
         \hline
         \texttt{real} & \makecell{
            \texttt{open price}\\
            \texttt{high price}\\
            \texttt{low price}\\
            \texttt{close price}\\
            \texttt{volume}\\
            \texttt{returns}\\
            \texttt{vix}\\
            \texttt{fed rates}\\
            \texttt{fear and greed}\\
            \texttt{open to close price}\\
            \texttt{high to close price}, \\
            \texttt{low to close price}\\
            \texttt{vol 1h}\\
            \texttt{vol 1d}\\
            \texttt{vol 7d}\\
            \texttt{sma 1h to close price}\\
            \texttt{sma 1d to close price}\\
            \texttt{sma 7d to close price}\\
            \texttt{ema 1h to close price}\\
            \texttt{ema 1d to close price}\\
            \texttt{macd}\\
            \texttt{macd signal}\\
            \texttt{rsi}\\
            \texttt{low bband to close price}\\ \texttt{up bband to close price}\\
            \texttt{mid bband to close price}}\\
\hline
         \texttt{categorical} &\makecell{\texttt{hour}\\ \texttt{weekday}}\\
    \end{tabular}
    \label{tab:input_vars}
\end{table}

Table \ref{tab:input_vars} presents the variables that were passed to the model as an input. The \texttt{open price}, \texttt{high price}, \texttt{low price}, \texttt{close price} and \texttt{volume} come directly from the dataset described in Chapter \ref{chapter:data}. The \texttt{returns} were computed according to \eqref{eq:ret} and the \texttt{vix}, \texttt{fed rates}, \texttt{fear and greed} are additional data that the dataset was augmented with (Section \ref{section:additional_data}). The other variables were derived from the variables in the data set: \texttt{open to close price}, \texttt{high to close price}, \texttt{low to close price}, \texttt{high to low price} are the ratios of the corresponding price values. \texttt{vol 1h}, \texttt{vol 1d}, \texttt{vol 7d} variables are volatility computed on the 1 hour, 1 day and 7 days past data, respectively. \texttt{sma 1h to close price}, \texttt{sma 1d to close price} and \texttt{sma 7d to close price} are ratios of Simple Moving Average computed on 1 hour, 1 day, and 7 day windows to \texttt{close price}. \texttt{ema 1h to close price} and \texttt{ema 1d to close price} are ratios of Exponential Moving Average computed on data from past 1 hour and 1 day to \texttt{close price}. The \texttt{macd} and \texttt{macd} comes from the MACD technical indicator computed using the \texttt{close price} value and default values for window sizes\footnote{The window sizes used for computing MACD were: 12 for fast window, 26 for slow window and 9 for signal window.}. The \texttt{rsi} is a RSI technical indicator, computed using the 14 past data points. Finally \texttt{low bband to close price}, \texttt{up bband to close price} and \texttt{mid bband to close price} are ratios of Bollinger Bands \parencite{bollinger2002bollinger} to \texttt{close price}.

The \texttt{categorical} variables have values of 0 to 23 for \texttt{hour} and 0 to 6 for \texttt{weekday} and represent an hour-of-the-day and a day-of-the-week of the interval (in respect to interval's \texttt{close time})\footnote{E.g. an interval with close time at 2023-10-18 01:29:59 will have values of 1 (second hour of the day) for the \texttt{hour} variable and 2 (third day of the week - Wednesday) for the \texttt{weekday} variable}.

The hyperparameters of the model were determined by randomly sampling the hyperparameter space. The random search was carried out due to a large number of possible combinations and the training time, which took approximately 1 hour per training run\footnote{The training was done using a single Nvidia GTX 1080TI card.}. A sample of 30 combinations (out of 1166400) was tested and the combination of hyperparameters with the lowest validation loss value was selected. Note that since two values of the two different loss functions cannot be directly compared, this procedure had to be carried out separately for each loss function and for every frequency, totaling in $\textit{num frequencies} \times \textit{num loss functions} \times \textit{sample size} = 3 \times  3 \times 30 = 270$ training runs. Furthermore, as this process is extremely heavy, the hyperparameter search was done only on the first data window, and the exact selected combination of training and model hyperparameters were later used to train the model on other data windows for the given time interval. The space of Informer model hyperparameters is presented in Table \ref{tab:informer_hspace}.
where \textit{past window} is a length of the input sequence, \textit{batch size} and \textit{learning rate} are the standard machine learning training parameters. Other hyperparameters are specific to Informer architecture and are described in Section \ref{section:informer}.
\begin{table}[h]
    \centering
    \caption{Informer model hyperparameter space.}
    \begin{tabular}{l  c}
         \textbf{Parameter} & \textbf{Values} \\
         \hline
         past window & [20, 21, ..., 119, 120] \\
         batch size & [64, 128, 256] \\
         learning rate & [0.001, 0.0005, 0.0001] \\    
         model dimensionality ($d$) & [256, 512, 1024] \\
         fully connected dimensionality ($f$) & [256, 512, 1024] \\
         attention heads ($h$) & [1, 2, 4, 6] \\
         dropout & [0.05, 0.1, 0.2, 0.3] \\
         number of encoder layers & [1, 2, 3] \\
         number of decoder layers & [1, 2, 3] \\
         \hline
         \multicolumn{2}{p{0.8\textwidth}}{\tiny Note: Table presents all hyperparameter values that were considered. The dimension of the grid is 1166400, so the search was done by randomly sampling 30 combinations.}
    \end{tabular}
    \label{tab:informer_hspace}
\end{table}

The Table \ref{tab:informer_hparams} lists all the hyperparameter combinations selected for the Informer model for different data frequency and loss functions. Interestingly, it seems that the better models were the smaller ones, having shorter input sequence (20-30) and smaller dimensionality. This might have been caused by still relatively small size of the training dataset (Fig \ref{fig:data-windows-stats}). The different values of \textit{batch size} and \textit{learning rate} from the tested range did not seem to have significant impact on the model training. 
\begin{table}[h]
    \vspace{-0.5cm}
    \caption{Selected hyperparameters for Informer model.}
    \centering
    \small
    \tabcolsep2pt
    \begin{tabular}{l c c c  c c c  c c c}
        & \multicolumn{3}{c}{\textbf{Quantile RMSE}} & \multicolumn{3}{c}{\textbf{Quantile Informer}} &\multicolumn{3}{c}{\textbf{GMADL Informer}} \\
         \textbf{Parameter} & \textbf{5min} & \textbf{15min} & \textbf{30min} & \textbf{5min} & \textbf{15min} & \textbf{30min} & \textbf{5min} & \textbf{15min} & \textbf{30min} \\
         \hline
         past window & 102 & 110 & 115 & 22 & 28 & 72 & 28 & 30 & 26 \\
         batch size & 64 & 64 & 64 & 64 & 64 & 128 & 256 & 128 & 128 \\
         model dimensionality ($d$) & 512 & 512 & 256 & 256 & 256 & 1024 & 256 & 256 & 256\\
         fully connected layer dim & 512 & 256 & 512 &  512 & 256 & 1024 & 256 & 256 & 256\\
         attention heads $h$ & 4 & 4 & 2 & 2 & 6 & 2 & 2 & 2 & 4\\
         dropout & 0.3 & 0.1 & 0.2 & 0.05 & 0.1 & 0.05 & 0.01 & 0.01 & 0.01\\
         number of encoder layers & 2 & 3 & 2 & 1 & 1 & 1 & 1 & 1 & 1\\
         number of decoder layers & 3 & 2 & 2 & 1 & 3 & 3 & 3 & 1 & 3 \\
         learning rate & 0.0005 & 0.0005 & 0.0005 &  0.0001 & 0.001 & 0.0001 & 0.0001 & 0.001 & 0.0001\\ 
        \hline
        \multicolumn{10}{p{0.8\textwidth}}{\tiny Note: Table presents hyperparameter values selected for Informer model trained on 5min, 15min and 30min interval data with RMSE loss, Quantile Loss and GMADL loss functions.}
    \end{tabular}
    \label{tab:informer_hparams}
\end{table}

\paragraph{RMSE Informer strategy hyperparameters}
For each data window, once the model was trained, its predictions were used to create a trading strategy. To select the optimal hyperparameters, \textit{validation} part of the data window was used. The hyperparameters of this strategy are \textit{enter long}, \textit{exit long}, \textit{enter short} and \textit{exit short}. The tested values range from $0.0001$ to $0.007$. "-" indicates that the position will not be taken regardless of the model prediction. The possible values of the hyperparameters are listed in Table \ref{tab:rmse_hspace}.

\begin{table}[h!]
    \centering
    \caption{Hyperparameter space for RMSE Informer Strategy}
    \begin{tabular}{l c}
         \textbf{Parameter} & \textbf{Values} \\
         \hline
        \textit{enter long} & [-, 0.001, 0.002, 0.003, 0.004, 0.005, 0.006, 0.007] \\
        \textit{exit long} & [-, -0.001, -0.002, -0.003, -0.004, -0.005, -0.006, -0.007]\\
        \textit{enter short} & [-, -0.001, -0.002, -0.003, -0.004, -0.005, -0.006, -0.007]\\
        \textit{exit short} & [-, 0.001, 0.002, 0.003, 0.004, 0.005, 0.006, 0.007]\\
        \hline
         \multicolumn{2}{p{0.9\textwidth}}{\tiny Note: Considered values of hyperparameters for RMSE Informer strategy, all 4096 combinations were searched exhaustively.}
    \end{tabular}
    \label{tab:rmse_hspace}
\end{table}

The selected hyperparameter values are presented in Table \ref{tab:rmse_hparams}. It can be observed that the meaningful values could be selected only for the 30 minute interval. For the 15 and 5 minute interval the predictions from the model were close to 0 (smaller than the transaction fee), so to do any trades the strategy had to pick the smallest value possible for the threshold which was 0.001.

\begin{table}[h]
	\begin{center}
            \caption{Selected hyperparameter values for RMSE Informer Strategy}
		\begin{tabular}{lcccc}
			\textbf{Window} & \textbf{\textit{enter long}} & \textbf{\textit{exit Long}} & \textbf{\textit{enter Short}} & \textbf{\textit{exit Short}} \\
			\hline
			W1-5min & 0.002 & - & -0.001 & 0.001 \\
			W2-5min & - & - & -0.001 & 0.001 \\
			W3-5min & - & - & -0.001 & 0.001 \\
			W4-5min & - & - & -0.001 & 0.001 \\
			W5-5min & - & - & -0.001 & 0.001 \\
			W6-5min & - & - & -0.001 & 0.001 \\
			\hline
			W1-15min & 0.002 & - & -0.001 & - \\
			W2-15min & - & - & -0.002 & 0.002 \\
			W3-15min & - & - & -0.001 & 0.002 \\
			W4-15min & 0.001 & - & -0.002 & - \\
			W5-15min & - & - & -0.001 & 0.001 \\
			W6-15min & - & - & -0.002 & 0.001 \\
			\hline
			W1-30min & 0.007 & - & -0.001 & - \\
			W2-30min & 0.003 & - & -0.007 & - \\
			W3-30min & - & - & -0.001 & 0.005 \\
			W4-30min & 0.002 & - & -0.006 & - \\
			W5-30min & 0.005 & -0.001 & - & - \\
			W6-30min & 0.003 & - & -0.004 & - \\
                \hline
                \multicolumn{5}{p{0.8\textwidth}}{\tiny Note: Table presents hyperparameter values selected for the RMSE Informer strategy for each window of 5min, 15min and 30min interval data.}
		\end{tabular}
        \label{tab:rmse_hparams}
	\end{center}
\end{table}

\paragraph{Quantile Informer strategy hyperparameters}
For model with Quantile loss, the parameters consisted of \textit{enter long}, \textit{exit long}, \textit{enter short}, \textit{exit short} and \textit{threshold} hyperparameters described in Section \ref{paragraph:quantile_strat}. Here the possible hyperparameter values correspond to the quantiles of the loss function, with "-" meaning that the position won't be opened/closed regardless of the model prediction. Three values for the \textit{threshold} were tested with the smallest one equal to the \textit{exchange fee}. All the possible values of the hyperparameters are listed in Table \ref{tab:quantile_hspace}.
\begin{table}[h]
    \centering
    \caption{Hyperparameter space for Quantile Informer Strategy}
    \begin{tabular}{l c}
         \textbf{Parameter} & \textbf{Values} \\
         \hline
        \textit{enter long} & [-, 0.75, 0.9, 0.95, 0.97, 0.98, 0.99] \\
        \textit{exit long} & [-, 0.75, 0.9, 0.95, 0.97, 0.98, 0.99]\\
        \textit{enter short} & [-, 0.75, 0.9, 0.95, 0.97, 0.98, 0.99]\\
        \textit{exit short} & [-, 0.75, 0.9, 0.95, 0.97, 0.98, 0.99]\\
        \textit{threshold} & [0.001, 0.002, 0.003] \\
        \hline
         \multicolumn{2}{p{0.9\textwidth}}{\tiny Note: Considered values of hyperparameters for Quantile Informer strategy, all 12288 combinations were searched exhaustively.}
    \end{tabular}
    \label{tab:quantile_hspace}
\end{table}

Selected hyperparameters are presented in Table \ref{tab:quantile_hparams}. It can be observed that mostly the parameters that require high model confidence to open/close the position were selected. This may be caused by the fact, that each position incurs a fee and each mistake is extremely costly. With transaction fee equal to $0.1\%$ even if the asset price stays the same, already with $log_{0.999}(0.5) \approx 692$ incorrect position changes the portfolio value is halved - which is not that many considering the minute frequency data. 

\begin{table}[h]
	\begin{center}
        \caption{Selected hyperparameter values for Quantile Informer strategy.}
		\begin{tabular}{lccccc}
			\textbf{Window} & \textbf{\textit{enter long}} & \textbf{\textit{exit long}} & \textbf{\textit{enter short}} & \textbf{\textit{exit short}} & \textbf{\textit{threshold}} \\
			\hline
			W1-5min & - & - & 0.980 & 0.970 & 0.003 \\
			W2-5min & 0.970 & - & 0.950 & - & 0.003 \\
			W3-5min & 0.990 & - & 0.950 & - & 0.002 \\
			W4-5min & 0.970 & 0.990 & - & - & 0.003 \\
			W5-5min & 0.900 & 0.900 & - & - & 0.002 \\
			W6-5min & 0.950 & 0.950 & - & - & 0.003 \\
			\hline
			W1-15min & 0.980 & - & 0.990 & - & 0.003 \\
			W2-15min & 0.990 & 0.950 & - & - & 0.003 \\
			W3-15min & - & - & 0.990 & 0.990 & 0.003 \\
			W4-15min & 0.990 & 0.970 & - & - & 0.003 \\
			W5-15min & 0.950 & 0.990 & - & - & 0.001 \\
			W6-15min & 0.990 & 0.950 & - & - & 0.002 \\
			\hline
			W1-30min & 0.980 & - & 0.950 & 0.970 & 0.003 \\
			W2-30min & - & - & 0.980 & 0.990 & 0.001 \\
			W3-30min & - & - & 0.990 & 0.900 & 0.003 \\
			W4-30min & 0.950 & 0.990 & - & - & 0.001 \\
			W5-30min & 0.900 & 0.990 & - & - & 0.001 \\
			W6-30min & 0.970 & 0.980 & - & - & 0.003 \\
                \hline
                \multicolumn{6}{p{0.9\textwidth}}{\tiny Note: Table presents hyperparameter values selected for the Quantile Informer strategy for each window of 5min, 15min and 30min interval data.}
		\end{tabular}
        \label{tab:quantile_hparams}
	\end{center}
\end{table}

\paragraph{GMADL Informer strategy hyperparameters}
The last strategy was created using the Informer trained with GMADL loss funtion. The hyperparameters of this strategy are \textit{enter long}, \textit{exit long}, \textit{enter short} and \textit{exit short}. They are described in Section \ref{paragraph:gmadl_strat}. The tested values range from $0.0001$ to $0.007$ - the same as with RMSE Informer strategy. "-" indicates that the position will not be taken regardless of the model prediction. The possible values of the hyperparameters are listed in Table \ref{tab:gmadl_hspace}.

\begin{table}[h!]
    \centering
    \caption{Hyperparameter space for GMADL Informer Strategy}
    \begin{tabular}{l c}
         \textbf{Parameter} & \textbf{Values} \\
         \hline
        \textit{enter long} & [-, 0.001, 0.002, 0.003, 0.004, 0.005, 0.006, 0.007] \\
        \textit{exit long} & [-, -0.001, -0.002, -0.003, -0.004, -0.005, -0.006, -0.007]\\
        \textit{enter short} & [-, -0.001, -0.002, -0.003, -0.004, -0.005, -0.006, -0.007]\\
        \textit{exit short} & [-, 0.001, 0.002, 0.003, 0.004, 0.005, 0.006, 0.007]\\
        \hline
         \multicolumn{2}{p{0.9\textwidth}}{\tiny Note: Considered values of hyperparameters for GMADL Informer strategy, all 4096 combinations were searched exhaustively.}
    \end{tabular}
    \label{tab:gmadl_hspace}
\end{table}

The selected hyperparameter values are presented in Table \ref{tab:rmse_hparams}. Similarly as with the RSI strategy, the values for \textit{exit long} and \textit{exit short} were rarely selected, meaning that throughout the period, the \texttt{long position} or \texttt{short position} was always open. There seems to be no bias towards larger return threshold in the lower frequency data, indicating that the sudden price changes happen and can be leveraged even in such frequent intervals as 5 min. In case of this strategy, the larger threshold could be selected even for smaller intervals.

\begin{table}[h]
	\begin{center}
            \caption{Selected hyperparameter values for GMADL Informer Strategy}
		\begin{tabular}{lcccc}
			\textbf{Window} & \textbf{\textit{enter long}} & \textbf{\textit{exit Long}} & \textbf{\textit{enter Short}} & \textbf{\textit{exit Short}} \\
			\hline
			W1-5min & 0.004 & - & -0.005 & - \\
			W2-5min & 0.002 & - & -0.001 & - \\
			W3-5min & - & - & -0.006 & 0.003 \\
			W4-5min & 0.002 & - & -0.005 & - \\
			W5-5min & 0.002 & - & -0.003 & - \\
			W6-5min & 0.001 & - & -0.007 & - \\
			\hline
			W1-15min & 0.005 & - & -0.002 & - \\
			W2-15min & 0.006 & - & -0.002 & - \\
			W3-15min & - & - & -0.001 & 0.005 \\
			W4-15min & 0.007 & - & -0.005 & - \\
			W5-15min & 0.001 & - & -0.004 & - \\
			W6-15min & 0.002 & -0.002 & - & - \\
			\hline
			W1-30min & - & - & -0.007 & 0.005 \\
			W2-30min & - & - & -0.007 & 0.004 \\
			W3-30min & - & - & -0.004 & 0.007 \\
			W4-30min & 0.003 & -0.007 & - & - \\
			W5-30min & 0.006 & - & -0.004 & - \\
			W6-30min & 0.001 & - & -0.005 & - \\
                \hline
                \multicolumn{5}{p{0.8\textwidth}}{\tiny Note: Table presents hyperparameter values selected for the GMADL Informer strategy for each window of 5min, 15min and 30min interval data.}
		\end{tabular}
        \label{tab:gmadl_hparams}
	\end{center}
\end{table}
\clearpage

\subsection{Evaluation Results}\label{section:results}
After the hyperparameters were selected, all strategies were evaluated on the corresponding test window parts. The metrics were computed for each window and the whole period by concatenating the positions selected by each strategy for every period. In this section, first, the evaluation results are analyzed separately for each frequency, then the strategies that achieved the best overall results are compared and discussed.

\subsubsection{Evaluation on 30 min data}

\begin{figure}[h!]
        \centering
        \caption{Evaluation results on 30 min data}
        \label{fig:results-30min}
        \includegraphics[width=\linewidth]{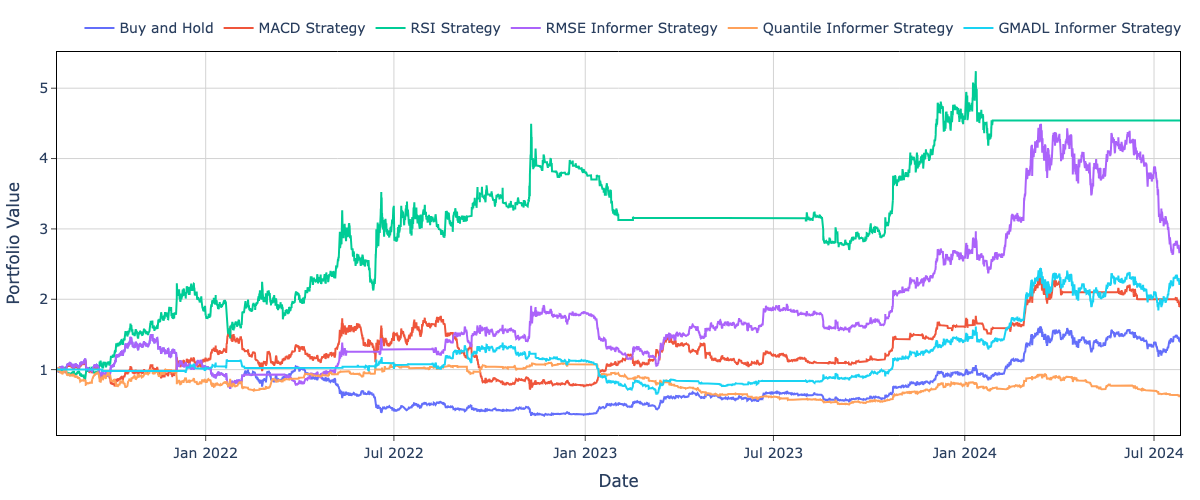}
	\begin{center}
            \small
		\begin{tabular}{lccccccccc}
			\textbf{Strategy} & \textbf{VAL} & \textbf{ARC} & \textbf{ASD} & \textbf{IR*} & \textbf{MD} & \textbf{IR**} & \textbf{N} & \textbf{LONG} & \textbf{SHORT} \\
			\hline
			Buy and Hold & 1.440 & 13.12\% & 55.95\% & 0.235 & 77.20\% & 0.040 & 2 & 100.00\% & 0.00\% \\
			MACD Strategy & 1.952 & 25.37\% & 52.36\% & 0.485 & 59.24\% & 0.207 & 327 & 52.30\% & 28.30\% \\
			RSI Strategy & 4.542 & 66.77\% & 46.25\% & 1.444 & 39.91\% & 2.415 & 377 & 30.79\% & 28.03\% \\
			RMSE Informer & 2.727 & 40.37\% & 50.47\% & 0.800 & 51.75\% & 0.624 & 34 & 64.40\% & 24.67\% \\
			Quantile Informer & 0.629 & -14.51\% & 36.91\% & -0.393 & 55.09\% & -0.104 & 1783 & 30.24\% & 15.27\% \\
			GMADL Informer & 2.263 & 31.79\% & 36.70\% & 0.866 & 53.35\% & 0.516 & 811 & 35.51\% & 19.59\% \\
   \hline
   \multicolumn{10}{p{\textwidth}}{\tiny
     Note: Evaluation results on the whole testing period of 30min interval BTC/USDT data. The presented metrics are: portfolio value at the end of the evaluation period (VAL), Annualized Return Compound (ARD), Annualized Standard Deviation (ASD), Information Ratio (IR*), Maximum Drawdown (MD), Modified Information Ratio (IR**), number of trades (N) and percent of the long/short positions (LONG/SHORT).}
		\end{tabular}
	\end{center}
\end{figure}

With the largest  30 minute interval dataset, the best strategy turned out to be the one based on RSI indicator. The results presented in Figure \ref{fig:results-30min}, show that all strategies except the one based on Quantile Informer signal achieved better results than buy-and-hold strategy. While RSI strategy overall achieved the best results, there were long periods during which this strategy took no position. The RMSE Informer, GMADL Informer and MACD strategy achieved similar results, though it is worth noting that GMADL strategy made much more trades 811 while RMSE Informer strategy changed its position only 34 times.

\begin{figure}[h!]
\caption{Evaluation results for individual windows on 30 min data}
\label{fig:results-30min-windows}
\vspace{0.5cm}
\hspace{-0.09\linewidth}
\begin{minipage}[b]{0.59\linewidth}
\includegraphics[width=\linewidth]{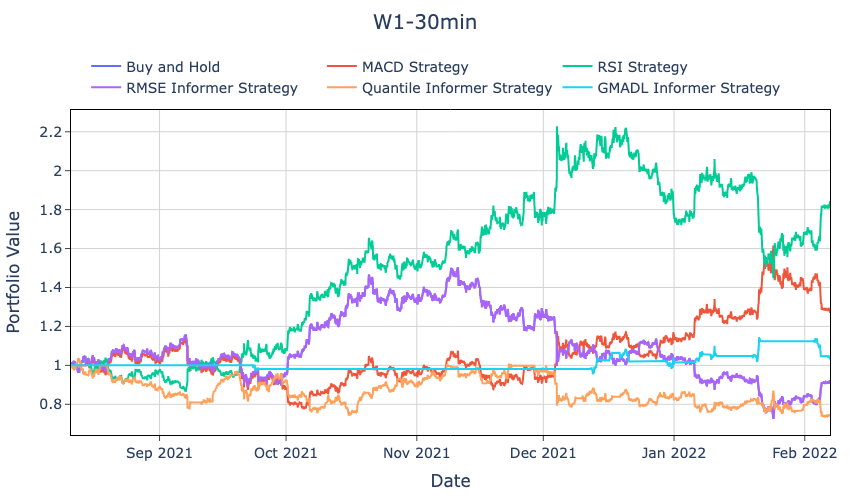}
\tiny
\tabcolsep1pt
\begin{tabular}{lccccccccc}
    \textbf{Strategy} & \textbf{VAL} & \textbf{ARC} & \textbf{ASD} & \textbf{IR*} & \textbf{MD} & \textbf{IR**} & \textbf{N} & \textbf{LONG} & \textbf{SHORT} \\
    \hline
    Buy and Hold & 0.924 & -14.84\% & 66.12\% & -0.225 & 51.75\% & -0.064 & 2 & 100.00\% & 0.00\% \\
    MACD Strategy & 1.269 & 62.19\% & 65.90\% & 0.944 & 31.81\% & 1.845 & 15 & 50.36\% & 47.71\% \\
    RSI Strategy & 1.843 & 245.59\% & 66.51\% & 3.693 & 35.02\% & 25.897 & 26 & 42.40\% & 57.60\% \\
    RMSE Informer & 0.924 & -14.84\% & 66.12\% & -0.225 & 51.75\% & -0.064 & 2 & 100.00\% & 0.00\% \\
    Quantile Informer & 0.743 & -45.20\% & 60.57\% & -0.746 & 29.11\% & -1.159 & 443 & 30.43\% & 53.98\% \\
    GMADL Informer & 1.030 & 6.23\% & 19.98\% & 0.312 & 10.01\% & 0.194 & 39 & 0.00\% & 8.72\% \\
\end{tabular}
\includegraphics[width=\linewidth]{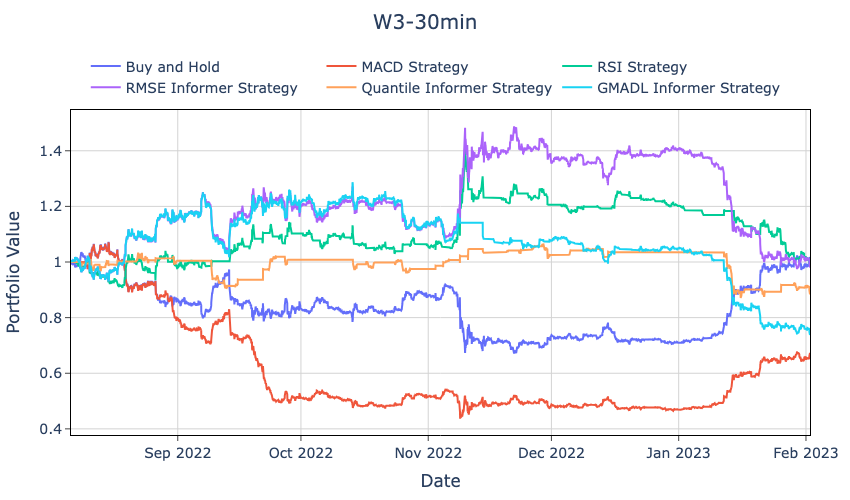}
\begin{tabular}{lccccccccc}
    \textbf{Strategy} & \textbf{VAL} & \textbf{ARC} & \textbf{ASD} & \textbf{IR*} & \textbf{MD} & \textbf{IR**} & \textbf{N} & \textbf{LONG} & \textbf{SHORT} \\
    \hline
    Buy and Hold & 1.018 & 3.75\% & 51.25\% & 0.073 & 37.47\% & 0.007 & 2 & 100.00\% & 0.00\% \\
    MACD Strategy & 0.673 & -55.16\% & 51.01\% & -1.081 & 59.24\% & -1.007 & 95 & 81.60\% & 16.48\% \\
    RSI Strategy & 0.988 & -2.50\% & 38.04\% & -0.066 & 30.42\% & -0.005 & 238 & 2.34\% & 50.54\% \\
    RMSE Informer & 0.980 & -4.00\% & 51.36\% & -0.078 & 34.51\% & -0.009 & 2 & 0.00\% & 100.00\% \\
    Quantile Informer & 0.884 & -22.19\% & 23.41\% & -0.948 & 18.39\% & -1.144 & 171 & 0.00\% & 18.83\% \\
    GMADL Informer & 0.737 & -46.11\% & 42.58\% & -1.083 & 42.71\% & -1.169 & 302 & 0.00\% & 81.21\% \\
\end{tabular}
\includegraphics[width=\linewidth]{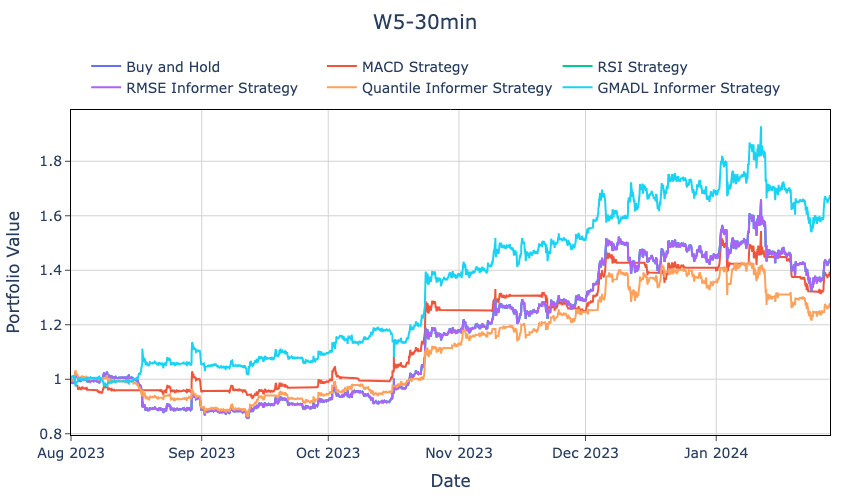}
\begin{tabular}{lccccccccc}
    \textbf{Strategy} & \textbf{VAL} & \textbf{ARC} & \textbf{ASD} & \textbf{IR*} & \textbf{MD} & \textbf{IR**} & \textbf{N} & \textbf{LONG} & \textbf{SHORT} \\
    \hline
    Buy and Hold & 1.439 & 109.30\% & 41.49\% & 2.634 & 20.18\% & 14.265 & 2 & 100.00\% & 0.00\% \\
    MACD Strategy & 1.393 & 95.71\% & 31.47\% & 3.041 & 15.00\% & 19.407 & 90 & 48.66\% & 0.00\% \\
    RSI Strategy & 1.439 & 109.30\% & 41.49\% & 2.634 & 20.18\% & 14.265 & 2 & 100.00\% & 0.00\% \\
    RMSE Informer & 1.439 & 109.30\% & 41.49\% & 2.634 & 20.18\% & 14.265 & 2 & 100.00\% & 0.00\% \\
    Quantile Informer & 1.277 & 64.08\% & 34.28\% & 1.869 & 16.61\% & 7.212 & 311 & 76.90\% & 0.00\% \\
    GMADL Informer & 1.672 & 183.71\% & 41.56\% & 4.420 & 20.18\% & 40.230 & 22 & 78.48\% & 21.52\% \\
\end{tabular}
\end{minipage}
\begin{minipage}[b]{0.59\linewidth}
\raggedleft
\includegraphics[width=\linewidth]{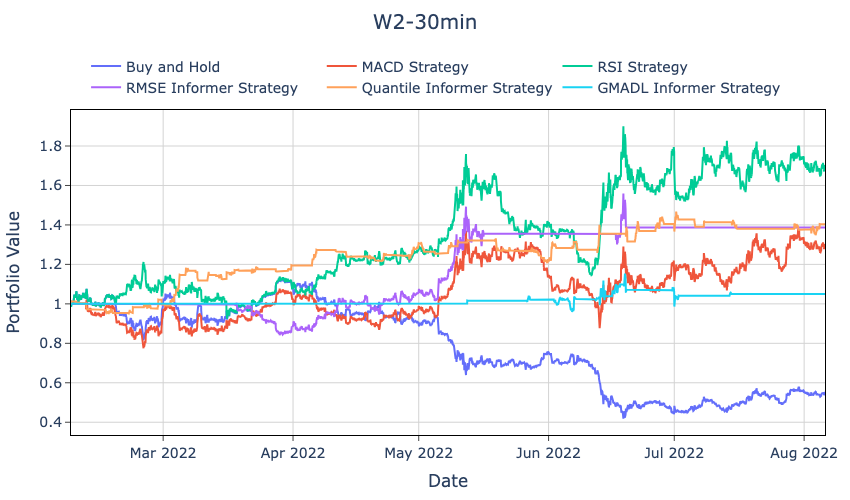}
\tiny
\tabcolsep1pt
\begin{tabular}{lccccccccc}
    \textbf{Strategy} & \textbf{VAL} & \textbf{ARC} & \textbf{ASD} & \textbf{IR*} & \textbf{MD} & \textbf{IR**} & \textbf{N} & \textbf{LONG} & \textbf{SHORT} \\
    \hline
    Buy and Hold & 0.548 & -70.51\% & 72.38\% & -0.974 & 63.18\% & -1.087 & 2 & 100.00\% & 0.00\% \\
    MACD Strategy & 1.303 & 70.93\% & 71.82\% & 0.988 & 36.06\% & 1.943 & 19 & 57.32\% & 40.76\% \\
    RSI Strategy & 1.703 & 194.24\% & 72.36\% & 2.684 & 34.89\% & 14.945 & 106 & 39.96\% & 60.04\% \\
    RMSE Informer & 1.387 & 94.09\% & 41.55\% & 2.264 & 16.34\% & 13.037 & 8 & 0.00\% & 34.41\% \\
    Quantile Informer & 1.403 & 98.74\% & 27.62\% & 3.575 & 9.44\% & 37.393 & 255 & 0.00\% & 18.79\% \\
    GMADL Informer & 1.050 & 10.44\% & 23.40\% & 0.446 & 14.45\% & 0.322 & 90 & 0.00\% & 5.98\% \\
\end{tabular}
\raggedleft
\includegraphics[width=\linewidth]{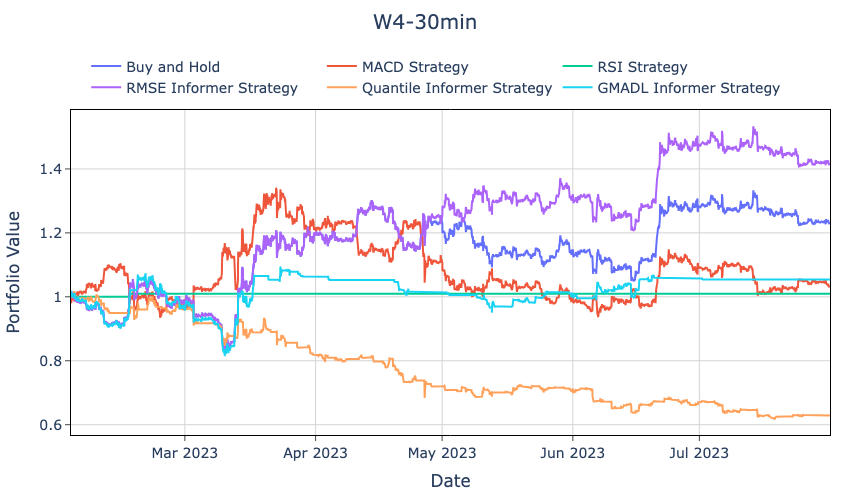}
\begin{tabular}{lccccccccc}
    \textbf{Strategy} & \textbf{VAL} & \textbf{ARC} & \textbf{ASD} & \textbf{IR*} & \textbf{MD} & \textbf{IR**} & \textbf{N} & \textbf{LONG} & \textbf{SHORT} \\
    \hline
    Buy and Hold & 1.229 & 51.90\% & 45.01\% & 1.153 & 21.74\% & 2.753 & 2 & 100.00\% & 0.00\% \\
    MACD Strategy & 1.031 & 6.45\% & 45.08\% & 0.143 & 30.15\% & 0.031 & 94 & 35.18\% & 64.82\% \\
    RSI Strategy & 1.010 & 2.03\% & 1.87\% & 1.084 & 0.30\% & 7.410 & 4 & 0.02\% & 0.00\% \\
    RMSE Informer & 1.414 & 101.80\% & 45.03\% & 2.260 & 21.74\% & 10.584 & 10 & 92.95\% & 7.05\% \\
    Quantile Informer & 0.629 & -60.99\% & 28.76\% & -2.121 & 39.23\% & -3.297 & 448 & 36.32\% & 0.00\% \\
    GMADL Informer & 1.054 & 11.31\% & 28.25\% & 0.400 & 24.18\% & 0.187 & 345 & 34.69\% & 0.00\% \\
\end{tabular}
\raggedleft
\includegraphics[width=\linewidth]{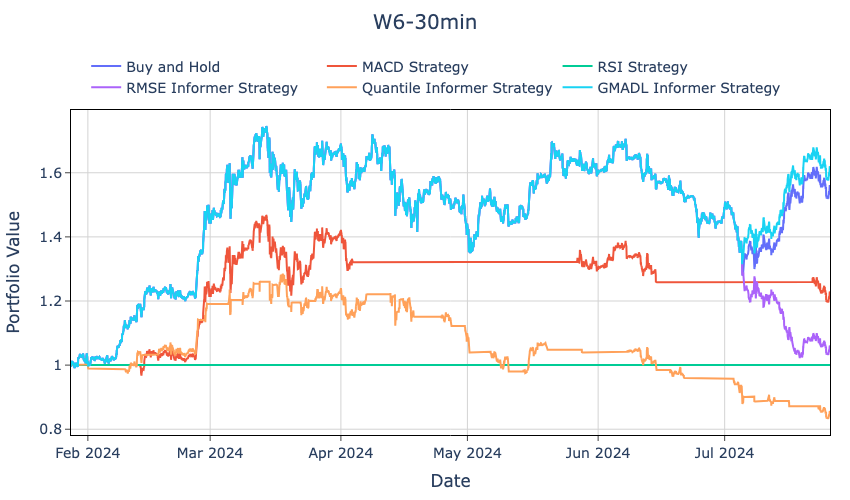}
\tabcolsep1.1pt
\begin{tabular}{lccccccccc}
    \textbf{Strategy} & \textbf{VAL} & \textbf{ARC} & \textbf{ASD} & \textbf{IR*} & \textbf{MD} & \textbf{IR**} & \textbf{N} & \textbf{LONG} & \textbf{SHORT} \\
    \hline
    Buy and Hold & 1.559 & 145.95\% & 52.85\% & 2.762 & 26.69\% & 15.103 & 2 & 100.00\% & 0.00\% \\
    MACD Strategy & 1.227 & 51.44\% & 36.38\% & 1.414 & 18.55\% & 3.921 & 11 & 40.69\% & 0.00\% \\
    RSI Strategy & 1 & 0.00\% & 0.00\% & 0 & 0.00\% & 0 & 0 & 0.00\% & 0.00\% \\
    RMSE Informer & 1.059 & 12.32\% & 52.86\% & 0.233 & 41.44\% & 0.069 & 10 & 93.44\% & 6.56\% \\
    Quantile Informer & 0.855 & -27.14\% & 34.50\% & -0.787 & 35.23\% & -0.606 & 150 & 37.79\% & 0.00\% \\
    GMADL Informer & 1.617 & 165.10\% & 52.85\% & 3.124 & 24.64\% & 20.929 & 10 & 99.88\% & 0.12\% \\
\end{tabular}
\end{minipage}
\tiny
\begin{tabular}{c}
    \hline
    \multicolumn{1}{p{\textwidth}}{
    \vspace{0.02cm}
     Note: Evaluation results for each of the out-of-sample data window of 30min interval BTC/USDT data. The presented metrics are: portfolio value at the end of the evaluation period (VAL), Annualized Return Compound (ARD), Annualized Standard Deviation (ASD), Information Ratio (IR*), Maximum Drawdown (MD), Modified Information Ratio (IR**), number of trades (N) and percent of the long/short positions (LONG/SHORT).} \\
\end{tabular}
\end{figure}

Looking at the results in the individual windows (Figure \ref{fig:results-30min-windows}) it can be observed that the periods when RSI strategy did not take any posisions happened in 4th and 6th window. In fact, during the first of these periods, the RSI strategy took a position for only 0.02\% of the time in the fourth window, and no position in the second. This may suggest that the distribution of validation data on which the hyperparameters were selected differed from the testing data or that there was high uncertainty during those periods.  
MACD strategy achieved better results than buy-and-hold benchmark in majority of periods, though overall performance was hindered by the poor result during the 3rd period.
Informer based strategies with GMADL and RMSE achieved very similar results, with GMADL Informer beased strategy achieving better result when the Bitcoin was in upward trend (5th window) and RMSE Informer based strategy achieving better result when it was in downward trend (2nd window). Quantile Informer strategy had poor performance in 4th and 6th window wich greatly influenced the overall performance. The fact that these were the windows in which the best strategy RSI took no position and that the second best strategy in both of these periods was buy-and-hold approach may suggest that they were the most difficult periods.

\clearpage

\subsubsection{Evaluation on 15 min data}
\begin{figure}[h!]
        \centering
        \caption{Evaluation results on 15 min data}
        \label{fig:results-15min}
        \includegraphics[width=\linewidth]{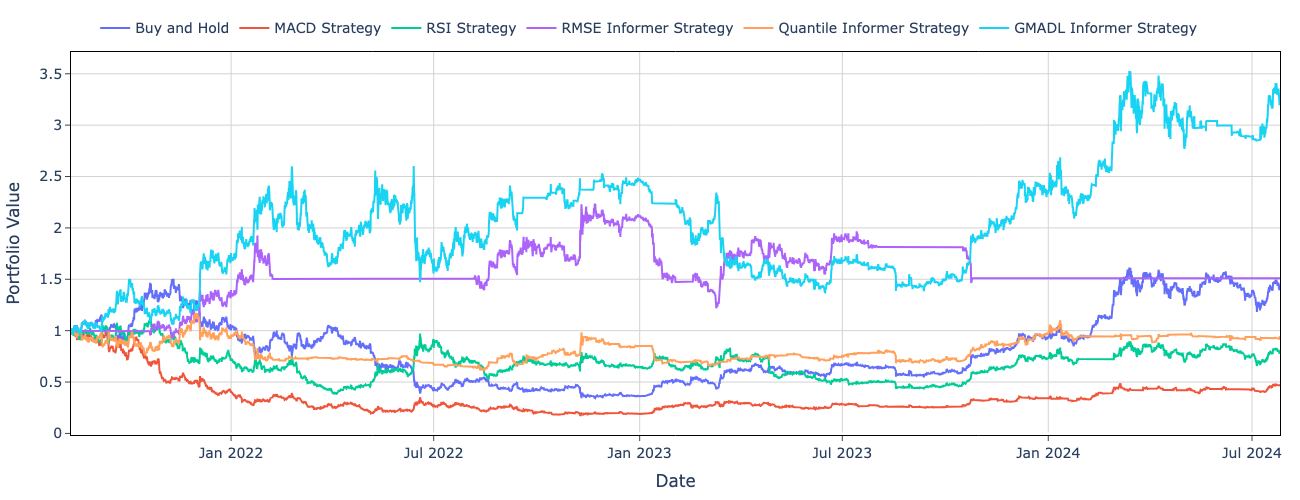}
        \vspace{-1cm}
	\begin{center}
             \small
		\begin{tabular}{lccccccccc}
			\textbf{Strategy} & \textbf{VAL} & \textbf{ARC} & \textbf{ASD} & \textbf{IR*} & \textbf{MD} & \textbf{IR**} & \textbf{N} & \textbf{LONG} & \textbf{SHORT} \\
			\hline
    Buy and Hold & 1.440 & 13.10\% & 56.03\% & 0.234 & 77.23\% & 0.040 & 2 & 100.00\% & 0.00\% \\
    MACD Strategy & 0.468 & -22.64\% & 52.43\% & -0.432 & 83.18\% & -0.118 & 1311 & 51.80\% & 31.33\% \\
    RSI Strategy & 0.800 & -7.28\% & 55.66\% & -0.131 & 66.67\% & -0.014 & 1206 & 54.02\% & 43.08\% \\
    RMSE Informer & 1.509 & 14.93\% & 34.90\% & 0.428 & 45.54\% & 0.140 & 16 & 15.24\% & 27.60\% \\
    Quantile Informer & 0.945 & -1.91\% & 37.77\% & -0.051 & 48.30\% & -0.002 & 824 & 28.48\% & 16.77\% \\
    GMADL Informer & 3.296 & 49.65\% & 52.70\% & 0.942 & 47.39\% & 0.987 & 362 & 49.37\% & 37.72\% \\
   \hline
   \multicolumn{10}{p{\textwidth}}{\tiny
     Note: Evaluation results on the whole testing period of 15min interval BTC/USDT data. The presented metrics are: portfolio value at the end of the evaluation period (VAL), Annualized Return Compound (ARD), Annualized Standard Deviation (ASD), Information Ratio (IR*), Maximum Drawdown (MD), Modified Information Ratio (IR**), number of trades (N) and percent of the long/short positions (LONG/SHORT).}
		\end{tabular}
	\end{center}
\end{figure}

The results for the 15 min data are presented in Figure \ref{fig:results-15min}. Here, the GMADL Informer strategy  yielded the best results. The modified information ratio was was the best one from the strategies evaluated on 15 min interval. The next best strategy was RMSE Informer and the 3rd best buy-and-hold benchmark. The strategies based on technical indicators as well as Quantile Informer performed poorly, achieving results worse than buy-and-hold. Quantile Informer Strategy had a position only for the 40\% of time suggesting the selection of the hyperparameters that too aggressively limit the event when the position is entered/closed. MACD and RSI Strategies had more position changes, which might have negatively contributed to their performance due to high exchange fees.

\begin{figure}[h!]
\caption{Evaluation results for individual windows on 15 min data}
\label{fig:results-15min-windows}
\vspace{0.5cm}
\hspace{-0.09\linewidth}
\begin{minipage}[b]{0.59\linewidth}
\includegraphics[width=\linewidth]{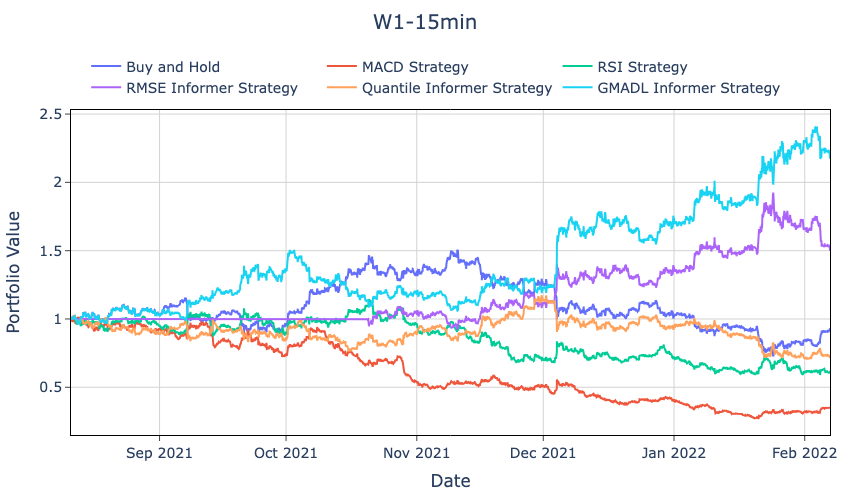}
\tiny
\tabcolsep1pt
\begin{tabular}{lccccccccc}
    \textbf{Strategy} & \textbf{VAL} & \textbf{ARC} & \textbf{ASD} & \textbf{IR*} & \textbf{MD} & \textbf{IR**} & \textbf{N} & \textbf{LONG} & \textbf{SHORT} \\
    \hline
    Buy and Hold & 0.933 & -13.15\% & 66.69\% & -0.197 & 51.81\% & -0.050 & 2 & 100.00\% & 0.00\% \\
    MACD Strategy & 0.357 & -87.59\% & 66.96\% & -1.308 & 73.46\% & -1.559 & 852 & 52.49\% & 47.51\% \\
    RSI Strategy & 0.621 & -61.93\% & 66.94\% & -0.925 & 48.44\% & -1.183 & 882 & 58.61\% & 41.39\% \\
    RMSE Informer & 1.498 & 127.03\% & 52.56\% & 2.417 & 22.20\% & 13.827 & 3 & 0.00\% & 61.05\% \\
    Quantile Informer & 0.715 & -49.34\% & 66.74\% & -0.739 & 40.17\% & -0.908 & 182 & 52.75\% & 47.25\% \\
    GMADL Informer & 2.173 & 382.27\% & 66.83\% & 5.720 & 29.76\% & 73.474 & 146 & 36.88\% & 63.12\% \\
\end{tabular}
\includegraphics[width=\linewidth]{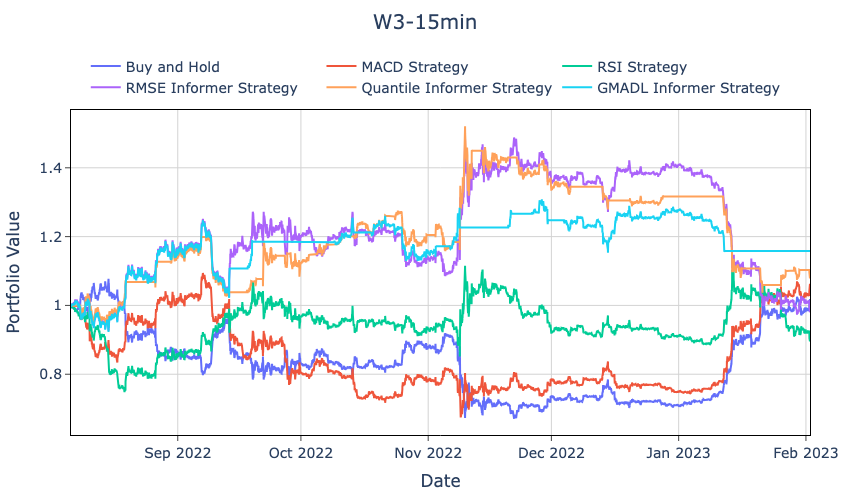}
\begin{tabular}{lccccccccc}
    \textbf{Strategy} & \textbf{VAL} & \textbf{ARC} & \textbf{ASD} & \textbf{IR*} & \textbf{MD} & \textbf{IR**} & \textbf{N} & \textbf{LONG} & \textbf{SHORT} \\
    \hline
    Buy and Hold & 1.016 & 3.26\% & 50.58\% & 0.065 & 37.76\% & 0.006 & 2 & 100.00\% & 0.00\% \\
    MACD Strategy & 1.060 & 12.62\% & 50.57\% & 0.250 & 38.35\% & 0.082 & 55 & 66.38\% & 32.66\% \\
    RSI Strategy & 0.898 & -19.62\% & 50.65\% & -0.387 & 25.20\% & -0.302 & 174 & 23.30\% & 76.70\% \\
    RMSE Informer & 0.982 & -3.55\% & 50.69\% & -0.070 & 34.60\% & -0.007 & 2 & 0.00\% & 100.00\% \\
    Quantile Informer & 1.081 & 17.17\% & 40.53\% & 0.424 & 31.82\% & 0.229 & 103 & 0.00\% & 53.36\% \\
    GMADL Informer & 1.158 & 34.60\% & 31.80\% & 1.088 & 18.06\% & 2.085 & 32 & 0.00\% & 59.07\% \\
\end{tabular}
\includegraphics[width=\linewidth]{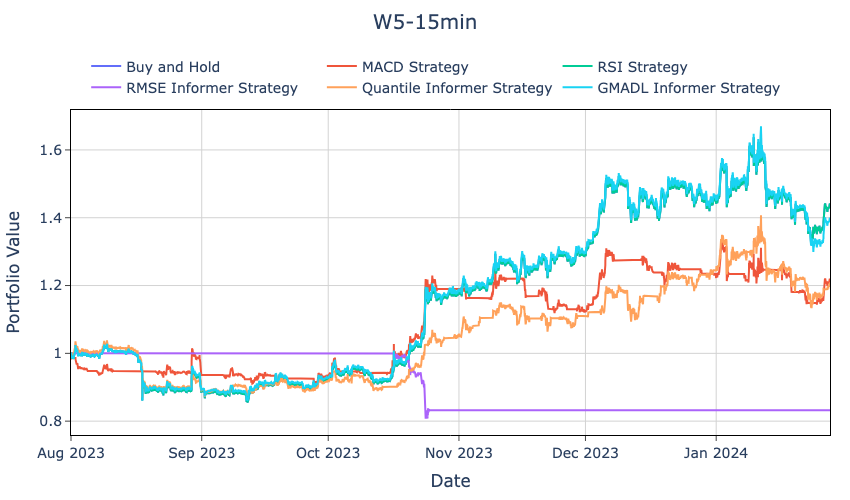}
\begin{tabular}{lccccccccc}
    \textbf{Strategy} & \textbf{VAL} & \textbf{ARC} & \textbf{ASD} & \textbf{IR*} & \textbf{MD} & \textbf{IR**} & \textbf{N} & \textbf{LONG} & \textbf{SHORT} \\
    \hline
    Buy and Hold & 1.440 & 109.38\% & 44.60\% & 2.452 & 20.31\% & 13.204 & 2 & 100.00\% & 0.00\% \\
    MACD Strategy & 1.218 & 49.24\% & 33.63\% & 1.464 & 13.88\% & 5.195 & 118 & 50.76\% & 0.00\% \\
    RSI Strategy & 1.440 & 109.38\% & 44.60\% & 2.452 & 20.31\% & 13.204 & 2 & 100.00\% & 0.00\% \\
    RMSE Informer & 0.832 & -31.06\% & 13.66\% & -2.274 & 19.29\% & -3.662 & 4 & 0.00\% & 4.56\% \\
    Quantile Informer & 1.206 & 46.21\% & 37.95\% & 1.218 & 19.49\% & 2.887 & 147 & 78.97\% & 0.00\% \\
    GMADL Informer & 1.398 & 97.31\% & 44.61\% & 2.181 & 22.22\% & 9.554 & 26 & 99.64\% & 0.36\% \\
\end{tabular}
\end{minipage}
\begin{minipage}[b]{0.59\linewidth}
\raggedleft
\includegraphics[width=\linewidth]{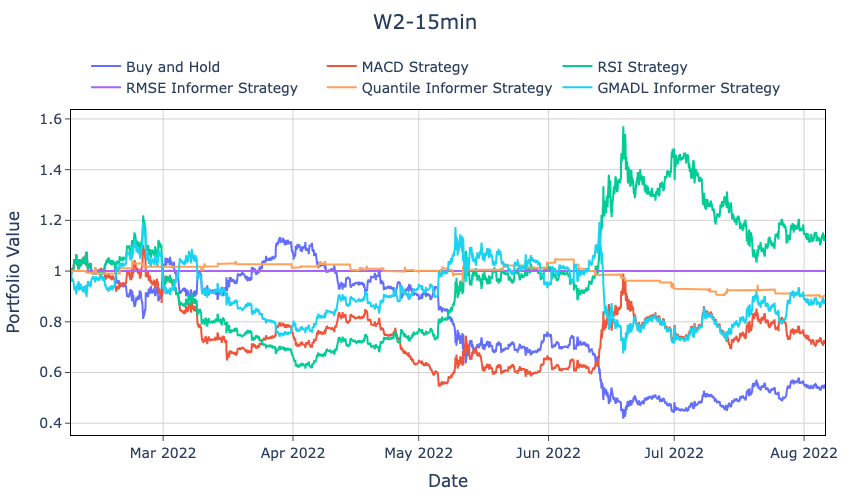}
\tiny
\tabcolsep1pt
\begin{tabular}{lccccccccc}
    \textbf{Strategy} & \textbf{VAL} & \textbf{ARC} & \textbf{ASD} & \textbf{IR*} & \textbf{MD} & \textbf{IR**} & \textbf{N} & \textbf{LONG} & \textbf{SHORT} \\
    \hline
    Buy and Hold & 0.548 & -70.48\% & 72.12\% & -0.977 & 63.18\% & -1.090 & 2 & 100.00\% & 0.00\% \\
    MACD Strategy & 0.714 & -49.48\% & 72.07\% & -0.687 & 50.67\% & -0.670 & 118 & 57.61\% & 42.39\% \\
    RSI Strategy & 1.141 & 30.61\% & 72.05\% & 0.425 & 49.51\% & 0.263 & 58 & 19.64\% & 80.36\% \\
    RMSE Informer & 1 & 0.00\% & 0.00\% & 0 & 0.00\% & 0 & 0 & 0.00\% & 0.00\% \\
    Quantile Informer & 0.903 & -18.73\% & 15.66\% & -1.196 & 14.91\% & -1.503 & 202 & 4.74\% & 0.00\% \\
    GMADL Informer & 0.885 & -21.88\% & 72.10\% & -0.303 & 43.48\% & -0.153 & 54 & 35.85\% & 64.15\% \\
\end{tabular}
\raggedleft
\includegraphics[width=\linewidth]{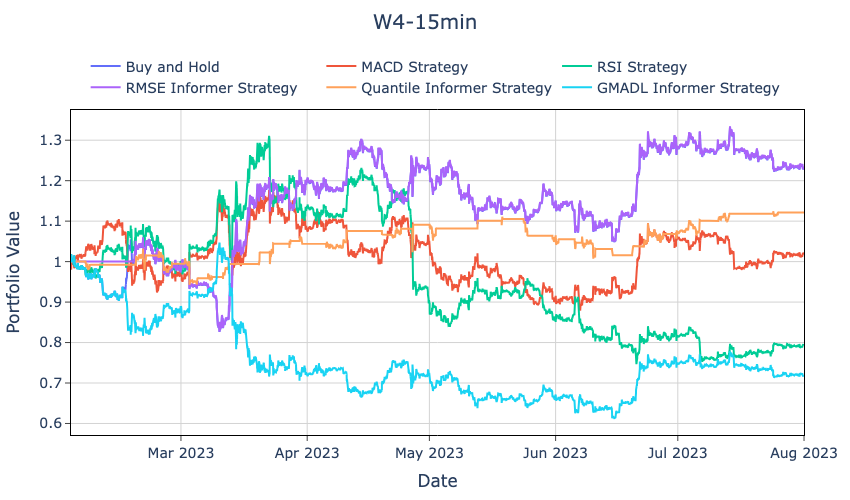}
\begin{tabular}{lccccccccc}
    \textbf{Strategy} & \textbf{VAL} & \textbf{ARC} & \textbf{ASD} & \textbf{IR*} & \textbf{MD} & \textbf{IR**} & \textbf{N} & \textbf{LONG} & \textbf{SHORT} \\
    \hline
    Buy and Hold & 1.231 & 52.34\% & 44.11\% & 1.186 & 22.01\% & 2.822 & 2 & 100.00\% & 0.00\% \\
    MACD Strategy & 1.020 & 4.03\% & 44.11\% & 0.091 & 25.36\% & 0.015 & 74 & 34.59\% & 65.41\% \\
    RSI Strategy & 0.794 & -37.41\% & 44.15\% & -0.848 & 43.03\% & -0.737 & 86 & 39.98\% & 60.02\% \\
    RMSE Informer & 1.232 & 52.63\% & 42.30\% & 1.244 & 22.01\% & 2.975 & 3 & 91.43\% & 0.00\% \\
    Quantile Informer & 1.122 & 26.21\% & 21.18\% & 1.238 & 9.32\% & 3.481 & 106 & 24.72\% & 0.00\% \\
    GMADL Informer & 0.718 & -48.86\% & 44.08\% & -1.108 & 41.58\% & -1.302 & 10 & 60.34\% & 39.66\% \\
\end{tabular}
\raggedleft
\includegraphics[width=\linewidth]{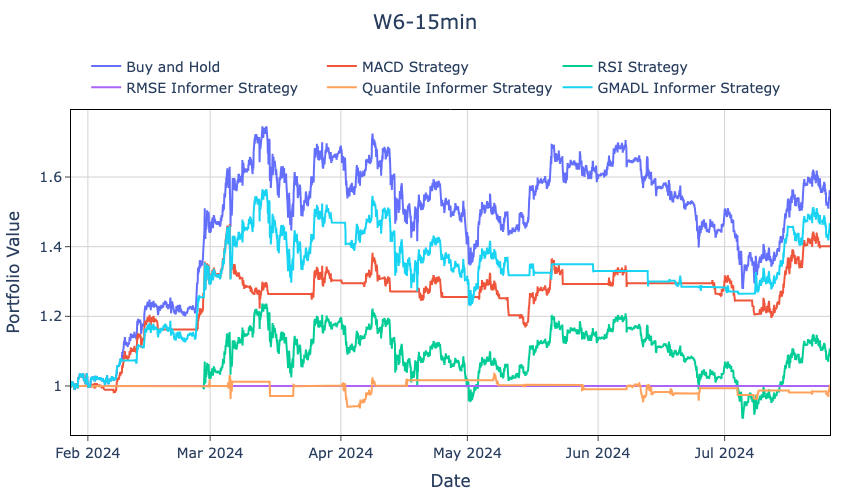}
\tabcolsep1.1pt
\begin{tabular}{lccccccccc}
    \textbf{Strategy} & \textbf{VAL} & \textbf{ARC} & \textbf{ASD} & \textbf{IR*} & \textbf{MD} & \textbf{IR**} & \textbf{N} & \textbf{LONG} & \textbf{SHORT} \\
    \hline
    Buy and Hold & 1.558 & 145.73\% & 51.90\% & 2.808 & 26.76\% & 15.290 & 2 & 100.00\% & 0.00\% \\
    MACD Strategy & 1.402 & 98.29\% & 34.32\% & 2.864 & 20.22\% & 13.925 & 93 & 48.96\% & 0.00\% \\
    RSI Strategy & 1.104 & 22.18\% & 49.12\% & 0.452 & 26.76\% & 0.374 & 3 & 82.57\% & 0.00\% \\
    RMSE Informer & 1 & 0.00\% & 0.00\% & 0 & 0.00\% & 0 & 0 & 0.00\% & 0.00\% \\
    Quantile Informer & 1.000 & 0.04\% & 18.06\% & 0.002 & 9.32\% & 0.000 & 81 & 9.73\% & 0.00\% \\
    GMADL Informer & 1.463 & 116.41\% & 45.31\% & 2.569 & 21.45\% & 13.942 & 94 & 63.52\% & 0.00\% \\
\end{tabular}
\end{minipage}
\tiny
\begin{tabular}{c}
    \hline
    \multicolumn{1}{p{\textwidth}}{
    \vspace{0.02cm}
     Note: Evaluation results for each of the out-of-sample data window of 15min interval BTC/USDT data. The presented metrics are: portfolio value at the end of the evaluation period (VAL), Annualized Return Compound (ARD), Annualized Standard Deviation (ASD), Information Ratio (IR*), Maximum Drawdown (MD), Modified Information Ratio (IR**), number of trades (N) and percent of the long/short positions (LONG/SHORT).} \\
\end{tabular}
\end{figure}

The results of the 15-minute individual windows presented in Figure \ref{fig:results-15min-windows}, show that the GMADL Informer strategy was the best only in two windows (first and third) and the majority of the value was accumulated during the first period. What is more it was the worst performing strategy in the fourth period. 
The RMSE Informer strategy also accumulated most of the value during the first period. Also it did not perform almost any trades during 2nd, 5th and 6th periods. This may suggest that the signal from Informer model was too weak - all predictions of returns were too close to 0 and the larger price movements that would incur position change were not detected correctly.
 Although the Quantile Informer strategy again poor overall results, it was the best performing strategy in the fourth window. However, during the other periods the positions either were changed too frequently or were missing on the high price swings. We can again observe that the last period proved to be the most difficult, and no strategy beat the buy-and-hold benchmark during that period. Despite poor overall results, MACD Strategy wasn't significantly worse than the other strategies except on the first window, on which a significant loss of portfolio value (value at the end of this period was 0.35) influenced the result on the whole testing period.

\clearpage

\subsubsection{Evaluation on 5 min data}
\begin{figure}[h!]
        \centering
        \caption{Evaluation results on 5 min data}
        \label{fig:results-5min}
        \includegraphics[width=\linewidth]{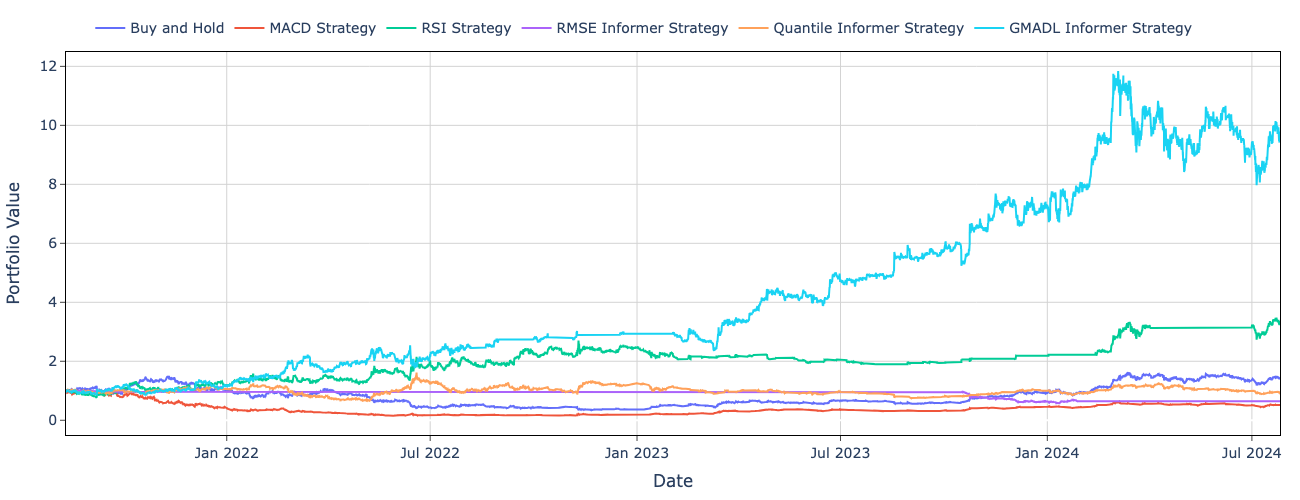}
        \vspace{-1cm}
	\begin{center}
            \small
		\begin{tabular}{lccccccccc}
			\textbf{Strategy} & \textbf{VAL} & \textbf{ARC} & \textbf{ASD} & \textbf{IR*} & \textbf{MD} & \textbf{IR**} & \textbf{N} & \textbf{LONG} & \textbf{SHORT} \\
			\hline
			Buy and Hold & 1.441 & 13.14\% & 57.74\% & 0.228 & 77.31\% & 0.039 & 2 & 100.00\% & 0.00\% \\
			MACD Strategy & 0.516 & -20.04\% & 54.14\% & -0.370 & 85.77\% & -0.087 & 2535 & 50.39\% & 32.38\% \\
			RSI Strategy & 3.341 & 50.34\% & 50.41\% & 0.999 & 29.99\% & 1.676 & 846 & 28.29\% & 33.47\% \\
			RMSE Informer & 0.643 & -13.88\% & 15.13\% & -0.917 & 44.61\% & -0.285 & 16 & 0.00\% & 9.58\% \\
			Quantile Informer & 0.956 & -1.52\% & 47.91\% & -0.032 & 53.96\% & -0.001 & 3395 & 40.24\% & 27.84\% \\
			GMADL Informer & 9.747 & 115.88\% & 54.44\% & 2.129 & 32.66\% & 7.552 & 846 & 44.80\% & 41.51\% \\
            \hline
            \multicolumn{10}{p{\textwidth}}{\tiny
             Note: Evaluation results on the whole testing period of 5min interval BTC/USDT data. The presented metrics are: portfolio value at the end of the evaluation period (VAL), Annualized Return Compound (ARD), Annualized Standard Deviation (ASD), Information Ratio (IR*), Maximum Drawdown (MD), Modified Information Ratio (IR**), number of trades (N) and percent of the long/short positions (LONG/SHORT).}
		\end{tabular}
	\end{center}
\end{figure}

Finally evaluation results for the 5 min interval throughout the testing period are presented in Figure \ref{fig:results-5min}, a GMADL Informer strategy significantly outperformed the other strategies. Throughout the testing period, it achieved annualized returns of $115\%$, while maintaining annualized standard division at a level similar to the buy-and-hold benchmark. What is more, the maximum drawdown of this strategy was only 32.7\% which is significantly lower than the 77.3\% of Buy and Hold. The strategy kept the \text{long position} and \textit{short position} for approximately a similar period of time. 
The next best strategy was an RSI based strategy, which however for almost 40\% of time did not hold a position. These are the only two strategies that beat the benchmark. It is worth noting that they had much fewer positions changes than the other strategies that performed worse, suggesting that they were better at detecting the trend changes and filtering out the insignificant signal. Both Quantile Informer and RMSE Informer achieved poor results, suggesting that these loss functions are not fit for higher frequency data. Though each for a different reason. RMSE Informer based strategy took only 16 trades during the whole period further showing the effect initially observed on 15 min interval. That with the many smaller returns model trained with this loss function has difficulty correctly predicting large returns (since when predicted incorrectly they are heavily penalized) and thus most of the model predictions are close to zero, falling below the exchange fee making it impossible to build strategy upon. On the other hand, Quantile informer performed the most trades from all the strategy suggesting the difficulty with filtering out insignificant signals, and in process loosing lots of value due to transaction fees.

\begin{figure}[h!]
\caption{Evaluation results for individual windows on 5 min data}
\label{fig:results-5min-windows}
\vspace{0.5cm}
\hspace{-0.09\linewidth}
\begin{minipage}[b]{0.59\linewidth}
\includegraphics[width=\linewidth]{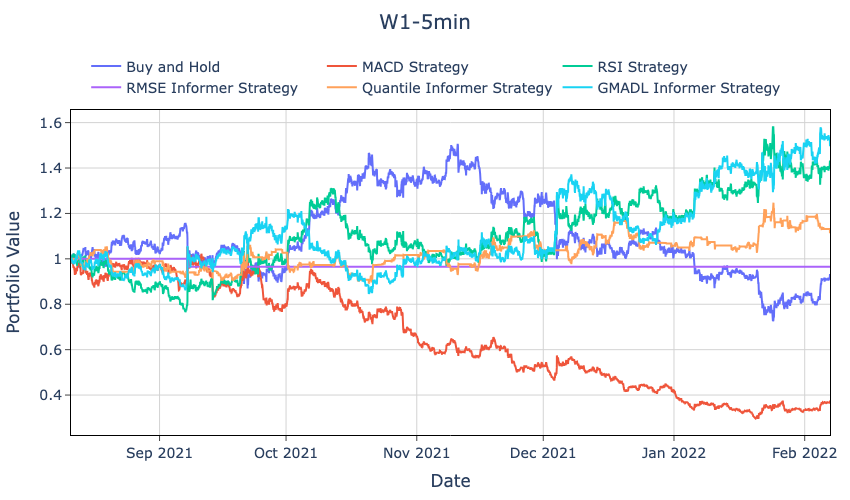}
\tiny
\tabcolsep1pt
\begin{tabular}{lccccccccc}
    \textbf{Strategy} & \textbf{VAL} & \textbf{ARC} & \textbf{ASD} & \textbf{IR*} & \textbf{MD} & \textbf{IR**} & \textbf{N} & \textbf{LONG} & \textbf{SHORT} \\
    \hline
    Buy and Hold & 0.930 & -13.76\% & 70.24\% & -0.196 & 51.87\% & -0.052 & 2 & 100.00\% & 0.00\% \\
    MACD Strategy & 0.375 & -86.35\% & 70.71\% & -1.221 & 71.90\% & -1.467 & 1542 & 52.05\% & 47.95\% \\
    RSI Strategy & 1.428 & 106.05\% & 70.55\% & 1.503 & 26.24\% & 6.074 & 114 & 26.31\% & 73.69\% \\
    RMSE Informer & 0.965 & -6.93\% & 3.11\% & -2.227 & 3.54\% & -4.361 & 12 & 0.02\% & 0.00\% \\
    Quantile Informer & 1.117 & 25.15\% & 43.69\% & 0.576 & 16.27\% & 0.889 & 365 & 0.00\% & 45.14\% \\
    GMADL Informer & 1.502 & 128.10\% & 70.54\% & 1.816 & 30.70\% & 7.576 & 82 & 41.54\% & 58.46\% \\
\end{tabular}
\includegraphics[width=\linewidth]{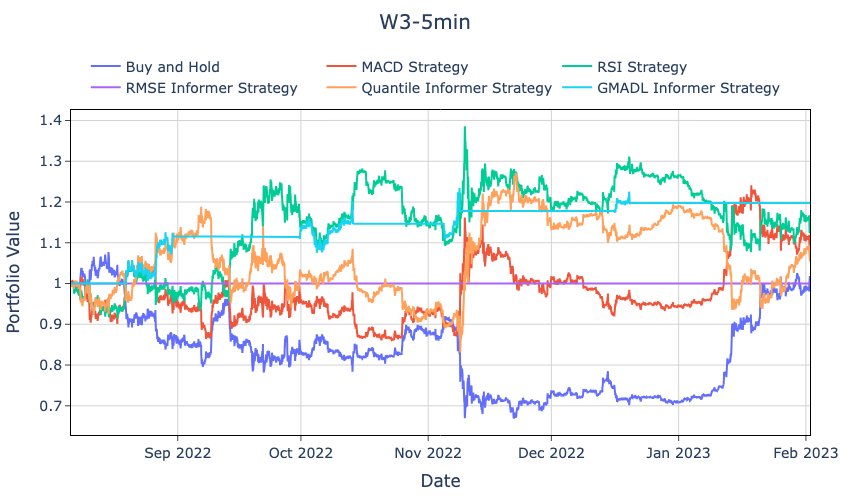}
\tabcolsep1pt
\begin{tabular}{lccccccccc}
    \textbf{Strategy} & \textbf{VAL} & \textbf{ARC} & \textbf{ASD} & \textbf{IR*} & \textbf{MD} & \textbf{IR**} & \textbf{N} & \textbf{LONG} & \textbf{SHORT} \\
    \hline
    Buy and Hold & 1.016 & 3.23\% & 51.77\% & 0.062 & 37.98\% & 0.005 & 2 & 100.00\% & 0.00\% \\
    MACD Strategy & 1.081 & 17.10\% & 51.79\% & 0.330 & 20.87\% & 0.270 & 154 & 58.24\% & 41.76\% \\
    RSI Strategy & 1.127 & 27.41\% & 51.92\% & 0.528 & 22.30\% & 0.649 & 398 & 11.48\% & 88.52\% \\
    RMSE Informer & 1 & 0.00\% & 0.00\% & 0 & 0.00\% & 0 & 0 & 0.00\% & 0.00\% \\
    Quantile Informer & 1.057 & 11.93\% & 51.90\% & 0.230 & 30.10\% & 0.091 & 894 & 27.62\% & 72.38\% \\
    GMADL Informer & 1.198 & 44.15\% & 19.18\% & 2.302 & 7.68\% & 13.227 & 16 & 0.00\% & 17.86\% \\
\end{tabular}
\includegraphics[width=\linewidth]{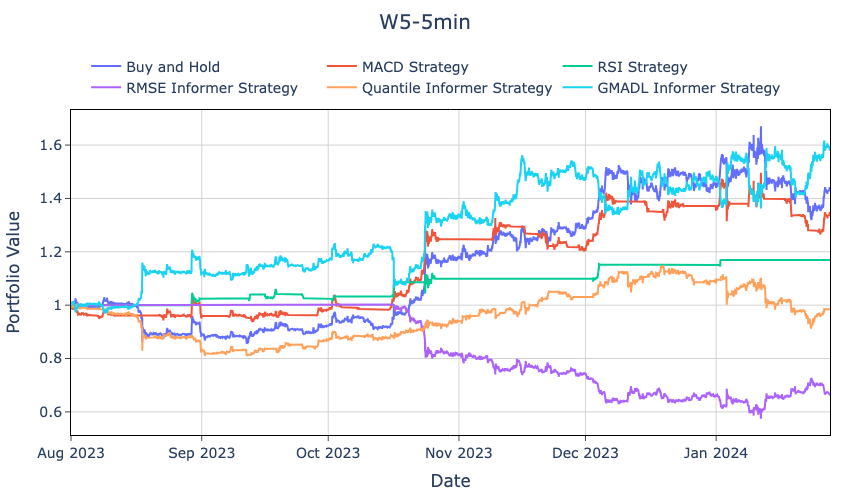}
\tabcolsep1pt
\begin{tabular}{lccccccccc}
    \textbf{Strategy} & \textbf{VAL} & \textbf{ARC} & \textbf{ASD} & \textbf{IR*} & \textbf{MD} & \textbf{IR**} & \textbf{N} & \textbf{LONG} & \textbf{SHORT} \\
    \hline
    Buy and Hold & 1.439 & 109.23\% & 44.72\% & 2.443 & 21.07\% & 12.664 & 2 & 100.00\% & 0.00\% \\
    MACD Strategy & 1.346 & 82.57\% & 32.85\% & 2.514 & 15.44\% & 13.447 & 94 & 49.06\% & 0.00\% \\
    RSI Strategy & 1.169 & 37.33\% & 13.86\% & 2.692 & 4.73\% & 21.237 & 42 & 5.67\% & 0.00\% \\
    RMSE Informer & 0.667 & -56.06\% & 36.94\% & -1.518 & 42.77\% & -1.989 & 3 & 0.00\% & 57.45\% \\
    Quantile Informer & 0.984 & -3.12\% & 33.97\% & -0.092 & 20.63\% & -0.014 & 578 & 59.26\% & 0.00\% \\
    GMADL Informer & 1.582 & 153.60\% & 45.11\% & 3.405 & 14.37\% & 36.394 & 130 & 44.70\% & 55.30\% \\
\end{tabular}
\end{minipage}
\begin{minipage}[b]{0.59\linewidth}
\raggedleft
\includegraphics[width=\linewidth]{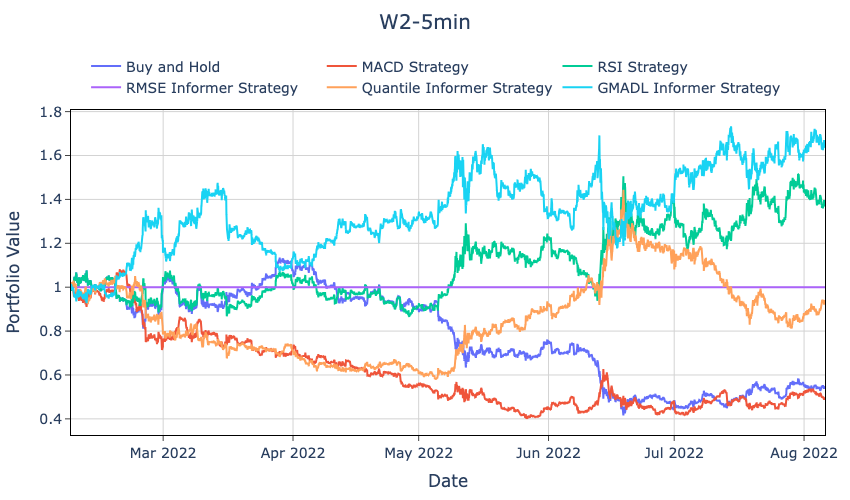}
\tiny
\tabcolsep1pt
\begin{tabular}{lccccccccc}
			\textbf{Strategy} & \textbf{VAL} & \textbf{ARC} & \textbf{ASD} & \textbf{IR*} & \textbf{MD} & \textbf{IR**} & \textbf{N} & \textbf{LONG} & \textbf{SHORT} \\
			\hline
			Buy and Hold & 0.549 & -70.37\% & 74.61\% & -0.943 & 63.35\% & -1.048 & 2 & 100.00\% & 0.00\% \\
			MACD Strategy & 0.498 & -75.63\% & 74.63\% & -1.013 & 62.97\% & -1.217 & 446 & 50.72\% & 49.28\% \\
			RSI Strategy & 1.368 & 88.76\% & 74.58\% & 1.190 & 27.34\% & 3.864 & 78 & 61.40\% & 38.60\% \\
			RMSE Informer & 1 & 0.00\% & 0.00\% & 0 & 0.00\% & 0 & 0 & 0.00\% & 0.00\% \\
			Quantile Informer & 0.941 & -11.54\% & 74.69\% & -0.154 & 44.84\% & -0.040 & 990 & 50.50\% & 49.50\% \\
			GMADL Informer & 1.635 & 170.94\% & 74.67\% & 2.289 & 30.47\% & 12.844 & 522 & 20.20\% & 79.80\% \\
\end{tabular}
\raggedleft
\includegraphics[width=\linewidth]{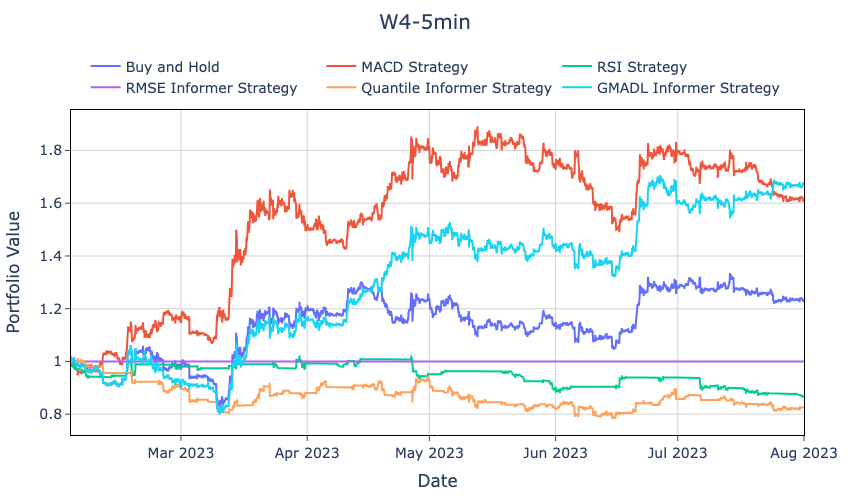}
\tabcolsep1pt
\begin{tabular}{lccccccccc}
    \textbf{Strategy} & \textbf{VAL} & \textbf{ARC} & \textbf{ASD} & \textbf{IR*} & \textbf{MD} & \textbf{IR**} & \textbf{N} & \textbf{LONG} & \textbf{SHORT} \\
    \hline
    Buy and Hold & 1.231 & 52.35\% & 45.25\% & 1.157 & 22.29\% & 2.718 & 2 & 100.00\% & 0.00\% \\
    MACD Strategy & 1.612 & 163.18\% & 45.34\% & 3.599 & 21.23\% & 27.660 & 190 & 44.69\% & 55.31\% \\
    RSI Strategy & 0.866 & -25.30\% & 19.03\% & -1.330 & 15.65\% & -2.150 & 205 & 23.63\% & 0.00\% \\
    RMSE Informer & 1 & 0.00\% & 0.00\% & 0 & 0.00\% & 0 & 0 & 0.00\% & 0.00\% \\
    Quantile Informer & 0.825 & -32.24\% & 27.88\% & -1.156 & 22.72\% & -1.641 & 290 & 44.45\% & 0.00\% \\
    GMADL Informer & 1.673 & 183.80\% & 45.26\% & 4.061 & 24.78\% & 30.125 & 62 & 65.14\% & 34.86\% \\
\end{tabular}
\raggedleft
\includegraphics[width=\linewidth]{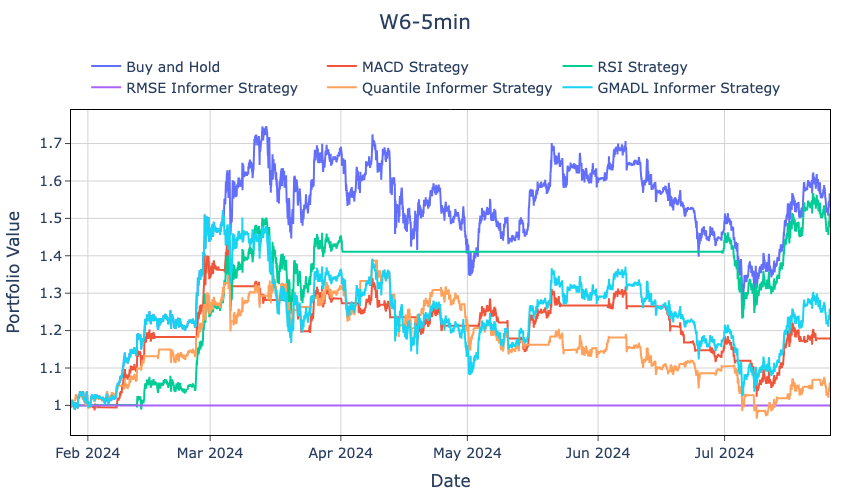}
\tabcolsep1pt
\begin{tabular}{lccccccccc}
    \textbf{Strategy} & \textbf{VAL} & \textbf{ARC} & \textbf{ASD} & \textbf{IR*} & \textbf{MD} & \textbf{IR**} & \textbf{N} & \textbf{LONG} & \textbf{SHORT} \\
    \hline
    Buy and Hold & 1.560 & 146.27\% & 52.72\% & 2.775 & 26.96\% & 15.054 & 2 & 100.00\% & 0.00\% \\
    MACD Strategy & 1.179 & 39.63\% & 34.65\% & 1.144 & 28.87\% & 1.570 & 111 & 47.58\% & 0.00\% \\
    RSI Strategy & 1.507 & 129.58\% & 38.15\% & 3.396 & 17.97\% & 24.493 & 7 & 41.29\% & 0.00\% \\
    RMSE Informer & 1 & 0.00\% & 0.00\% & 0 & 0.00\% & 0 & 0 & 0.00\% & 0.00\% \\
    Quantile Informer & 1.056 & 11.76\% & 40.73\% & 0.289 & 30.60\% & 0.111 & 280 & 59.60\% & 0.00\% \\
    GMADL Informer & 1.253 & 57.92\% & 52.71\% & 1.099 & 32.66\% & 1.949 & 34 & 97.21\% & 2.79\% \\
\end{tabular}
\end{minipage}
\tiny
\begin{tabular}{c}
    \hline
    \multicolumn{1}{p{\textwidth}}{
   \vspace{0.02cm}
     Note: Evaluation results for each of the out-of-sample data window of 5min interval BTC/USDT data. The presented metrics are: portfolio value at the end of the evaluation period (VAL), Annualized Return Compound (ARD), Annualized Standard Deviation (ASD), Information Ratio (IR*), Maximum Drawdown (MD), Modified Information Ratio (IR**), number of trades (N) and percent of the long/short positions (LONG/SHORT).} \\
\end{tabular}
\end{figure}

Looking at the evaluation results for each of the data windows separately, presented in Figure \ref{fig:results-5min-windows}, confirms the superiority of the GMADL informer strategy. It was the top performing strategy on all the evaluation periods but the last one, on which the best performing strategy was Buy and Hold, suggesting this was the most difficult period to trade. Most of the GMADL Strategy trades were done during the second window period, while during the other periods the positions were changed sparingly. In the third period the number of transactions was the smallest, only 16 and the strategy did not hold any position for 80\% of the time; however, it still achieved the best results compared to the other strategies, as it correctly predicted and leveraged the sudden price swings.

The RSI strategy performed quite well on the first three periods; however, on the next three the quality suddenly dropped, again suggesting the market being more unpredictable during past two windows. 

\clearpage

\subsubsection{Best strategies}
\begin{figure}[h!]
        \centering
        \caption{Best strategies}
        \label{fig:results-top}
        \includegraphics[trim={0 0 0 1.5cm},clip,width=\linewidth]{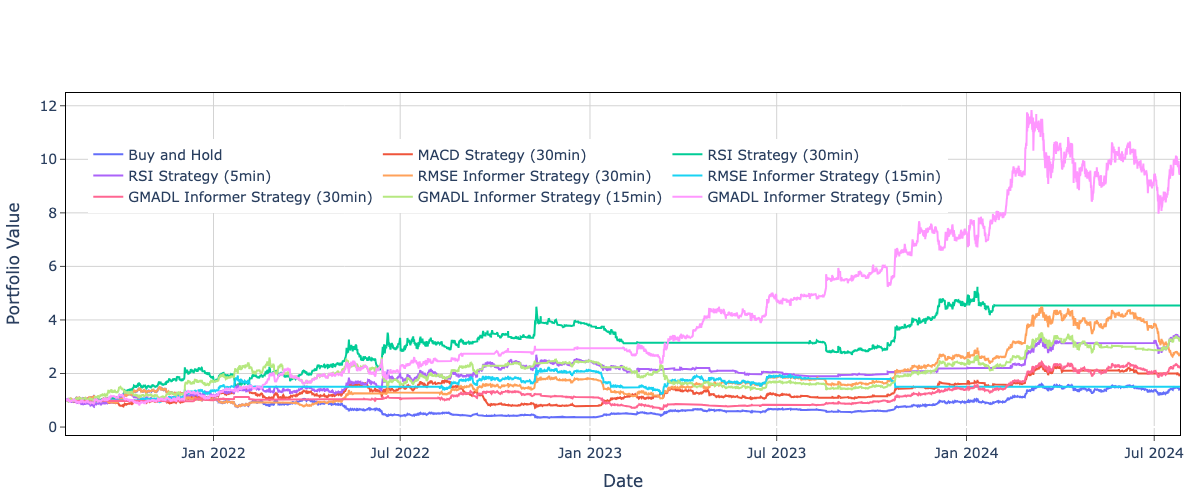}
        \vspace{-1cm}
	\begin{center}
            \small
            \tabcolsep7pt
		\begin{tabular}{lccccccccc}
			\textbf{Strategy} & \textbf{VAL} & \textbf{ARC} & \textbf{ASD} & \textbf{IR*} & \textbf{MD} & \textbf{IR**} & \textbf{N} & \textbf{LONG} & \textbf{SHORT} \\
			\hline
			Buy and Hold & 1.441 & 0.131 & 0.577 & 0.228 & 0.773 & 0.039 & 2 & 100.00\% & 0.00\% \\
			MACD Strategy (30min) & 1.952 & 0.254 & 0.524 & 0.485 & 0.592 & 0.207 & 327 & 52.30\% & 28.30\% \\
			RSI Strategy (30min) & 4.542 & 0.668 & 0.462 & 1.444 & 0.399 & 2.415 & 377 & 30.79\% & 28.03\% \\
			RSI Strategy (5min) & 3.341 & 0.503 & 0.504 & 0.999 & 0.300 & 1.676 & 846 & 28.29\% & 33.47\% \\
			RMSE Informer (30min) & 2.727 & 0.404 & 0.505 & 0.800 & 0.518 & 0.624 & 34 & 64.40\% & 24.67\% \\
			RMSE Informer (15min) & 1.509 & 0.149 & 0.349 & 0.428 & 0.455 & 0.140 & 16 & 15.24\% & 27.60\% \\
			GMADL Informer (30min) & 2.263 & 0.318 & 0.367 & 0.866 & 0.533 & 0.516 & 811 & 35.51\% & 19.59\% \\
			GMADL Informer (15min) & 3.296 & 0.496 & 0.527 & 0.942 & 0.474 & 0.987 & 362 & 49.37\% & 37.72\% \\
			GMADL Informer (5min) & 9.747 & 1.159 & 0.544 & 2.129 & 0.327 & 7.552 & 846 & 44.80\% & 41.51\% \\
   \hline
   \multicolumn{10}{p{\textwidth}}{\tiny
     Note: Strategies that achieved better performance than the buy-and-hold benchmark evaluated on testing period with BTC/USDT data. The presented metrics are: portfolio value at the end of the evaluation period (VAL), Annualized Return Compound (ARD), Annualized Standard Deviation (ASD), Information Ratio (IR*), Maximum Drawdown (MD), Modified Information Ratio (IR**), number of trades (N) and percent of the long/short positions (LONG/SHORT).}
		\end{tabular}
	\end{center}
\end{figure}

The strategies that performed better than the buy-and-hold benchmark are again presented in Figure \ref{fig:results-top}.

GMADL Informer strategy seems to be benefiting from the higher frequency data, with the IR** value monotonically increasing when the data frequency is increased. An opposite relation can be observed with the RMSE Informer, where the performance of the strategy decreases as the data interval becomes smaller. These relations seem consistent with assumptions that were made when designing the GMADL loss function, that it is better at pushing the model towards correctly predicting the direction of the return (positive vs negative), especially with sudden large changes, and that this signal can be better utilized for constructing automated trading strategies.
 
The RSI-based strategy turned out to give quite good results, both for 5min and 30min data. This indicates that despite the long-standing presence of technical indicators and the availability of more advanced and intricate models for generating trading signals, these indicators remain valuable and can be employed to develop effective trading strategies. MACD strategy was performing poorly for the higher frequency data, however, for the 30 min interval data the strategy was in the top five best. This might indicate that the MACD oscilator might not be applicable for the higher frequencies but is still useful for the lower frequencies. This hypothesis seems to be confirmed by the hyperparameters selected for the strategies that had a bias towards longer sequences.

The Quantile Informer strategy did not beat the Buy and Hold benchmark. In fact, it was one of the worst performing strategies. This may suggests that Quantile loss function is not the best choice for training a model for trading strategy or the proposed strategy do not use the model predictions effectively. However, more in depth analysis should be carried out, testing the training with various quantile combinations to confirm this hypothesis. Specifically, in this study a decision has been made to use a Quantile Loss with many different quantiles and then selecting the quantile value for entering/exiting the position as a part of strategy hyperparameter search. This method provided the advantage of eliminating the requirement to retrain the model every time a new quantile pair is selected. However, training the model once with many quantiles could have impact on the predictions quality.
It may also suggest that the strategy that uses signal from the Quantile Informer model was build improperly and should be revisited. A shape of the distribution e.g. by including information about the variance, could be utilised in the strategy, possibly making it better in selecting the relevant signals.

To verify statistical significance of the results, a probabilistic t-test that compares the Information Ratios of the best strategies against the buy-and-hold benchmark was conducted. The test uses a simplified formula for the standard error of Sharpie ratio, expressed as $\frac{\sigma}{\sqrt{N}}$, where $N$ is the sample size of the returns and $\sigma$ is the standard deviation of the excess of the strategy return over the benchmark return. The full test statistic is then as follows:
$$
t = \frac{IR^{*}_{\text{strategy}} - IR^{*}_{\text{benchmark}}}{\frac{\sigma}{\sqrt{N}}}
$$
The test assumes the differences in returns follow a normal distribution. The results are presented in Table \ref{tab:best_trategies_stat_test}. The null hypothesis that the Information Ratio of the top performing strategies is not greater than the buy-and-hold Modified Information Ratio was rejected for all the tested strategies.

\begin{table}[h]
    \centering
    \caption{Statistical t-test for comparing the performance of strategies over buy-and-hold.}
    \label{tab:best_trategies_stat_test}
    \vspace{0.5cm}
    \tabcolsep12pt
    \begin{tabular}{lcccc}
        \textbf{Strategy} & \textbf{N} & $\mathbf{\sigma}$ & \textbf{t-statistic} & \textbf{p-value} \\
        \hline
             MACD (30 min) & 51840 & 0.326504 & 174.34 & 0.000000*** \\
            RSI (30 min) & 51840 & 1.065079 & 258.49 & 0.000000*** \\
            RSI (5 min) & 311040 & 0.648741 & 662.77 & 0.000000*** \\
            RMSE (30 min) & 51840 & 0.796647 & 161.58 & 0.000000*** \\
            RMSE (15 min) & 103680 & 0.541611 & 115.29 & 0.000000*** \\
            GMADL (30 min) & 51840 & 0.320689 & 448.49 & 0.000000*** \\
            GMADL (15 min) & 103680 & 0.558743 & 408.13 & 0.000000*** \\
            GMADL (5min) & 311040 & 2.820834 & 375.84 & 0.000000*** \\
        \hline
        \multicolumn{5}{p{0.8\textwidth}}{\tiny
     Note: T-test H0: The information Ratio of the strategy is not greater than the buy-and-hold information ratio. The values marked with *** indicate the p-value lower than the critical value $\alpha = 0.01$, rejecting the null hypothesis.}
    \end{tabular}
\end{table}

\section{Sensitivity Analysis}\label{section:sensitivity}

The sensitivity analysis of the parameters of the Informer model requires retraining of the Informer model for each modified parameter and is outside the scope of this study due to the high computational cost. However, this section presents a brief analysis of how factors that impact the selection of strategy hyperparameters influence the strategy performance.

\subsection{Validation part size}
The hyperparameter selection for the strategies (described in Section \ref{section:hparams}) was carried out using the validation part of the data windows. The size of this part was arbitrarily selected to be 0.2 of the out-of-sample data size, which, in terms of the length of the period, is roughly equal to 6 months. The length of this period influences what strategy hyperparamers are selected. If the period is too short, the data sample would not be representative enough to select meaningful values for the hyperparameters. On the other hand, if the period is too long, the strategy might not be relevant to the current market conditions. Thus, it is important to perform an analysis of how this parameter influence the strategy performance.

The analysis has been carried out only for the best strategies, that is GMADL Informer with 5min data.

\begin{figure}[h!]
        \centering
        \caption{GMADL Informer strategies with parameters selected using different lengths of validation windows}
        \label{fig:sensitivity-gmadl}
        \includegraphics[width=\linewidth]{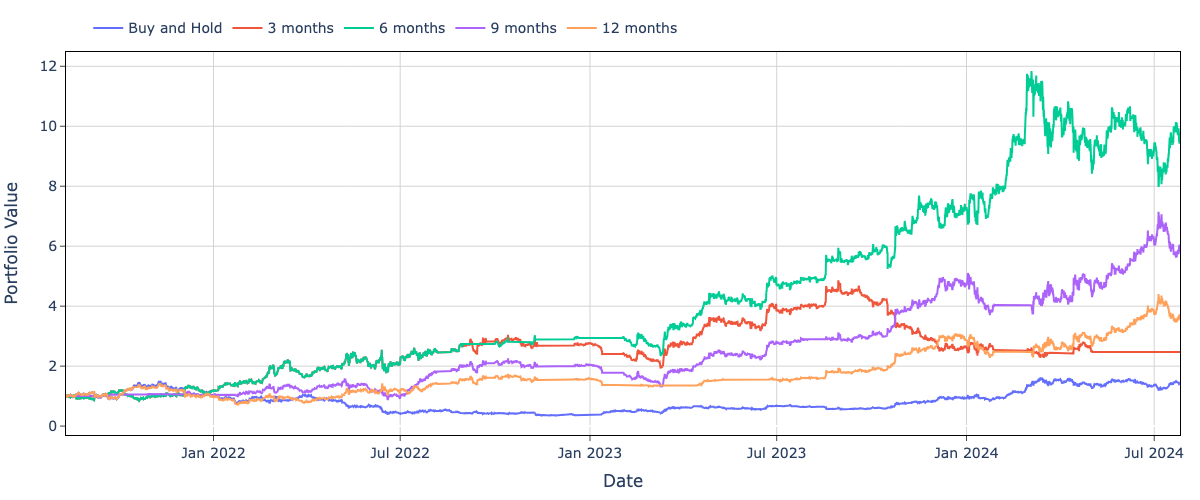}
        \vspace{-1.2cm}
	\begin{center}
            \small
		\begin{tabular}{lccccccccc}
			\textbf{Strategy} & \textbf{VAL} & \textbf{ARC} & \textbf{ASD} & \textbf{IR*} & \textbf{MD} & \textbf{IR**} & \textbf{N} & \textbf{LONG} & \textbf{SHORT} \\
			\hline
			Buy and Hold & 1.441 & 13.1\% & 57.7\% & 0.228 & 77.3\% & 0.039 & 2 & 100.0\% & 0.0\% \\
			3 months & 2.473 & 35.8\% & 52.5\% & 0.682 & 53.3\% & 0.458 & 912 & 27.0\% & 51.6\% \\
			6 months & 9.747 & 115.9\% & 54.4\% & 2.129 & 32.7\% & 7.552 & 846 & 44.8\% & 41.5\% \\
			9 months & 5.847 & 81.6\% & 48.5\% & 1.683 & 45.1\% & 3.048 & 300 & 38.7\% & 35.5\% \\
			12 months & 3.602 & 54.2\% & 52.2\% & 1.038 & 51.7\% & 1.088 & 214 & 38.1\% & 39.7\% \\
   \hline
   \multicolumn{10}{p{1\textwidth}}{\tiny
     Note: Best GMADL Informer strategy with hyperparameters selected on various lengths of validation windows. The presented metrics are: portfolio value at the end of the evaluation period (VAL), Annualized Return Compound (ARD), Annualized Standard Deviation (ASD), Information Ratio (IR*), Maximum Drawdown (MD), Modified Information Ratio (IR**), number of trades (N) and percent of the long/short positions (LONG/SHORT).}
		\end{tabular}
	\end{center}
\end{figure}

The results for evaluating GMADL Informer on 5 minute interval dataset when selecting hyperparameters using various validation part sizes are presented on Figure \ref{fig:sensitivity-gmadl}. For this strategy, selecting a longer or shorter validation period than 6 months resulted in decreased strategy performance. However, regardless of the length of the validation part, the strategy outperformed the benchmark, and Modified Information Ratio stayed on relatively high level. 
It can be seen that the number of trades decreases monotonically when the validation window is elongated. This suggests that the longer validation periods require the strategy to adapt to varying market conditions and more aggressively reduce the signals to change positions.
The worst performing variant was with the validation windows length of three months; in this case the strategy recorded significant losses during the last two windows of the testing period.
\clearpage

\subsection{Number of data windows}
Another implicit parameter that may affect the evaluation results is the number of data windows in a period in which the evaluation is carried out. In the original experiment, six data windows, with 6-month out-of-sample testing periods were considered. This section explores how the result of the best strategy - GMADL Informer with 5 min data - is affected, if the testing period is split differently: into three or twelve windows. Note that while selecting a different number of windows affects the length of the testing period of each window, the total testing period does not change and the in-sample (training and validation parts) length was kept constant. 

\begin{figure}[h!]
        \centering
        \vspace{-0.5cm}
        \caption{GMADL Informer strategies evaluated on different number of windows}
        \label{fig:sensitivity-w_gmadl}
        \includegraphics[width=\linewidth]{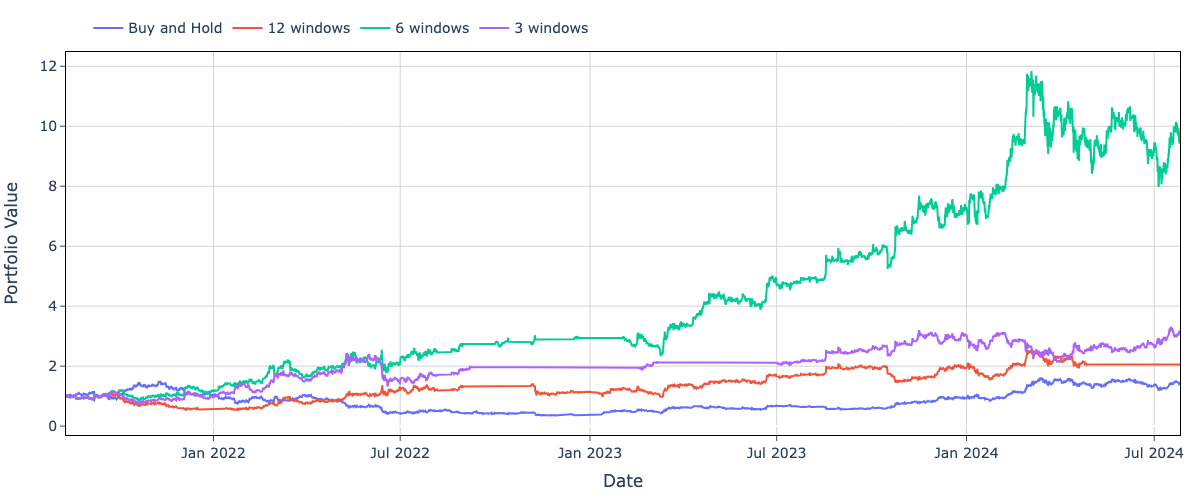}
        \vspace{-1cm}
	\begin{center}
            \small
		\begin{tabular}{lccccccccc}
			\textbf{Strategy} & \textbf{VAL} & \textbf{ARC} & \textbf{ASD} & \textbf{IR*} & \textbf{MD} & \textbf{IR**} & \textbf{N} & \textbf{LONG} & \textbf{SHORT} \\
			\hline
			Buy and Hold & 1.44 & 13.1\% & 57.7\% & 0.23 & 77.3\% & 0.04 & 2 & 100.0\% & 0.0\% \\
			12 windows & 2.06 & 27.6\% & 53.1\% & 0.52 & 53.3\% & 0.27 & 568 & 43.0\% & 37.9\% \\
			6 windows & 9.75 & 115.9\% & 54.4\% & 2.13 & 32.7\% & 7.55 & 846 & 44.8\% & 41.5\% \\
			3 windows & 3.08 & 46.2\% & 51.6\% & 0.90 & 45.9\% & 0.90 & 426 & 27.1\% & 45.9\% \\
   \hline
   \multicolumn{10}{p{\textwidth}}{\tiny
     Note: Best GMADL Informer strategy evaluated on different number of windows. The presented metrics are: portfolio value at the end of the evaluation period (VAL), Annualized Return Compound (ARD), Annualized Standard Deviation (ASD), Information Ratio (IR*), Maximum Drawdown (MD), Modified Information Ratio (IR**), number of trades (N) and percent of the long/short positions (LONG/SHORT).}
		\end{tabular}
	\end{center}
\end{figure}

Evaluation of the GMADL Informer strategy with a different number of windows required retraining of the underlying Informer model with the new data. For each of the cases considered: three and twelve windows, a new model was trained for each window, keeping the values of all the other hyperparameters the same as in the base scenario. The evaluation results are presented in Figure \ref{fig:sensitivity-w_gmadl}. It can be seen that the performance of the strategy was worse both when the number of windows increased and decreased. Although the Modified Information Ratio was still higher than the benchmark, it was significantly lower than the one in the six-window evaluation. This may suggest that such an excellent performance of the GMADL Informer strategy might have been an isolated event caused by the selection of a concrete number of evaluation windows. However, this conclusion should not belittle the fact that the GMADL Informer strategy, evaluated on different numbers of windows, still achieves impressive results compared to the benchmark. 

\clearpage

\subsection{Top n-th strategy}
During the selection of hyperparameters, it was implicitly assumed that the best hyperparameters for a strategy come from the strategy that performs the best in the validation period. This might not be true, as the best-performing strategy might be overfitted to said period. 

\begin{figure}[h!]
    \vspace{-0.1cm}
    \caption{Strategies with top 10 hyperparameter combinations for 5min data }
    \label{fig:sensitivity_top10}
    \centering
    \includegraphics[width=0.9\linewidth]{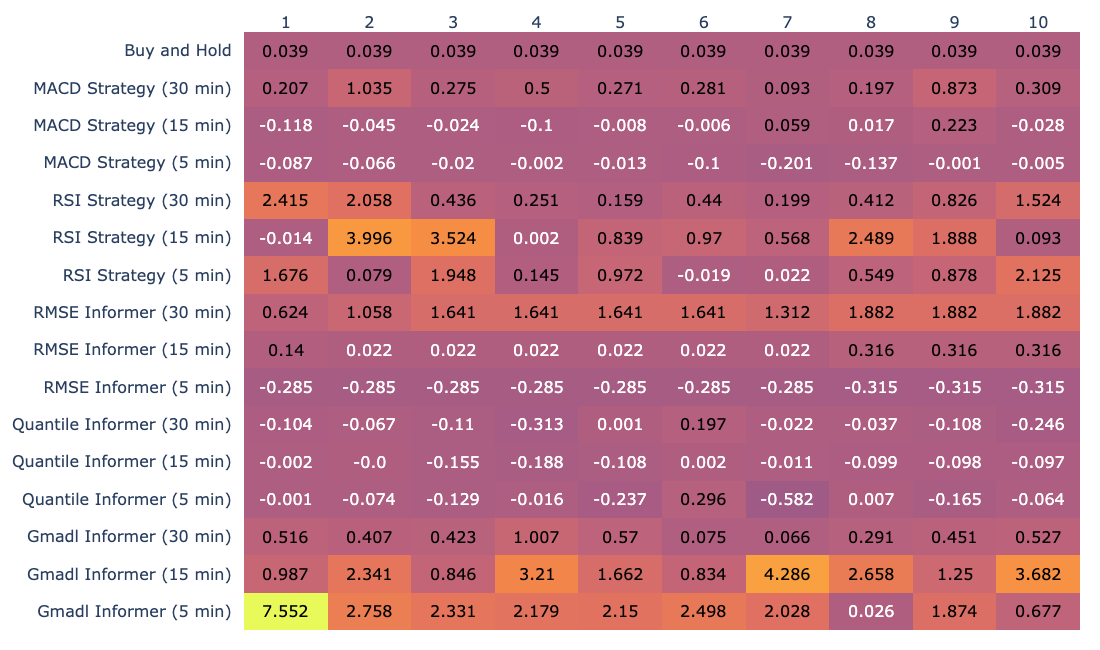}    
    
    \raggedright \tiny Note: Modified Information Ratio(IR**) of the strategies evaluated on the whole testing period, with the strategies using the top 10 hyperparameter sets according to IR** from the evaluation on the validation set.
\end{figure}

Figure \ref{fig:sensitivity_top10} present the modified information ratio after evaluating strategies with the top 10 hyperparameter sets from validation windows. In case of GMADL Strategy with 5 min data, the best overall performing strategy is indeed the one with the first set of hyperparameters, and the consecutive sets gradually decrease the results of the strategy. However, in case of 15min and 30min data, the second best would turn out to be better. Similar situation can be observed with the best RSI strategy, the overall best RSI strategy with 30min windows was the first one from the validation, but in case of 15min data the second best was significantly better. For RMSE Informer with 30min data the first two sets of hyperparameters seemed overfitted, with the performance greatly increasing past second set. In case of strategies that had overall poor performance, selection of top n-th set of the hyperparameters wouldn't have significant impact.
\newpage
    
    

\section{Conclusion}\label{chapter:conclusion}

The research aimed to explore different methods to create automated investment strategies for trading Bitcoin. Five strategies were proposed, two using technical indicators signal and three using predictions of the Informer machine learning model. The strategies were evaluated independently over six periods on BTC/USDT data of various frequencies: 5 minutes, 15 minutes, and 30 minutes. For each data window and every frequency, the strategies that use a machine learning model
were first trained using the training part of the data window. The hyperparameters were then selected using the validation part of the data window. Finally, a strategy with the best set of hyperparameters according to the Modified Information Ratio was evaluated on the test part of the data window. Eight strategies beat the Buy and Hold benchmark, when compared throughout the testing period, i.e., combined results from all data windows. The best performing strategy was the one based on predictions of the Informer model trained on 5-minute data with the GMADL loss function. In addition, it significantly outperformed all the other strategies, proving to be best in almost all testing periods. A sensitivity analysis was performed to measure the impact of the selection of hyperparameters on strategy performance. It showed that the size of the validation window proved to have an impact on the selection of hyperparameters and overall strategy performance. This was also the case when modifying the number of testing periods, changing that parameter affected results and decreased the performance of the strategy, although the GMADL Informer strategy in all cases still managed to beat the benchmark.

Based on the results of the research an attempt to answer the research questions can be made:
\begin{questions}
    \item \textit{Is it possible to create an algorithmic strategy for trading Bitcoin, that is more efficient than Buy\&Hold approach?}
\end{questions}
 The research showed that it is possible to create an algorithmic strategy that outperforms the Buy\&Hold benchmark on the Bitcoin data. Eight strategies achieved better results during the testing period than the benchmark. The strategies were evaluated in six different test windows, through an overall testing period of almost three years.

\begin{questions}
\item \textit{Does signal from Informer model allow to create strategies that are more efficient on trading Bitcoin than strategies based on technical indicators?}
\end{questions}
While the strategies based on Informer trained with RMSE and GMADL loss functions turned out to be more efficient than the strategy based on MACD indicator, the strategy based on RSI indicator achieved comparable results. Moreover, strategy based on RSI indicator on 30 min windows was the second best overall performing strategy. This shows that even though the technical indicators have been around for a long time and are well established in the literature, they still provide a valid alternative to expensive, machine learning based approaches.

\begin{questions}
\item \textit{How does selection of the machine learning model loss function influence the strategy performance?}
\end{questions}
Different behaviours of the strategies based on the model trained with different loss functions could be observed. Especially when analyzed together with looking at the different data frequencies. Predictions of the Informer trained with RMSE loss function were worse for the sudden, large returns, making it difficult to cunstruct the strategy for higher frequency data. In contrast Informer trained with GMADL loss function benefited from higher frequency. Approach utilizing a model with the Quantile loss function underperformed relative to a benchmark, showing the difficulty for constructing the strategy based on the predictions of model trained with such loss. From the three tested loss function, GMADL loss function seem to be best fitted for creating models that are to be used in automated trading strategies. What is more important, the selection of the loss function seem to be one of the deciding factors when training the model that generates the signal for algorithmic trading systems.

\begin{questions}
\item \textit{Does usage of higher frequency data allow to create more efficient strategies ?} 
\end{questions}
The strategies with GMADL and RMSE loss functions achieved comparable results when evaluated on 30 min data, though usage of high frequency data  improved the performance of the GMADL Informer strategy, while deteriorated the performance of the strategy based on the Informer with RMSE loss function. Strategies based on technical indicators did not seem to benefit much from the higher frequencies.

The main contribution of the study was the analysis of the novel application of loss functions: Quantile and GMADL loss, to train Informer model to predict returns of Bitcoin and compare results with the better established approach of training the machine learning model with RMSE loss. Predictions of all three model types, were used to construct automated trading strategies which were compared against buy-and-hold benchmark and the two benchmark strategies based on MACD and RSI technical indicators. The exhaustive analysis included usage of different testing periods, different data intervals and sensitivity analysis. It showed that the GMADL loss function allows to effectively train the machine learning model, which is able to provide meaningful signal for the trading strategy. Such a strategy can be deployed to provide higher yields than the buy-and-hold approach.

There are various directions in which the research could be extended. A more extensive sensitivity analysis could be carried to better understand the impact of various hyperparameters on the strategies, especially in case of Informer based strategies and loss functions used for training the model. Another direction would be to test even higher data frequencies and explore whether the performance of the Informer-based model with GMADL loss function would be further improved. Finally, datasets of other financial instruments could be used, to see whether the same strategies perform well on other data than Bitcoin.

\printbibliography

@Article{s22030917,
AUTHOR = {Michańków, Jakub and Sakowski, Paweł and Ślepaczuk, Robert},
TITLE = {LSTM in Algorithmic Investment Strategies on BTC and S\&P500 Index},
JOURNAL = {Sensors},
VOLUME = {22},
YEAR = {2022},
NUMBER = {3},
ARTICLE-NUMBER = {917},
URL = {https://www.mdpi.com/1424-8220/22/3/917},
PubMedID = {35161663},
ISSN = {1424-8220},
DOI = {10.3390/s22030917}
}

@article{btc2009,
author = {Nakamoto, Satoshi},
year = {2009},
month = {03},
pages = {},
title = {Bitcoin: A Peer-to-Peer Electronic Cash System},
journal = {Cryptography Mailing list at https://metzdowd.com}
}

@book{Malkiel1973,
  author = {Malkiel, Burton. G.},
  comment = {Malkiel defends his random position well, explaining
                 how the market eventually corrects any
                 irrationality--albeit in its own slow, inexorable
                 fashion. Anomalies can crop up, markets can get
                 irrationally optimistic, and often they attract unwary
                 investors. But eventually, true value is recognized by
                 the market, and this is the main lesson investors must
                 heed.},
  publisher = {Norton, New York},
  title = {A Random Walk Down Wall Street},
  year = 1973
}

@article{fama1970,
  author = {Fama, Eugene},
  journal = {Journal of Finance},
  pages = {383--417},
  title = {Efficient Capital Markets: A Review of Theory and Empirical Work },
  volume = 25,
  year = 1970
}

@Article{RePEc:bla:jfinan:v:46:y:1991:i:5:p:1575-617,
  author={ Fama, Eugene},
  title={{Efficient Capital Markets: II}},
  journal={Journal of Finance},
  year=1991,
  volume={46},
  number={5},
  pages={1575-1617},
  month={12},
  doi={},
  url={https://ideas.repec.org/a/bla/jfinan/v46y1991i5p1575-617.html}
}

@article{malkiel2005,
author = {Malkiel, Burton},
year = {2005},
month = {02},
pages = {1-9},
title = {Reflections on the Efficient Market Hypothesis: 30 Years Later},
volume = {40},
journal = {The Financial Review},
doi = {10.1111/j.0732-8516.2005.00090.x}
}

@Article{RePEc:bla:jecsur:v:25:y:2011:i:1:p:69-108,
  author={Kian‐Ping Lim and Robert Brooks},
  title={{The Evolution Of Stock Market Efficiency Over Time: A Survey Of The Empirical Literature}},
  journal={Journal of Economic Surveys},
  year=2011,
  volume={25},
  number={1},
  pages={69-108},
  month={2},
  doi={},
  url={https://ideas.repec.org/a/bla/jecsur/v25y2011i1p69-108.html}
}

@Article{RePEc:eee:empfin:v:18:y:2011:i:5:p:868-879,
  author={Kim, Jae H. and Shamsuddin, Abul and Lim, Kian-Ping},
  title={{Stock return predictability and the adaptive markets hypothesis: Evidence from century-long U.S. data}},
  journal={Journal of Empirical Finance},
  year=2011,
  volume={18},
  number={5},
  pages={868-879},
  month={},
  doi={10.1016/j.jempfin.2011.08},
  url={https://ideas.repec.org/a/eee/empfin/v18y2011i5p868-879.html}
}

@article{fischoff78,
author = {Fischhoff, Baruch and Slovic, Paul},
year = {1978},
month = {06},
pages = {58},
title = {A Little Learning...: Confidence in Multicue Judgment Tasks}
}

@ARTICLE{RePEc:oup:qjecon:v:116:y:2001:i:1:p:261-292.,
title = {Boys will be Boys: Gender, Overconfidence, and Common Stock Investment},
author = {Barber, Brad and Odean, Terrance},
year = {2001},
journal = {The Quarterly Journal of Economics},
volume = {116},
number = {1},
pages = {261-292},
url = {https://EconPapers.repec.org/RePEc:oup:qjecon:v:116:y:2001:i:1:p:261-292.}
}

@article{odean2001,
author = {Gervais, Simon and Odean, Terrance},
year = {2001},
month = {02},
pages = {1-27},
title = {Learning To Be Overconfident},
volume = {14},
journal = {Review of Financial Studies},
doi = {10.2139/ssrn.36313}
}

@article{28e39847-87a7-3356-8a2d-ed0ea430ab56,
 ISSN = {00221082, 15406261},
 URL = {http://www.jstor.org/stable/2327804},
 author = {Werner F. M. De Bondt and Richard Thaler},
 journal = {The Journal of Finance},
 number = {3},
 pages = {793--805},
 publisher = {[American Finance Association, Wiley]},
 title = {Does the Stock Market Overreact?},
 urldate = {2024-10-05},
 volume = {40},
 year = {1985}
}

@Article{RePEc:ecm:emetrp:v:47:y:1979:i:2:p:263-91,
  author={Kahneman, Daniel and Tversky, Amos},
  title={{Prospect Theory: An Analysis of Decision under Risk}},
  journal={Econometrica},
  year=1979,
  volume={47},
  number={2},
  pages={263-291},
  month={3},
  doi={},
  url={https://ideas.repec.org/a/ecm/emetrp/v47y1979i2p263-91.html}
}

@TechReport{RePEc:ucb:calbrf:rpf-269,
  author={Terrance Odean.},
  title={{Are Investors Reluctant to Realize Their Losses?}},
  year=1996,
  month=Nov,
  institution={University of California at Berkeley},
  type={Research Program in Finance Working Papers},
  url={https://ideas.repec.org/p/ucb/calbrf/rpf-269.html},
  number={RPF-269},
  doi={},
}

@ARTICLE{RePEc:bla:jfinan:v:56:y:2001:i:1:p:387-396,
title = {Contagious Speculation and a Cure for Cancer: A Nonevent that Made Stock Prices Soar},
author = {Huberman, Gur and Regev, Tomer},
year = {2001},
journal = {Journal of Finance},
volume = {56},
number = {1},
pages = {387-396},
url = {https://EconPapers.repec.org/RePEc:bla:jfinan:v:56:y:2001:i:1:p:387-396}
}

@article{
doi:10.1126/science.7455683,
author = {Amos Tversky  and Daniel Kahneman },
title = {The Framing of Decisions and the Psychology of Choice},
journal = {Science},
volume = {211},
number = {4481},
pages = {453-458},
year = {1981},
doi = {10.1126/science.7455683},
URL = {https://www.science.org/doi/abs/10.1126/science.7455683},
eprint = {https://www.science.org/doi/pdf/10.1126/science.7455683}}

@book{boxjen76,
  author = {Box, George.E.P. and Jenkins, Gwilym M.},
  publisher = {Holden-Day},
  title = {Time Series Analysis: Forecasting and Control},
  year = 1976
}

@inproceedings{ayo2014,
author = {Adebiyi, Ayodele and Adewumi, Aderemi and Ayo, Charles},
year = {2014},
month = {03},
pages = {},
title = {Stock price prediction using the ARIMA model},
journal = {Proceedings - UKSim-AMSS 16th International Conference on Computer Modelling and Simulation, UKSim 2014},
doi = {10.1109/UKSim.2014.67}
}

@article{PAI2005497,
title = {A hybrid ARIMA and support vector machines model in stock price forecasting},
journal = {Omega},
volume = {33},
number = {6},
pages = {497-505},
year = {2005},
issn = {0305-0483},
doi = {https://doi.org/10.1016/j.omega.2004.07.024},
url = {https://www.sciencedirect.com/science/article/pii/S0305048304001082},
author = {Ping-Feng Pai and Chih-Sheng Lin}
}

@article{ZHANG2003159,
title = {Time series forecasting using a hybrid ARIMA and neural network model},
journal = {Neurocomputing},
volume = {50},
pages = {159-175},
year = {2003},
issn = {0925-2312},
doi = {https://doi.org/10.1016/S0925-2312(01)00702-0},
url = {https://www.sciencedirect.com/science/article/pii/S0925231201007020},
author = {G.Peter Zhang}
}

@unknown{azari2018,
author = {Azari, Amin},
year = {2018},
month = {10},
pages = {},
title = {Bitcoin Price Prediction: An ARIMA Approach},
doi = {10.48550/arXiv.1904.05315}
}

@article{WANG2012758,
title = {Stock index forecasting based on a hybrid model},
journal = {Omega},
volume = {40},
number = {6},
pages = {758-766},
year = {2012},
note = {Special Issue on Forecasting in Management Science},
issn = {0305-0483},
doi = {https://doi.org/10.1016/j.omega.2011.07.008},
url = {https://www.sciencedirect.com/science/article/pii/S0305048311001435},
author = {Ju-Jie Wang and Jian-Zhou Wang and Zhe-George Zhang and Shu-Po Guo}
}

@article{billah2006,
author = {Billah, Baki and King, Maxwell and Snyder, Ralph and Koehler, Anne},
year = {2006},
month = {02},
pages = {239-247},
title = {Exponential smoothing model selection for forecasting},
volume = {22},
journal = {International Journal of Forecasting},
doi = {10.1016/j.ijforecast.2005.08.002}
}

@article{KHASHEI20112664,
title = {A novel hybridization of artificial neural networks and ARIMA models for time series forecasting},
journal = {Applied Soft Computing},
volume = {11},
number = {2},
pages = {2664-2675},
year = {2011},
note = {The Impact of Soft Computing for the Progress of Artificial Intelligence},
issn = {1568-4946},
doi = {https://doi.org/10.1016/j.asoc.2010.10.015},
url = {https://www.sciencedirect.com/science/article/pii/S1568494610002759},
author = {Mehdi Khashei and Mehdi Bijari}
}

@article{10.1162/neco.1997.9.8.1735,
author = {Hochreiter, Sepp and Schmidhuber, J\"{u}rgen},
title = {Long Short-Term Memory},
year = {1997},
issue_date = {November 15, 1997},
publisher = {MIT Press},
address = {Cambridge, MA, USA},
volume = {9},
number = {8},
issn = {0899-7667},
url = {https://doi.org/10.1162/neco.1997.9.8.1735},
doi = {10.1162/neco.1997.9.8.1735},
journal = {Neural Comput.},
month = nov,
pages = {1735–1780},
numpages = {46}
}

@article{DBLP:journals/corr/ChoMGBSB14,
  author       = {Kyunghyun Cho and
                  Bart van Merrienboer and
                  {\c{C}}aglar G{\"{u}}l{\c{c}}ehre and
                  Fethi Bougares and
                  Holger Schwenk and
                  Yoshua Bengio},
  title        = {Learning Phrase Representations using {RNN} Encoder-Decoder for Statistical
                  Machine Translation},
  journal      = {CoRR},
  volume       = {abs/1406.1078},
  year         = {2014},
  url          = {http://arxiv.org/abs/1406.1078},
  eprinttype    = {arXiv},
  eprint       = {1406.1078},
  bibsource    = {dblp computer science bibliography, https://dblp.org}
}

@article{lecun_deep_2015,
	title = {Deep learning},
	volume = {521},
	issn = {0028-0836, 1476-4687},
	url = {https://www.nature.com/articles/nature14539},
	doi = {10.1038/nature14539},
	language = {en},
	number = {7553},
	urldate = {2024-10-06},
	journal = {Nature},
	author = {LeCun, Yann and Bengio, Yoshua and Hinton, Geoffrey},
	month = may,
	year = {2015},
	pages = {436--444},
}

@ARTICLE{Siami‐Namini_2018,title={Forecasting Economics and Financial Time Series: ARIMA vs. LSTM.},year={2018},author={Sima Siami‐Namini and Sima Siami-Namini and Akbar Siami Namin and Akbar Siami Namin},doi={null},pmid={null},pmcid={null},mag_id={2791884409},journal={arXiv: Learning}}

@article{RePEc:war:wpaper:2021-23,
title = {Application of machine learning in algorithmic investment strategies on global stock markets},
journal = {Research in International Business and Finance},
volume = {66},
pages = {102052},
year = {2023},
issn = {0275-5319},
doi = {https://doi.org/10.1016/j.ribaf.2023.102052},
url = {https://www.sciencedirect.com/science/article/pii/S0275531923001782},
author = {Jan Grudniewicz and Robert Ślepaczuk}
}

@article{honchar2016,
author = {Honchar, Oleksandr and Di Persio, Luca},
year = {2016},
month = {01},
pages = {158-162},
title = {Artificial neural networks approach to the forecast of stock market price movements},
volume = {Vol.1}
}

@ARTICLE{Persio_2017,title={Recurrent Neural Networks Approach to the Financial Forecast of Google Assets},year={2017},author={Luca Di Persio and Luca Di Persio and Oleksandr Honchar and Oleksandr Honchar},doi={null},pmid={null},pmcid={null},mag_id={2603235440},journal={null}}

@inproceedings{RePEc:war:wpaper:2020-27,
author = {Kijewski, Mateusz and Slepaczuk, Robert and Wysocki, Maciej},
year = {2024},
month = {09},
booktitle = {Proceedings of the 32nd International Conference on Information Systems Development (ISD 2024)},
pages = {},
title = {Predicting Prices Of S\&P 500 Index Using Classical Methods and Recurrent Neural Networks},
doi = {10.62036/ISD.2024.89}
}

@inproceedings{hossain2018,
author = {Hossain, Mohammad and Karim, Rezaul and Thulasiram, Ruppa and Bruce, Neil and Wang, Yang},
year = {2018},
month = {11},
pages = {1837-1844},
title = {Hybrid Deep Learning Model for Stock Price Prediction},
doi = {10.1109/SSCI.2018.8628641}
}

@ARTICLE{Fischer_2017,title={Deep learning with long short-term memory networks for financial market predictions},year={2017},author={Thomas Fischer and Thomas G. Fischer and Thomas Fischer and Thomas Fischer and Thomas Fischer and Christopher Krauß and Christopher Krauss},doi={10.1016/j.ejor.2017.11.054},pmid={null},pmcid={null},mag_id={2624385633},journal={European Journal of Operational Research}}

@ARTICLE{Selvin_2017,title={Stock price prediction using LSTM, RNN and CNN-sliding window model},year={2017},author={Sreelekshmy Selvin and Sreelekshmy Selvin and Vinayakumar Ravi and R. Vinayakumar and E. A. Gopalakrishnan and E. A. Gopalakrishnan and E. A. Gopalakrishnan and Vijay Menon and Vijay Krishna Menon and K. P. Soman and K. P. Soman and K. P. Soman},doi={10.1109/icacci.2017.8126078},pmid={null},pmcid={null},mag_id={2774513877},journal={null}}

@ARTICLE{Chen_2015,title={A LSTM-based method for stock returns prediction: A case study of China stock market},year={2015},author={Kai Chen and Kai Chen and Kai Chen and Yi Zhou and Yi Zhou and Fengwei Dai and Fangyan Dai},doi={10.1109/bigdata.2015.7364089},pmid={null},pmcid={null},mag_id={2209610041},journal={null}}

@ARTICLE{Nelson_2017,
title={Stock market's price movement prediction with LSTM neural networks},
year={2017},
author={David M. Nelson and David M. Q. Nelson and Adriano C. M. Pereira and Adriano C. M. Pereira and Renato Arantes de Oliveira and Renato A. de Oliveira},
doi={10.1109/ijcnn.2017.7966019}
}

@ARTICLE{9142152,
  author={Basodi, Sunitha and Ji, Chunyan and Zhang, Haiping and Pan, Yi},
  journal={Big Data Mining and Analytics}, 
  title={Gradient amplification: An efficient way to train deep neural networks}, 
  year={2020},
  volume={3},
  number={3},
  pages={196-207},
  doi={10.26599/BDMA.2020.9020004}}

@misc{bahdanau2014neural,
  author = {Bahdanau, Dzmitry and Cho, Kyunghyun and Bengio, Yoshua},
  description = {[1409.0473] Neural Machine Translation by Jointly Learning to Align and Translate},
  note = {cite arxiv:1409.0473Comment: Accepted at ICLR 2015 as oral presentation},
  title = {Neural Machine Translation by Jointly Learning to Align and Translate},
  url = {http://arxiv.org/abs/1409.0473},
  year = 2014
}

@inproceedings{luong-etal-2015-effective,
    title = "Effective Approaches to Attention-based Neural Machine Translation",
    author = "Luong, Thang  and
      Pham, Hieu  and
      Manning, Christopher D.",
    editor = "M{\`a}rquez, Llu{\'\i}s  and
      Callison-Burch, Chris  and
      Su, Jian",
    booktitle = "Proceedings of the 2015 Conference on Empirical Methods in Natural Language Processing",
    month = sep,
    year = "2015",
    address = "Lisbon, Portugal",
    publisher = "Association for Computational Linguistics",
    url = "https://aclanthology.org/D15-1166",
    doi = "10.18653/v1/D15-1166",
    pages = "1412--1421",
}

@article{DBLP:journals/corr/VaswaniSPUJGKP17,
  author       = {Ashish Vaswani and
                  Noam Shazeer and
                  Niki Parmar and
                  Jakob Uszkoreit and
                  Llion Jones and
                  Aidan N. Gomez and
                  Lukasz Kaiser and
                  Illia Polosukhin},
  title        = {Attention Is All You Need},
  journal      = {CoRR},
  volume       = {abs/1706.03762},
  year         = {2017},
  url          = {http://arxiv.org/abs/1706.03762},
  eprinttype    = {arXiv},
  eprint       = {1706.03762},
  bibsource    = {dblp computer science bibliography, https://dblp.org}
}

@misc{openai2024gpt4technicalreport,
      title={GPT-4 Technical Report}, 
      author={OpenAI and Josh Achiam and Steven Adler and Sandhini Agarwal and Lama Ahmad and Ilge Akkaya and Florencia Leoni Aleman and Diogo Almeida and Janko Altenschmidt and Sam Altman and Shyamal Anadkat and Red Avila and Igor Babuschkin and Suchir Balaji and Valerie Balcom and Paul Baltescu and Haiming Bao and Mohammad Bavarian and Jeff Belgum and Irwan Bello and Jake Berdine and Gabriel Bernadett-Shapiro and Christopher Berner and Lenny Bogdonoff and Oleg Boiko and Madelaine Boyd and Anna-Luisa Brakman and Greg Brockman and Tim Brooks and Miles Brundage and Kevin Button and Trevor Cai and Rosie Campbell and Andrew Cann and Brittany Carey and Chelsea Carlson and Rory Carmichael and Brooke Chan and Che Chang and Fotis Chantzis and Derek Chen and Sully Chen and Ruby Chen and Jason Chen and Mark Chen and Ben Chess and Chester Cho and Casey Chu and Hyung Won Chung and Dave Cummings and Jeremiah Currier and Yunxing Dai and Cory Decareaux and Thomas Degry and Noah Deutsch and Damien Deville and Arka Dhar and David Dohan and Steve Dowling and Sheila Dunning and Adrien Ecoffet and Atty Eleti and Tyna Eloundou and David Farhi and Liam Fedus and Niko Felix and Simón Posada Fishman and Juston Forte and Isabella Fulford and Leo Gao and Elie Georges and Christian Gibson and Vik Goel and Tarun Gogineni and Gabriel Goh and Rapha Gontijo-Lopes and Jonathan Gordon and Morgan Grafstein and Scott Gray and Ryan Greene and Joshua Gross and Shixiang Shane Gu and Yufei Guo and Chris Hallacy and Jesse Han and Jeff Harris and Yuchen He and Mike Heaton and Johannes Heidecke and Chris Hesse and Alan Hickey and Wade Hickey and Peter Hoeschele and Brandon Houghton and Kenny Hsu and Shengli Hu and Xin Hu and Joost Huizinga and Shantanu Jain and Shawn Jain and Joanne Jang and Angela Jiang and Roger Jiang and Haozhun Jin and Denny Jin and Shino Jomoto and Billie Jonn and Heewoo Jun and Tomer Kaftan and Łukasz Kaiser and Ali Kamali and Ingmar Kanitscheider and Nitish Shirish Keskar and Tabarak Khan and Logan Kilpatrick and Jong Wook Kim and Christina Kim and Yongjik Kim and Jan Hendrik Kirchner and Jamie Kiros and Matt Knight and Daniel Kokotajlo and Łukasz Kondraciuk and Andrew Kondrich and Aris Konstantinidis and Kyle Kosic and Gretchen Krueger and Vishal Kuo and Michael Lampe and Ikai Lan and Teddy Lee and Jan Leike and Jade Leung and Daniel Levy and Chak Ming Li and Rachel Lim and Molly Lin and Stephanie Lin and Mateusz Litwin and Theresa Lopez and Ryan Lowe and Patricia Lue and Anna Makanju and Kim Malfacini and Sam Manning and Todor Markov and Yaniv Markovski and Bianca Martin and Katie Mayer and Andrew Mayne and Bob McGrew and Scott Mayer McKinney and Christine McLeavey and Paul McMillan and Jake McNeil and David Medina and Aalok Mehta and Jacob Menick and Luke Metz and Andrey Mishchenko and Pamela Mishkin and Vinnie Monaco and Evan Morikawa and Daniel Mossing and Tong Mu and Mira Murati and Oleg Murk and David Mély and Ashvin Nair and Reiichiro Nakano and Rajeev Nayak and Arvind Neelakantan and Richard Ngo and Hyeonwoo Noh and Long Ouyang and Cullen O'Keefe and Jakub Pachocki and Alex Paino and Joe Palermo and Ashley Pantuliano and Giambattista Parascandolo and Joel Parish and Emy Parparita and Alex Passos and Mikhail Pavlov and Andrew Peng and Adam Perelman and Filipe de Avila Belbute Peres and Michael Petrov and Henrique Ponde de Oliveira Pinto and Michael and Pokorny and Michelle Pokrass and Vitchyr H. Pong and Tolly Powell and Alethea Power and Boris Power and Elizabeth Proehl and Raul Puri and Alec Radford and Jack Rae and Aditya Ramesh and Cameron Raymond and Francis Real and Kendra Rimbach and Carl Ross and Bob Rotsted and Henri Roussez and Nick Ryder and Mario Saltarelli and Ted Sanders and Shibani Santurkar and Girish Sastry and Heather Schmidt and David Schnurr and John Schulman and Daniel Selsam and Kyla Sheppard and Toki Sherbakov and Jessica Shieh and Sarah Shoker and Pranav Shyam and Szymon Sidor and Eric Sigler and Maddie Simens and Jordan Sitkin and Katarina Slama and Ian Sohl and Benjamin Sokolowsky and Yang Song and Natalie Staudacher and Felipe Petroski Such and Natalie Summers and Ilya Sutskever and Jie Tang and Nikolas Tezak and Madeleine B. Thompson and Phil Tillet and Amin Tootoonchian and Elizabeth Tseng and Preston Tuggle and Nick Turley and Jerry Tworek and Juan Felipe Cerón Uribe and Andrea Vallone and Arun Vijayvergiya and Chelsea Voss and Carroll Wainwright and Justin Jay Wang and Alvin Wang and Ben Wang and Jonathan Ward and Jason Wei and CJ Weinmann and Akila Welihinda and Peter Welinder and Jiayi Weng and Lilian Weng and Matt Wiethoff and Dave Willner and Clemens Winter and Samuel Wolrich and Hannah Wong and Lauren Workman and Sherwin Wu and Jeff Wu and Michael Wu and Kai Xiao and Tao Xu and Sarah Yoo and Kevin Yu and Qiming Yuan and Wojciech Zaremba and Rowan Zellers and Chong Zhang and Marvin Zhang and Shengjia Zhao and Tianhao Zheng and Juntang Zhuang and William Zhuk and Barret Zoph},
      year={2024},
      eprint={2303.08774},
      archivePrefix={arXiv},
      primaryClass={cs.CL},
      url={https://arxiv.org/abs/2303.08774}, 
}

@misc{geminiteam2024geminifamilyhighlycapable,
      title={Gemini: A Family of Highly Capable Multimodal Models}, 
      author={Gemini Team and Rohan Anil and Sebastian Borgeaud and Jean-Baptiste Alayrac and Jiahui Yu and Radu Soricut and Johan Schalkwyk and Andrew M. Dai and Anja Hauth and Katie Millican and David Silver and Melvin Johnson and Ioannis Antonoglou and Julian Schrittwieser and Amelia Glaese and Jilin Chen and Emily Pitler and Timothy Lillicrap and Angeliki Lazaridou and Orhan Firat and James Molloy and Michael Isard and Paul R. Barham and Tom Hennigan and Benjamin Lee and Fabio Viola and Malcolm Reynolds and Yuanzhong Xu and Ryan Doherty and Eli Collins and Clemens Meyer and Eliza Rutherford and Erica Moreira and Kareem Ayoub and Megha Goel and Jack Krawczyk and Cosmo Du and Ed Chi and Heng-Tze Cheng and Eric Ni and Purvi Shah and Patrick Kane and Betty Chan and Manaal Faruqui and Aliaksei Severyn and Hanzhao Lin and YaGuang Li and Yong Cheng and Abe Ittycheriah and Mahdis Mahdieh and Mia Chen and Pei Sun and Dustin Tran and Sumit Bagri and Balaji Lakshminarayanan and Jeremiah Liu and Andras Orban and Fabian Güra and Hao Zhou and Xinying Song and Aurelien Boffy and Harish Ganapathy and Steven Zheng and HyunJeong Choe and Ágoston Weisz and Tao Zhu and Yifeng Lu and Siddharth Gopal and Jarrod Kahn and Maciej Kula and Jeff Pitman and Rushin Shah and Emanuel Taropa and Majd Al Merey and Martin Baeuml and Zhifeng Chen and Laurent El Shafey and Yujing Zhang and Olcan Sercinoglu and George Tucker and Enrique Piqueras and Maxim Krikun and Iain Barr and Nikolay Savinov and Ivo Danihelka and Becca Roelofs and Anaïs White and Anders Andreassen and Tamara von Glehn and Lakshman Yagati and Mehran Kazemi and Lucas Gonzalez and Misha Khalman and Jakub Sygnowski and Alexandre Frechette and Charlotte Smith and Laura Culp and Lev Proleev and Yi Luan and Xi Chen and James Lottes and Nathan Schucher and Federico Lebron and Alban Rrustemi and Natalie Clay and Phil Crone and Tomas Kocisky and Jeffrey Zhao and Bartek Perz and Dian Yu and Heidi Howard and Adam Bloniarz and Jack W. Rae and Han Lu and Laurent Sifre and Marcello Maggioni and Fred Alcober and Dan Garrette and Megan Barnes and Shantanu Thakoor and Jacob Austin and Gabriel Barth-Maron and William Wong and Rishabh Joshi and Rahma Chaabouni and Deeni Fatiha and Arun Ahuja and Gaurav Singh Tomar and Evan Senter and Martin Chadwick and Ilya Kornakov and Nithya Attaluri and Iñaki Iturrate and Ruibo Liu and Yunxuan Li and Sarah Cogan and Jeremy Chen and Chao Jia and Chenjie Gu and Qiao Zhang and Jordan Grimstad and Ale Jakse Hartman and Xavier Garcia and Thanumalayan Sankaranarayana Pillai and Jacob Devlin and Michael Laskin and Diego de Las Casas and Dasha Valter and Connie Tao and Lorenzo Blanco and Adrià Puigdomènech Badia and David Reitter and Mianna Chen and Jenny Brennan and Clara Rivera and Sergey Brin and Shariq Iqbal and Gabriela Surita and Jane Labanowski and Abhi Rao and Stephanie Winkler and Emilio Parisotto and Yiming Gu and Kate Olszewska and Ravi Addanki and Antoine Miech and Annie Louis and Denis Teplyashin and Geoff Brown and Elliot Catt and Jan Balaguer and Jackie Xiang and Pidong Wang and Zoe Ashwood and Anton Briukhov and Albert Webson and Sanjay Ganapathy and Smit Sanghavi and Ajay Kannan and Ming-Wei Chang and Axel Stjerngren and Josip Djolonga and Yuting Sun and Ankur Bapna and Matthew Aitchison and Pedram Pejman and Henryk Michalewski and Tianhe Yu and Cindy Wang and Juliette Love and Junwhan Ahn and Dawn Bloxwich and Kehang Han and Peter Humphreys and Thibault Sellam and James Bradbury and Varun Godbole and Sina Samangooei and Bogdan Damoc and Alex Kaskasoli and Sébastien M. R. Arnold and Vijay Vasudevan and Shubham Agrawal and Jason Riesa and Dmitry Lepikhin and Richard Tanburn and Srivatsan Srinivasan and Hyeontaek Lim and Sarah Hodkinson and Pranav Shyam and Johan Ferret and Steven Hand and Ankush Garg and Tom Le Paine and Jian Li and Yujia Li and Minh Giang and Alexander Neitz and Zaheer Abbas and Sarah York and Machel Reid and Elizabeth Cole and Aakanksha Chowdhery and Dipanjan Das and Dominika Rogozińska and Vitaliy Nikolaev and Pablo Sprechmann and Zachary Nado and Lukas Zilka and Flavien Prost and Luheng He and Marianne Monteiro and Gaurav Mishra and Chris Welty and Josh Newlan and Dawei Jia and Miltiadis Allamanis and Clara Huiyi Hu and Raoul de Liedekerke and Justin Gilmer and Carl Saroufim and Shruti Rijhwani and Shaobo Hou and Disha Shrivastava and Anirudh Baddepudi and Alex Goldin and Adnan Ozturel and Albin Cassirer and Yunhan Xu and Daniel Sohn and Devendra Sachan and Reinald Kim Amplayo and Craig Swanson and Dessie Petrova and Shashi Narayan and Arthur Guez and Siddhartha Brahma and Jessica Landon and Miteyan Patel and Ruizhe Zhao and Kevin Villela and Luyu Wang and Wenhao Jia and Matthew Rahtz and Mai Giménez and Legg Yeung and James Keeling and Petko Georgiev and Diana Mincu and Boxi Wu and Salem Haykal and Rachel Saputro and Kiran Vodrahalli and James Qin and Zeynep Cankara and Abhanshu Sharma and Nick Fernando and Will Hawkins and Behnam Neyshabur and Solomon Kim and Adrian Hutter and Priyanka Agrawal and Alex Castro-Ros and George van den Driessche and Tao Wang and Fan Yang and Shuo-yiin Chang and Paul Komarek and Ross McIlroy and Mario Lučić and Guodong Zhang and Wael Farhan and Michael Sharman and Paul Natsev and Paul Michel and Yamini Bansal and Siyuan Qiao and Kris Cao and Siamak Shakeri and Christina Butterfield and Justin Chung and Paul Kishan Rubenstein and Shivani Agrawal and Arthur Mensch and Kedar Soparkar and Karel Lenc and Timothy Chung and Aedan Pope and Loren Maggiore and Jackie Kay and Priya Jhakra and Shibo Wang and Joshua Maynez and Mary Phuong and Taylor Tobin and Andrea Tacchetti and Maja Trebacz and Kevin Robinson and Yash Katariya and Sebastian Riedel and Paige Bailey and Kefan Xiao and Nimesh Ghelani and Lora Aroyo and Ambrose Slone and Neil Houlsby and Xuehan Xiong and Zhen Yang and Elena Gribovskaya and Jonas Adler and Mateo Wirth and Lisa Lee and Music Li and Thais Kagohara and Jay Pavagadhi and Sophie Bridgers and Anna Bortsova and Sanjay Ghemawat and Zafarali Ahmed and Tianqi Liu and Richard Powell and Vijay Bolina and Mariko Iinuma and Polina Zablotskaia and James Besley and Da-Woon Chung and Timothy Dozat and Ramona Comanescu and Xiance Si and Jeremy Greer and Guolong Su and Martin Polacek and Raphaël Lopez Kaufman and Simon Tokumine and Hexiang Hu and Elena Buchatskaya and Yingjie Miao and Mohamed Elhawaty and Aditya Siddhant and Nenad Tomasev and Jinwei Xing and Christina Greer and Helen Miller and Shereen Ashraf and Aurko Roy and Zizhao Zhang and Ada Ma and Angelos Filos and Milos Besta and Rory Blevins and Ted Klimenko and Chih-Kuan Yeh and Soravit Changpinyo and Jiaqi Mu and Oscar Chang and Mantas Pajarskas and Carrie Muir and Vered Cohen and Charline Le Lan and Krishna Haridasan and Amit Marathe and Steven Hansen and Sholto Douglas and Rajkumar Samuel and Mingqiu Wang and Sophia Austin and Chang Lan and Jiepu Jiang and Justin Chiu and Jaime Alonso Lorenzo and Lars Lowe Sjösund and Sébastien Cevey and Zach Gleicher and Thi Avrahami and Anudhyan Boral and Hansa Srinivasan and Vittorio Selo and Rhys May and Konstantinos Aisopos and Léonard Hussenot and Livio Baldini Soares and Kate Baumli and Michael B. Chang and Adrià Recasens and Ben Caine and Alexander Pritzel and Filip Pavetic and Fabio Pardo and Anita Gergely and Justin Frye and Vinay Ramasesh and Dan Horgan and Kartikeya Badola and Nora Kassner and Subhrajit Roy and Ethan Dyer and Víctor Campos Campos and Alex Tomala and Yunhao Tang and Dalia El Badawy and Elspeth White and Basil Mustafa and Oran Lang and Abhishek Jindal and Sharad Vikram and Zhitao Gong and Sergi Caelles and Ross Hemsley and Gregory Thornton and Fangxiaoyu Feng and Wojciech Stokowiec and Ce Zheng and Phoebe Thacker and Çağlar Ünlü and Zhishuai Zhang and Mohammad Saleh and James Svensson and Max Bileschi and Piyush Patil and Ankesh Anand and Roman Ring and Katerina Tsihlas and Arpi Vezer and Marco Selvi and Toby Shevlane and Mikel Rodriguez and Tom Kwiatkowski and Samira Daruki and Keran Rong and Allan Dafoe and Nicholas FitzGerald and Keren Gu-Lemberg and Mina Khan and Lisa Anne Hendricks and Marie Pellat and Vladimir Feinberg and James Cobon-Kerr and Tara Sainath and Maribeth Rauh and Sayed Hadi Hashemi and Richard Ives and Yana Hasson and Eric Noland and Yuan Cao and Nathan Byrd and Le Hou and Qingze Wang and Thibault Sottiaux and Michela Paganini and Jean-Baptiste Lespiau and Alexandre Moufarek and Samer Hassan and Kaushik Shivakumar and Joost van Amersfoort and Amol Mandhane and Pratik Joshi and Anirudh Goyal and Matthew Tung and Andrew Brock and Hannah Sheahan and Vedant Misra and Cheng Li and Nemanja Rakićević and Mostafa Dehghani and Fangyu Liu and Sid Mittal and Junhyuk Oh and Seb Noury and Eren Sezener and Fantine Huot and Matthew Lamm and Nicola De Cao and Charlie Chen and Sidharth Mudgal and Romina Stella and Kevin Brooks and Gautam Vasudevan and Chenxi Liu and Mainak Chain and Nivedita Melinkeri and Aaron Cohen and Venus Wang and Kristie Seymore and Sergey Zubkov and Rahul Goel and Summer Yue and Sai Krishnakumaran and Brian Albert and Nate Hurley and Motoki Sano and Anhad Mohananey and Jonah Joughin and Egor Filonov and Tomasz Kępa and Yomna Eldawy and Jiawern Lim and Rahul Rishi and Shirin Badiezadegan and Taylor Bos and Jerry Chang and Sanil Jain and Sri Gayatri Sundara Padmanabhan and Subha Puttagunta and Kalpesh Krishna and Leslie Baker and Norbert Kalb and Vamsi Bedapudi and Adam Kurzrok and Shuntong Lei and Anthony Yu and Oren Litvin and Xiang Zhou and Zhichun Wu and Sam Sobell and Andrea Siciliano and Alan Papir and Robby Neale and Jonas Bragagnolo and Tej Toor and Tina Chen and Valentin Anklin and Feiran Wang and Richie Feng and Milad Gholami and Kevin Ling and Lijuan Liu and Jules Walter and Hamid Moghaddam and Arun Kishore and Jakub Adamek and Tyler Mercado and Jonathan Mallinson and Siddhinita Wandekar and Stephen Cagle and Eran Ofek and Guillermo Garrido and Clemens Lombriser and Maksim Mukha and Botu Sun and Hafeezul Rahman Mohammad and Josip Matak and Yadi Qian and Vikas Peswani and Pawel Janus and Quan Yuan and Leif Schelin and Oana David and Ankur Garg and Yifan He and Oleksii Duzhyi and Anton Älgmyr and Timothée Lottaz and Qi Li and Vikas Yadav and Luyao Xu and Alex Chinien and Rakesh Shivanna and Aleksandr Chuklin and Josie Li and Carrie Spadine and Travis Wolfe and Kareem Mohamed and Subhabrata Das and Zihang Dai and Kyle He and Daniel von Dincklage and Shyam Upadhyay and Akanksha Maurya and Luyan Chi and Sebastian Krause and Khalid Salama and Pam G Rabinovitch and Pavan Kumar Reddy M and Aarush Selvan and Mikhail Dektiarev and Golnaz Ghiasi and Erdem Guven and Himanshu Gupta and Boyi Liu and Deepak Sharma and Idan Heimlich Shtacher and Shachi Paul and Oscar Akerlund and François-Xavier Aubet and Terry Huang and Chen Zhu and Eric Zhu and Elico Teixeira and Matthew Fritze and Francesco Bertolini and Liana-Eleonora Marinescu and Martin Bölle and Dominik Paulus and Khyatti Gupta and Tejasi Latkar and Max Chang and Jason Sanders and Roopa Wilson and Xuewei Wu and Yi-Xuan Tan and Lam Nguyen Thiet and Tulsee Doshi and Sid Lall and Swaroop Mishra and Wanming Chen and Thang Luong and Seth Benjamin and Jasmine Lee and Ewa Andrejczuk and Dominik Rabiej and Vipul Ranjan and Krzysztof Styrc and Pengcheng Yin and Jon Simon and Malcolm Rose Harriott and Mudit Bansal and Alexei Robsky and Geoff Bacon and David Greene and Daniil Mirylenka and Chen Zhou and Obaid Sarvana and Abhimanyu Goyal and Samuel Andermatt and Patrick Siegler and Ben Horn and Assaf Israel and Francesco Pongetti and Chih-Wei "Louis" Chen and Marco Selvatici and Pedro Silva and Kathie Wang and Jackson Tolins and Kelvin Guu and Roey Yogev and Xiaochen Cai and Alessandro Agostini and Maulik Shah and Hung Nguyen and Noah Ó Donnaile and Sébastien Pereira and Linda Friso and Adam Stambler and Adam Kurzrok and Chenkai Kuang and Yan Romanikhin and Mark Geller and ZJ Yan and Kane Jang and Cheng-Chun Lee and Wojciech Fica and Eric Malmi and Qijun Tan and Dan Banica and Daniel Balle and Ryan Pham and Yanping Huang and Diana Avram and Hongzhi Shi and Jasjot Singh and Chris Hidey and Niharika Ahuja and Pranab Saxena and Dan Dooley and Srividya Pranavi Potharaju and Eileen O'Neill and Anand Gokulchandran and Ryan Foley and Kai Zhao and Mike Dusenberry and Yuan Liu and Pulkit Mehta and Ragha Kotikalapudi and Chalence Safranek-Shrader and Andrew Goodman and Joshua Kessinger and Eran Globen and Prateek Kolhar and Chris Gorgolewski and Ali Ibrahim and Yang Song and Ali Eichenbaum and Thomas Brovelli and Sahitya Potluri and Preethi Lahoti and Cip Baetu and Ali Ghorbani and Charles Chen and Andy Crawford and Shalini Pal and Mukund Sridhar and Petru Gurita and Asier Mujika and Igor Petrovski and Pierre-Louis Cedoz and Chenmei Li and Shiyuan Chen and Niccolò Dal Santo and Siddharth Goyal and Jitesh Punjabi and Karthik Kappaganthu and Chester Kwak and Pallavi LV and Sarmishta Velury and Himadri Choudhury and Jamie Hall and Premal Shah and Ricardo Figueira and Matt Thomas and Minjie Lu and Ting Zhou and Chintu Kumar and Thomas Jurdi and Sharat Chikkerur and Yenai Ma and Adams Yu and Soo Kwak and Victor Ähdel and Sujeevan Rajayogam and Travis Choma and Fei Liu and Aditya Barua and Colin Ji and Ji Ho Park and Vincent Hellendoorn and Alex Bailey and Taylan Bilal and Huanjie Zhou and Mehrdad Khatir and Charles Sutton and Wojciech Rzadkowski and Fiona Macintosh and Konstantin Shagin and Paul Medina and Chen Liang and Jinjing Zhou and Pararth Shah and Yingying Bi and Attila Dankovics and Shipra Banga and Sabine Lehmann and Marissa Bredesen and Zifan Lin and John Eric Hoffmann and Jonathan Lai and Raynald Chung and Kai Yang and Nihal Balani and Arthur Bražinskas and Andrei Sozanschi and Matthew Hayes and Héctor Fernández Alcalde and Peter Makarov and Will Chen and Antonio Stella and Liselotte Snijders and Michael Mandl and Ante Kärrman and Paweł Nowak and Xinyi Wu and Alex Dyck and Krishnan Vaidyanathan and Raghavender R and Jessica Mallet and Mitch Rudominer and Eric Johnston and Sushil Mittal and Akhil Udathu and Janara Christensen and Vishal Verma and Zach Irving and Andreas Santucci and Gamaleldin Elsayed and Elnaz Davoodi and Marin Georgiev and Ian Tenney and Nan Hua and Geoffrey Cideron and Edouard Leurent and Mahmoud Alnahlawi and Ionut Georgescu and Nan Wei and Ivy Zheng and Dylan Scandinaro and Heinrich Jiang and Jasper Snoek and Mukund Sundararajan and Xuezhi Wang and Zack Ontiveros and Itay Karo and Jeremy Cole and Vinu Rajashekhar and Lara Tumeh and Eyal Ben-David and Rishub Jain and Jonathan Uesato and Romina Datta and Oskar Bunyan and Shimu Wu and John Zhang and Piotr Stanczyk and Ye Zhang and David Steiner and Subhajit Naskar and Michael Azzam and Matthew Johnson and Adam Paszke and Chung-Cheng Chiu and Jaume Sanchez Elias and Afroz Mohiuddin and Faizan Muhammad and Jin Miao and Andrew Lee and Nino Vieillard and Jane Park and Jiageng Zhang and Jeff Stanway and Drew Garmon and Abhijit Karmarkar and Zhe Dong and Jong Lee and Aviral Kumar and Luowei Zhou and Jonathan Evens and William Isaac and Geoffrey Irving and Edward Loper and Michael Fink and Isha Arkatkar and Nanxin Chen and Izhak Shafran and Ivan Petrychenko and Zhe Chen and Johnson Jia and Anselm Levskaya and Zhenkai Zhu and Peter Grabowski and Yu Mao and Alberto Magni and Kaisheng Yao and Javier Snaider and Norman Casagrande and Evan Palmer and Paul Suganthan and Alfonso Castaño and Irene Giannoumis and Wooyeol Kim and Mikołaj Rybiński and Ashwin Sreevatsa and Jennifer Prendki and David Soergel and Adrian Goedeckemeyer and Willi Gierke and Mohsen Jafari and Meenu Gaba and Jeremy Wiesner and Diana Gage Wright and Yawen Wei and Harsha Vashisht and Yana Kulizhskaya and Jay Hoover and Maigo Le and Lu Li and Chimezie Iwuanyanwu and Lu Liu and Kevin Ramirez and Andrey Khorlin and Albert Cui and Tian LIN and Marcus Wu and Ricardo Aguilar and Keith Pallo and Abhishek Chakladar and Ginger Perng and Elena Allica Abellan and Mingyang Zhang and Ishita Dasgupta and Nate Kushman and Ivo Penchev and Alena Repina and Xihui Wu and Tom van der Weide and Priya Ponnapalli and Caroline Kaplan and Jiri Simsa and Shuangfeng Li and Olivier Dousse and Fan Yang and Jeff Piper and Nathan Ie and Rama Pasumarthi and Nathan Lintz and Anitha Vijayakumar and Daniel Andor and Pedro Valenzuela and Minnie Lui and Cosmin Paduraru and Daiyi Peng and Katherine Lee and Shuyuan Zhang and Somer Greene and Duc Dung Nguyen and Paula Kurylowicz and Cassidy Hardin and Lucas Dixon and Lili Janzer and Kiam Choo and Ziqiang Feng and Biao Zhang and Achintya Singhal and Dayou Du and Dan McKinnon and Natasha Antropova and Tolga Bolukbasi and Orgad Keller and David Reid and Daniel Finchelstein and Maria Abi Raad and Remi Crocker and Peter Hawkins and Robert Dadashi and Colin Gaffney and Ken Franko and Anna Bulanova and Rémi Leblond and Shirley Chung and Harry Askham and Luis C. Cobo and Kelvin Xu and Felix Fischer and Jun Xu and Christina Sorokin and Chris Alberti and Chu-Cheng Lin and Colin Evans and Alek Dimitriev and Hannah Forbes and Dylan Banarse and Zora Tung and Mark Omernick and Colton Bishop and Rachel Sterneck and Rohan Jain and Jiawei Xia and Ehsan Amid and Francesco Piccinno and Xingyu Wang and Praseem Banzal and Daniel J. Mankowitz and Alex Polozov and Victoria Krakovna and Sasha Brown and MohammadHossein Bateni and Dennis Duan and Vlad Firoiu and Meghana Thotakuri and Tom Natan and Matthieu Geist and Ser tan Girgin and Hui Li and Jiayu Ye and Ofir Roval and Reiko Tojo and Michael Kwong and James Lee-Thorp and Christopher Yew and Danila Sinopalnikov and Sabela Ramos and John Mellor and Abhishek Sharma and Kathy Wu and David Miller and Nicolas Sonnerat and Denis Vnukov and Rory Greig and Jennifer Beattie and Emily Caveness and Libin Bai and Julian Eisenschlos and Alex Korchemniy and Tomy Tsai and Mimi Jasarevic and Weize Kong and Phuong Dao and Zeyu Zheng and Frederick Liu and Fan Yang and Rui Zhu and Tian Huey Teh and Jason Sanmiya and Evgeny Gladchenko and Nejc Trdin and Daniel Toyama and Evan Rosen and Sasan Tavakkol and Linting Xue and Chen Elkind and Oliver Woodman and John Carpenter and George Papamakarios and Rupert Kemp and Sushant Kafle and Tanya Grunina and Rishika Sinha and Alice Talbert and Diane Wu and Denese Owusu-Afriyie and Cosmo Du and Chloe Thornton and Jordi Pont-Tuset and Pradyumna Narayana and Jing Li and Saaber Fatehi and John Wieting and Omar Ajmeri and Benigno Uria and Yeongil Ko and Laura Knight and Amélie Héliou and Ning Niu and Shane Gu and Chenxi Pang and Yeqing Li and Nir Levine and Ariel Stolovich and Rebeca Santamaria-Fernandez and Sonam Goenka and Wenny Yustalim and Robin Strudel and Ali Elqursh and Charlie Deck and Hyo Lee and Zonglin Li and Kyle Levin and Raphael Hoffmann and Dan Holtmann-Rice and Olivier Bachem and Sho Arora and Christy Koh and Soheil Hassas Yeganeh and Siim Põder and Mukarram Tariq and Yanhua Sun and Lucian Ionita and Mojtaba Seyedhosseini and Pouya Tafti and Zhiyu Liu and Anmol Gulati and Jasmine Liu and Xinyu Ye and Bart Chrzaszcz and Lily Wang and Nikhil Sethi and Tianrun Li and Ben Brown and Shreya Singh and Wei Fan and Aaron Parisi and Joe Stanton and Vinod Koverkathu and Christopher A. Choquette-Choo and Yunjie Li and TJ Lu and Abe Ittycheriah and Prakash Shroff and Mani Varadarajan and Sanaz Bahargam and Rob Willoughby and David Gaddy and Guillaume Desjardins and Marco Cornero and Brona Robenek and Bhavishya Mittal and Ben Albrecht and Ashish Shenoy and Fedor Moiseev and Henrik Jacobsson and Alireza Ghaffarkhah and Morgane Rivière and Alanna Walton and Clément Crepy and Alicia Parrish and Zongwei Zhou and Clement Farabet and Carey Radebaugh and Praveen Srinivasan and Claudia van der Salm and Andreas Fidjeland and Salvatore Scellato and Eri Latorre-Chimoto and Hanna Klimczak-Plucińska and David Bridson and Dario de Cesare and Tom Hudson and Piermaria Mendolicchio and Lexi Walker and Alex Morris and Matthew Mauger and Alexey Guseynov and Alison Reid and Seth Odoom and Lucia Loher and Victor Cotruta and Madhavi Yenugula and Dominik Grewe and Anastasia Petrushkina and Tom Duerig and Antonio Sanchez and Steve Yadlowsky and Amy Shen and Amir Globerson and Lynette Webb and Sahil Dua and Dong Li and Surya Bhupatiraju and Dan Hurt and Haroon Qureshi and Ananth Agarwal and Tomer Shani and Matan Eyal and Anuj Khare and Shreyas Rammohan Belle and Lei Wang and Chetan Tekur and Mihir Sanjay Kale and Jinliang Wei and Ruoxin Sang and Brennan Saeta and Tyler Liechty and Yi Sun and Yao Zhao and Stephan Lee and Pandu Nayak and Doug Fritz and Manish Reddy Vuyyuru and John Aslanides and Nidhi Vyas and Martin Wicke and Xiao Ma and Evgenii Eltyshev and Nina Martin and Hardie Cate and James Manyika and Keyvan Amiri and Yelin Kim and Xi Xiong and Kai Kang and Florian Luisier and Nilesh Tripuraneni and David Madras and Mandy Guo and Austin Waters and Oliver Wang and Joshua Ainslie and Jason Baldridge and Han Zhang and Garima Pruthi and Jakob Bauer and Feng Yang and Riham Mansour and Jason Gelman and Yang Xu and George Polovets and Ji Liu and Honglong Cai and Warren Chen and XiangHai Sheng and Emily Xue and Sherjil Ozair and Christof Angermueller and Xiaowei Li and Anoop Sinha and Weiren Wang and Julia Wiesinger and Emmanouil Koukoumidis and Yuan Tian and Anand Iyer and Madhu Gurumurthy and Mark Goldenson and Parashar Shah and MK Blake and Hongkun Yu and Anthony Urbanowicz and Jennimaria Palomaki and Chrisantha Fernando and Ken Durden and Harsh Mehta and Nikola Momchev and Elahe Rahimtoroghi and Maria Georgaki and Amit Raul and Sebastian Ruder and Morgan Redshaw and Jinhyuk Lee and Denny Zhou and Komal Jalan and Dinghua Li and Blake Hechtman and Parker Schuh and Milad Nasr and Kieran Milan and Vladimir Mikulik and Juliana Franco and Tim Green and Nam Nguyen and Joe Kelley and Aroma Mahendru and Andrea Hu and Joshua Howland and Ben Vargas and Jeffrey Hui and Kshitij Bansal and Vikram Rao and Rakesh Ghiya and Emma Wang and Ke Ye and Jean Michel Sarr and Melanie Moranski Preston and Madeleine Elish and Steve Li and Aakash Kaku and Jigar Gupta and Ice Pasupat and Da-Cheng Juan and Milan Someswar and Tejvi M. and Xinyun Chen and Aida Amini and Alex Fabrikant and Eric Chu and Xuanyi Dong and Amruta Muthal and Senaka Buthpitiya and Sarthak Jauhari and Nan Hua and Urvashi Khandelwal and Ayal Hitron and Jie Ren and Larissa Rinaldi and Shahar Drath and Avigail Dabush and Nan-Jiang Jiang and Harshal Godhia and Uli Sachs and Anthony Chen and Yicheng Fan and Hagai Taitelbaum and Hila Noga and Zhuyun Dai and James Wang and Chen Liang and Jenny Hamer and Chun-Sung Ferng and Chenel Elkind and Aviel Atias and Paulina Lee and Vít Listík and Mathias Carlen and Jan van de Kerkhof and Marcin Pikus and Krunoslav Zaher and Paul Müller and Sasha Zykova and Richard Stefanec and Vitaly Gatsko and Christoph Hirnschall and Ashwin Sethi and Xingyu Federico Xu and Chetan Ahuja and Beth Tsai and Anca Stefanoiu and Bo Feng and Keshav Dhandhania and Manish Katyal and Akshay Gupta and Atharva Parulekar and Divya Pitta and Jing Zhao and Vivaan Bhatia and Yashodha Bhavnani and Omar Alhadlaq and Xiaolin Li and Peter Danenberg and Dennis Tu and Alex Pine and Vera Filippova and Abhipso Ghosh and Ben Limonchik and Bhargava Urala and Chaitanya Krishna Lanka and Derik Clive and Yi Sun and Edward Li and Hao Wu and Kevin Hongtongsak and Ianna Li and Kalind Thakkar and Kuanysh Omarov and Kushal Majmundar and Michael Alverson and Michael Kucharski and Mohak Patel and Mudit Jain and Maksim Zabelin and Paolo Pelagatti and Rohan Kohli and Saurabh Kumar and Joseph Kim and Swetha Sankar and Vineet Shah and Lakshmi Ramachandruni and Xiangkai Zeng and Ben Bariach and Laura Weidinger and Tu Vu and Alek Andreev and Antoine He and Kevin Hui and Sheleem Kashem and Amar Subramanya and Sissie Hsiao and Demis Hassabis and Koray Kavukcuoglu and Adam Sadovsky and Quoc Le and Trevor Strohman and Yonghui Wu and Slav Petrov and Jeffrey Dean and Oriol Vinyals},
      year={2024},
      eprint={2312.11805},
      archivePrefix={arXiv},
      primaryClass={cs.CL},
      url={https://arxiv.org/abs/2312.11805}, 
}

@article{benidis_deep_2023,
	title = {Deep {Learning} for {Time} {Series} {Forecasting}: {Tutorial} and {Literature} {Survey}},
	volume = {55},
	issn = {0360-0300, 1557-7341},
	shorttitle = {Deep {Learning} for {Time} {Series} {Forecasting}},
	url = {https://dl.acm.org/doi/10.1145/3533382},
	doi = {10.1145/3533382},
	language = {en},
	number = {6},
	urldate = {2023-03-29},
	journal = {ACM Computing Surveys},
	author = {Benidis, Konstantinos and Rangapuram, Syama Sundar and Flunkert, Valentin and Wang, Yuyang and Maddix, Danielle and Turkmen, Caner and Gasthaus, Jan and Bohlke-Schneider, Michael and Salinas, David and Stella, Lorenzo and Aubet, François-Xavier and Callot, Laurent and Januschowski, Tim},
	month = jul,
	year = {2023},
	pages = {1--36},
}

@misc{salinas2019deeparprobabilisticforecastingautoregressive,
      title={DeepAR: Probabilistic Forecasting with Autoregressive Recurrent Networks}, 
      author={David Salinas and Valentin Flunkert and Jan Gasthaus},
      year={2019},
      eprint={1704.04110},
      archivePrefix={arXiv},
      primaryClass={cs.AI},
      url={https://arxiv.org/abs/1704.04110}, 
}

@misc{lim2020temporalfusiontransformersinterpretable,
      title={Temporal Fusion Transformers for Interpretable Multi-horizon Time Series Forecasting}, 
      author={Bryan Lim and Sercan O. Arik and Nicolas Loeff and Tomas Pfister},
      year={2020},
      eprint={1912.09363},
      archivePrefix={arXiv},
      primaryClass={stat.ML},
      url={https://arxiv.org/abs/1912.09363}, 
}

@article{DBLP:journals/corr/abs-2009-14799,
  author       = {Carson Eisenach and
                  Yagna Patel and
                  Dhruv Madeka},
  title        = {MQTransformer: Multi-Horizon Forecasts with Context Dependent and
                  Feedback-Aware Attention},
  journal      = {CoRR},
  volume       = {abs/2009.14799},
  year         = {2020},
  url          = {https://arxiv.org/abs/2009.14799},
  eprinttype    = {arXiv},
  eprint       = {2009.14799},
  bibsource    = {dblp computer science bibliography, https://dblp.org}
}

@misc{zhou2021informerefficienttransformerlong,
      title={Informer: Beyond Efficient Transformer for Long Sequence Time-Series Forecasting}, 
      author={Haoyi Zhou and Shanghang Zhang and Jieqi Peng and Shuai Zhang and Jianxin Li and Hui Xiong and Wancai Zhang},
      year={2021},
      eprint={2012.07436},
      archivePrefix={arXiv},
      primaryClass={cs.LG},
      url={https://arxiv.org/abs/2012.07436}, 
}

@ARTICLE{Barez_2023,title={Exploring the Advantages of Transformers for High-Frequency Trading},year={2023},author={Fazl Barez and Paul Bilokon and Arthur Gervais and Nikita Lisitsyn},doi={10.2139/ssrn.4364833},pmid={null},pmcid={null},mag_id={4321766651},journal={Social Science Research Network}}

@ARTICLE{Zhao_2022,title={Attention! Transformer with Sentiment on Cryptocurrencies Price Prediction},year={2022},author={Huali Zhao and Martin Crane and Marija Bezbradica},doi={10.5220/0011103400003197},pmid={null},pmcid={null},mag_id={4226299725},journal={International Conference on Complex Information Systems}}

@ARTICLE{Hu_2021,title={Stock Price Prediction Based on Temporal Fusion Transformer},year={2021},author={Xiaokang Hu and Xiaokang Hu},doi={10.1109/mlbdbi54094.2021.00019},pmid={null},pmcid={null},mag_id={4226270189},journal={2021 3rd International Conference on Machine Learning, Big Data and Business Intelligence (MLBDBI)}}

@ARTICLE{Hájek_2024,title={Beyond Sentiment in Stock Price Prediction: Integrating News Sentiment and Investor Attention with Temporal Fusion Transformer},year={2024},author={Petr Hájek and Josef Novotny},doi={10.1007/978-3-031-63219-8_3},pmid={null},pmcid={null},mag_id={null},journal={Artificial Intelligence Applications and Innovations}}

@ARTICLE{Penmetsa_2023,title={Cryptocurrency Price Prediction with LSTM and Transformer Models Leveraging Momentum and Volatility Technical Indicators},year={2023},author={Siddharth Penmetsa and Maruthi Vemula},doi={10.1109/icdsca59871.2023.10393319},pmid={null},pmcid={null},mag_id={null},journal={2023 IEEE 3rd International Conference on Data Science and Computer Application (ICDSCA)}}

@ARTICLE{Wang_2023,title={A Stock Price Prediction Method Based on BiLSTM and Improved Transformer},year={2023},author={S. Wang},doi={10.1109/access.2023.3296308},pmid={null},pmcid={null},mag_id={4384518871},journal={IEEE Access}}

@INPROCEEDINGS{10280785,
  author={Ding, Ying and Zhang, Changsheng and Zhang, Chen},
  booktitle={2023 19th International Conference on Natural Computation, Fuzzy Systems and Knowledge Discovery (ICNC-FSKD)}, 
  title={An Informer Based Method for Stock Intraday Price Prediction}, 
  year={2023},
  volume={},
  number={},
  pages={1-6},
  doi={10.1109/ICNC-FSKD59587.2023.10280785}}

@ARTICLE{Duan_2024,title={Advanced Stock Price Prediction Using LSTM and Informer Models},year={2024},author={Chuyang Duan and Wenjun Ke},doi={10.60087/jaigs.v5i1.183},pmid={null},pmcid={null},mag_id={null},journal={Journal of Artificial Intelligence General science (JAIGS) ISSN:3006-4023}}

@misc{lu2023stockmarketindexprediction,
      title={Stock and market index prediction using Informer network}, 
      author={Yuze Lu and Hailong Zhang and Qiwen Guo},
      year={2023},
      eprint={2305.14382},
      archivePrefix={arXiv},
      primaryClass={q-fin.ST},
      url={https://arxiv.org/abs/2305.14382}, 
}

@article{DBLP:journals/corr/abs-1810-04805,
  author       = {Jacob Devlin and
                  Ming{-}Wei Chang and
                  Kenton Lee and
                  Kristina Toutanova},
  title        = {{BERT:} Pre-training of Deep Bidirectional Transformers for Language
                  Understanding},
  journal      = {CoRR},
  volume       = {abs/1810.04805},
  year         = {2018},
  url          = {http://arxiv.org/abs/1810.04805},
  eprinttype    = {arXiv},
  eprint       = {1810.04805},
  bibsource    = {dblp computer science bibliography, https://dblp.org}
}

@inbook{Sharpe+1998+169+178,
url = {https://doi.org/10.1515/9781400829408-022},
title = {The Sharpe Ratio (Fall 1994)},
booktitle = {Streetwise},
booktitle = {The Best of The Journal of Portfolio Management},
author = {William F. Sharpe},
editor = {Peter L. Bernstein and Frank J. Fabozzi},
publisher = {Princeton University Press},
address = {Princeton},
pages = {169--178},
doi = {doi:10.1515/9781400829408-022},
isbn = {9781400829408},
year = {1998},
lastchecked = {2024-10-10}
}

@article{RePEc:war:wpaper:2018-09,
title = {Momentum and contrarian effects on the cryptocurrency market},
journal = {Physica A: Statistical Mechanics and its Applications},
volume = {523},
pages = {691-701},
year = {2019},
issn = {0378-4371},
doi = {https://doi.org/10.1016/j.physa.2019.02.057},
url = {https://www.sciencedirect.com/science/article/pii/S037843711930216X},
author = {Krzysztof Kosc and Paweł Sakowski and Robert Ślepaczuk}
}

@book{10.5555/1408581,
author = {Appel, Gerald},
title = {Technical analysis: power tools for active investors},
year = {2005},
isbn = {0131479024},
publisher = {FT Press},
edition = {First}
}

@book{wilder1978new,
  title={New Concepts in Technical Trading Systems},
  author={Wilder, J.W.},
  isbn={9780894590276},
  lccn={78060759},
  url={https://books.google.pl/books?id=WesJAQAAMAAJ},
  year={1978},
  publisher={Trend Research}
}

@inproceedings{chollet2017,
author = {Chollet, Francois},
year = {2017},
month = {07},
pages = {1800-1807},
title = {Xception: Deep Learning with Depthwise Separable Convolutions},
doi = {10.1109/CVPR.2017.195}
}

@INBOOK{5264952,
  author={Kolen, John F. and Kremer, Stefan C.},
  booktitle={A Field Guide to Dynamical Recurrent Networks}, 
  title={Gradient Flow in Recurrent Nets: The Difficulty of Learning LongTerm Dependencies}, 
  year={2001},
  volume={},
  number={},
  pages={237-243},
  doi={10.1109/9780470544037.ch14}}

@misc{ba2016layernormalization,
      title={Layer Normalization}, 
      author={Jimmy Lei Ba and Jamie Ryan Kiros and Geoffrey E. Hinton},
      year={2016},
      eprint={1607.06450},
      archivePrefix={arXiv},
      primaryClass={stat.ML},
      url={https://arxiv.org/abs/1607.06450}, 
}

@misc{kiranyaz20191dconvolutionalneuralnetworks,
      title={1D Convolutional Neural Networks and Applications: A Survey}, 
      author={Serkan Kiranyaz and Onur Avci and Osama Abdeljaber and Turker Ince and Moncef Gabbouj and Daniel J. Inman},
      year={2019},
      eprint={1905.03554},
      archivePrefix={arXiv},
      primaryClass={eess.SP},
      url={https://arxiv.org/abs/1905.03554}, 
}

@misc{clevert2016fastaccuratedeepnetwork,
      title={Fast and Accurate Deep Network Learning by Exponential Linear Units (ELUs)}, 
      author={Djork-Arné Clevert and Thomas Unterthiner and Sepp Hochreiter},
      year={2016},
      eprint={1511.07289},
      archivePrefix={arXiv},
      primaryClass={cs.LG},
      url={https://arxiv.org/abs/1511.07289}, 
}

@INPROCEEDINGS{6144164,
  author={Nagi, Jawad and Ducatelle, Frederick and Di Caro, Gianni A. and Cireşan, Dan and Meier, Ueli and Giusti, Alessandro and Nagi, Farrukh and Schmidhuber, Jürgen and Gambardella, Luca Maria},
  booktitle={2011 IEEE International Conference on Signal and Image Processing Applications (ICSIPA)}, 
  title={Max-pooling convolutional neural networks for vision-based hand gesture recognition}, 
  year={2011},
  volume={},
  number={},
  pages={342-347},
  doi={10.1109/ICSIPA.2011.6144164}}

@misc{agarap2019deeplearningusingrectified,
      title={Deep Learning using Rectified Linear Units (ReLU)}, 
      author={Abien Fred Agarap},
      year={2019},
      eprint={1803.08375},
      archivePrefix={arXiv},
      primaryClass={cs.NE},
      url={https://arxiv.org/abs/1803.08375}, 
}

@article{michankow2024102375,
title = {Mean Absolute Directional Loss as a new loss function for machine learning problems in algorithmic investment strategies},
journal = {Journal of Computational Science},
volume = {81},
pages = {102375},
year = {2024},
issn = {1877-7503},
doi = {https://doi.org/10.1016/j.jocs.2024.102375},
url = {https://www.sciencedirect.com/science/article/pii/S1877750324001686},
author = {Jakub Michańków and Paweł Sakowski and Robert Ślepaczuk}
}

@article{Yadav2015,
author = {Yadav, Y.},
year = {2015},
month = {11},
pages = {1607-1671},
title = {How algorithmic trading undermines efficiency in capital markets},
volume = {68}
}

@article{ctx69563056230003681,
volume = {5},
author = {Vaserstein, L.N.},
year = {1969},
journal = {Probl. Peredachi Inf.},
title = {Markov processes over denumerable products of spaces, describing large systems of automata},
}

@book{bollinger2002bollinger,
  title={Bollinger on Bollinger Bands},
  author={Bollinger, J.},
  isbn={9780071373685},
  lccn={2001030666},
  url={https://books.google.pl/books?id=FLlxAz85iysC},
  year={2002},
  publisher={McGraw-Hill Education}
}

@article{nguyen2022,
author = {Võ, Thi Ái Nguyên and Ślepaczuk, Robert},
year = {2022},
month = {01},
pages = {158},
title = {Applying Hybrid ARIMA-SGARCH in Algorithmic Investment Strategies on S\&P500 Index},
volume = {24},
journal = {Entropy},
doi = {10.3390/e24020158}
}
\end{document}